\documentclass[aps,prb,twocolumn,10pt]{revtex4-2} 
\usepackage{subfiles}

\usepackage{amssymb,amsmath}
\usepackage{graphicx}
\usepackage{natbib}
\usepackage{multirow}

\usepackage{braket}
\usepackage{siunitx}

\usepackage[utf8]{inputenc}

\usepackage[unicode,raiselinks=false, hyperfootnotes=false,bookmarksopen=true,hyperfigures=false,bookmarksopenlevel=2]{hyperref} 
\hypersetup{colorlinks=true,
    linkcolor=black,
    filecolor=black,      
    urlcolor=blue,
    citecolor=blue
}
\usepackage[all]{hypcap} 
\usepackage{xargs}  
\usepackage[colorinlistoftodos,prependcaption,textsize=tiny]{todonotes}
\newcommandx{\note}[2][1=]{\todo[linecolor=red,backgroundcolor=blue!25,bordercolor=red,#1]{#2}}

\begin{document}
\title{Increasing the proximity induced spin-orbit coupling in bilayer graphene/WSe$_2$ heterostructures with pressure}
\author{Bálint Szentpéteri$^{1,2}$, Albin Márffy$^{1,2}$, Máté Kedves$^{1,2}$, Endre Tóvári$^{1,2}$,  Bálint Fülöp$^{1,2}$, István Kükemezey$^{4}$, András Magyarkuti$^{4}$, Kenji Watanabe$^5$, Takashi Taniguchi$^5$, Szabolcs Csonka$^{1,3,6}$, Péter Makk$^{1,2}$}
\affiliation{$^1$Department of Physics, Budapest University of Technology and Economics and
Nanoelectronics Momentum Research Group of the Hungarian Academy of Sciences,
Budafoki ut 8, 1111 Budapest, Hungary\\
$^2$MTA-BME Correlated van der Waals Structures Momentum Research Group, Műegyetem rkp. 3., H-1111 Budapest, Hungary\\
$^3$MTA-BME Superconducting Nanoelectronics Momentum Research Group, Műegyetem rkp. 3., H-1111 Budapest, Hungary\\
$^4$Semilab Co. Ltd. Prielle Kornélia u. 2, 1117 Budapest, Hungary\\
$^5$Research Center for Functional Materials, National Institute for Materials Science, 1-1 Namiki, Tsukuba 305-0044, Japan\\
$^6$HUN-REN Centre for Energy Research, Institute of Technical Physics and Materials Science, Konkoly Thege Miklós út 29-33., H-1121 Budapest, Hungary
}
\date{\today}

\begin{abstract}
Combining graphene with transition metal dichalcogenides (TMDs) leads to enhanced spin-orbit coupling (SOC) in the graphene. The induced SOC has a large effect on the low-energy part of the band structure leading to or stabilizing novel phases such as topological phases or superconductivity. Here, the pressure dependence of the SOC strength is investigated in bilayer graphene/WSe$_2$ heterostructures. We performed magnetoconductance studies, such as weak localization, quantum Hall, and Shubnikov-de Haas oscillation measurements to extract the different SOC terms which determine the low-energy band structure of BLG. We find the proximity-induced SOC strengths increased by more than 50\% as a result of applying 2\,GPa hydrostatic pressure. Our studies highlight the opportunity to increase the SOC coupling strength with pressure, which can be important for correlated phases or spin qubits in BLG/WSe$_2$ heterostructures.
\end{abstract}

\maketitle
\section{Introduction}
Graphene is an ideal material for spintronics due to its excellent properties, such as low spin-orbit coupling (SOC), and hyperfine coupling\cite{Han2014,Roche2015}. The long spin relaxation times combined with the large mobilities of the charge carriers lead to an exceptionally long spin relaxation length of several tens of micrometers\cite{Han2014, Roche2015, Droegeler2016, Bisswanger2022}. These properties also make graphene an ideal platform for the realization of spin qubits defined by quantum dots\cite{Eich2018, Banszerus2018, Eich2018a, Banszerus2020, Eich2020}. In bilayer graphene (BLG), where the electrons can be confined using electric fields, more than \SI{100}{\micro\second} to even half second of relaxation time have been demonstrated\cite{Gaechter2022, Banszerus2022, Garreis2024,Denisov2024}. However, to use graphene in spintronic devices, electrical control of the spin information is necessary\cite{Gmitra2015b, Gmitra2017}, for which large SOC is required. It was theoretically predicted that the SOC can be increased by bringing graphene in proximity to a transition metal dichalcogenide (TMD), which has a strong intrinsic SOC\cite{Gmitra2015b}. The enhancement of SOC was demonstrated with weak localization measurements\cite{Wang2015, Wang2016, Yang2016, Yang2017, Wakamura2018, Zihlmann2018, Amann2022}, Shubnikov-de Haas (SdH) oscillations\cite{Wang2015, Wang2016, Afzal2018, Tiwari2022, Rao2023}, spin valves\cite{Omar2017, Avsar2014, Benitez2017, Ghiasi2017, Yan2016, Dankert2017} and quasiparticle interference imaging\cite{Sun2023}. The control of the spin currents in graphene/TMD heterostructures was successfully performed with electrical gating\cite{Yan2016, Yang2016, Dankert2017, Omar2018} and with spin-to-charge conversion\cite{Benitez2019, Herling2020, Safeer2019, Yang2024}. Moreover, in Bernal-stacked and twisted structures the low-energy behavior was found to be greatly influenced by the SOC\cite{Gmitra2017, Zollner2021, Lin2022, Bhowmik2023, Chou2024, Zhang2024}. Therefore, tuning the SOC in these heterostructures is important to understand and control their behavior. 

Applying hydrostatic pressure is a reliable way to increase the proximity-induced SOC in these heterostructures as it was shown in our pioneering work with single-layer graphene\cite{Fueloep2021} and also with a BLG encapsulated between two layers of WSe$_2$, where the application of pressure led to the stabilization of a peculiar, band-inverted phase\cite{Kedves2023}. 

In this work, we focus on BLG/WSe$_2$ heterostructures, where the presence of SOC in the conduction or valence band is expected to be controllable using gate electrodes \cite{Gmitra2017, Khoo2017}. We show the presence of the SOC by performing weak localization measurements. From a detailed study of SdH oscillations and quantum Hall effect we extract the Rashba-type and Ising-type SOC strengths. By applying 2\,GPa hydrostatic pressure, we manage to achieve a sizable increase of the SOC strength. Surprisingly, we detect a significantly larger Rashba-type SOC than expected from theoretical predictions.

\section{Results}
\begin{figure}[!tb]
\begin{center}
\includegraphics[width=1\linewidth]{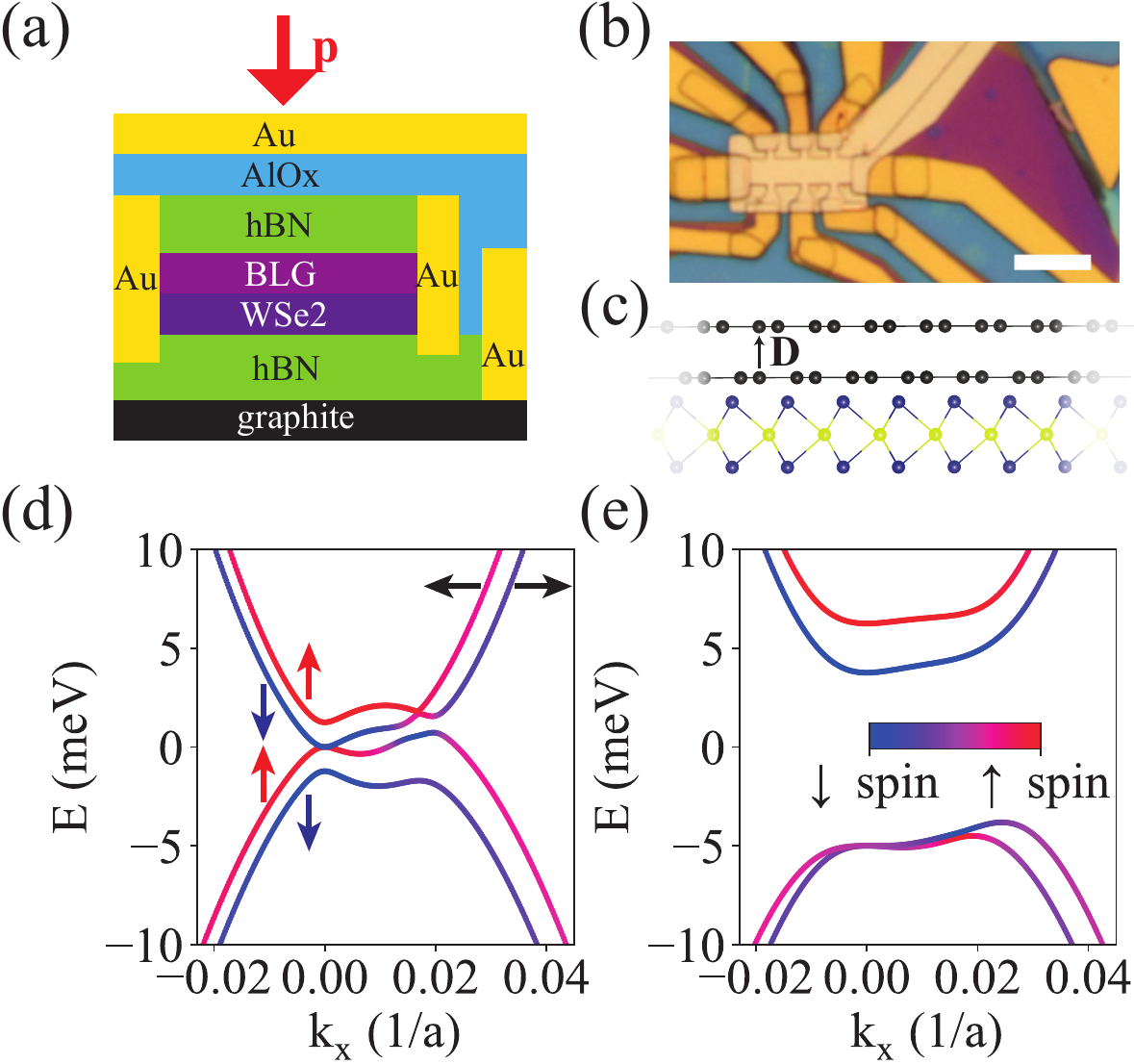}
\caption{Properties of the BLG/WSe2 heterostructure. (a) Schematic picture of the heterostructure. The BLG and the WSe$_2$ flakes are sandwiched between two hBN crystals. The heterostructure is contacted with Cr/Au edge contacts and a 30 nm thick ALD AlOx layer is deposited on the heterostructure to separate the gate from the device. A bottom graphite gate and a Cr/Au top gate is used to tune the charge density $n$ and displacement field $D$. (b) Optical microscope image of the measured device. The device is shaped into a Hall-bar and it is covered with a global top gate. The scale bar is \SI{5}{\micro\metre}. (c) Schematic structure of the BLG/WSe$_2$ heterostructure. The direction of the applied electric displacement is also shown between the graphene layers. (d) Low-energy band structure of the bilayer graphene including the spin-orbit coupling (SOC) with $\lambda_R=15$\,meV and $\lambda_I=2.5$\,meV. (e) An $u=10\,$meV interlayer potential difference opens a bandgap and helps spin-polarize the conduction bands. The bands in panel (d) and (e) are colored by their spin polarization shown by the color bar on panel (e). }
\label{fig1}
\end{center}
\end{figure}

The schematic of our device is shown in Fig.\,\ref{fig1}a, whereas an optical micrograph can be found in Fig.\,\ref{fig1}b. It consists of a bilayer graphene (BLG) flake placed on top of a WSe$_2$ flake, and encapsulated between two hexagonal boron nitride (hBN) layers. The graphene is contacted with etched side contacts. The graphite bottom gate and the metallic top gate allow to separately tune the charge carrier density, $n$ and the displacement field, $D$. Further details on the fabrication and conversion of gate voltages to $n$ and $D$ is given in the Supplemental Material\cite{supmat}. In the main text, measurements on the device with the highest quality are shown, and similar results on two other devices are detailed in the Supplemental Material\cite{supmat}. 

The schematic illustration of our BLG/WSe$_2$ heterostructure is shown in Fig.\,\ref{fig1}c. The low-energy band structure of pristine BLG near the $K$ and $K'$ points is described by the Hamiltonian $H_0$, which is given in the Supplemental Material\cite{supmat}, and exhibits parabolic bands touching near the $K$ points with small modifications from the remote hopping terms. The presence of the WSe$_2$ flake leads to the modification of the band structure of BLG, which is reflected by the appearance of additional SOC terms: a Rashba-type $H_R=\frac{\lambda_R}{2}(\xi\sigma_xs_y-\sigma_ys_x)$ and an Ising-type $H_I=\xi\frac{\lambda_I}{2}\sigma_z$ SOC term, also referred to as valley-Zeeman (VZ) SOC. They are assumed to be present only at the bottom carbon layer, which is in proximity with WSe$_2$\cite{McCann2013, Khoo2017}. Here, $\xi$ is the valley index, $s_i$ and $\sigma_i$ are Pauli-matrices acting on the spin and sub-lattice degree of freedom of the bottom layer and $\lambda_R$ and $\lambda_I$ are the SOC strengths of the Rashba-type and Ising-type SOC, respectively. We calculated the band structure based on the low-energy model described in the Supplemental Material\cite{supmat}. The band structure of the heterostructure around the $K$ point with $\lambda_R=15$\,meV and $\lambda_I=2.5$\,meV is shown in Fig.\,\ref{fig1}d. These parameters are plausible values based on our measurements (see later). The effect of $\lambda_R$ is depicted with black arrows, which splits the bands at higher $k$ values and the splitting becomes larger at higher energies. The effect of $\lambda_I$ on the low-energy spectrum is shown with red and blue arrows. At low energies it splits the spin degeneracy of the bands and makes them spin polarized in the out-of-plane directions, whereas for higher energies a more complicated, canted spin structure arises as the combination of the two types of spin-orbit interactions\cite{Gmitra2017}. The effect of electric field is modeled by introducing interlayer potential difference $u$ and it is illustrated in Fig.\,\ref{fig1}e with $u=10$\,meV. The electric field leads to layer polarization of the conduction and valence bands. For positive $u$, the conduction band is localized on the bottom layer neighboring the WSe$_2$ crystal. Hence the spin splitting caused by the proximity-induced SOC appears in this band (see panel e). For negative $u$, the layer polarization is reversed and the spin splitting appears in the valence band, as shown in the Supplemental Material Fig.\,S3c\cite{supmat}.

\begin{figure}[!tb]
\begin{center}
\includegraphics[width=1\linewidth]{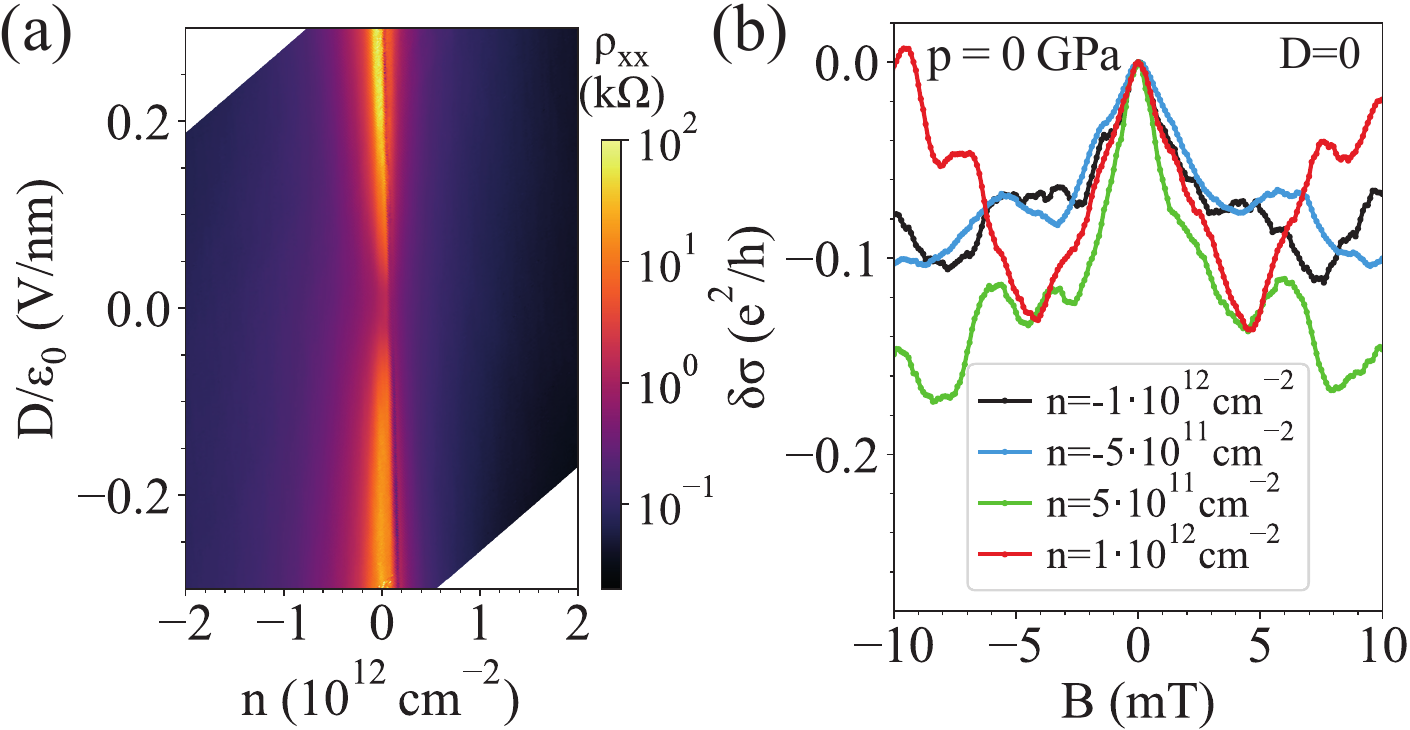}
\caption{(a) Four-probe resistance of the BLG/WSe$_2$ device as a function of  the charge density ($n$) and electric displacement field ($D$) measured at ambient pressure, $T=50$\,mK and $B=0$\,T. Near $n=0$ the charge neutrality line the resistance is high, and  increases with $|D|$ as it opens a gap in the BLG. (b) Magnetoconductivity at $D=0$, $T=1.5$\,K at $n=\pm1, \pm0.5 \cdot 10^{12}$cm$^{-2}$. Here, we averaged over 41 curves in a $0.2\cdot10^{12}$\,cm$^{-2}$ range at a fixed $D$ to average out the universal conductance fluctuations. A clear antilocalization  signal is visible in all densitiy range.}
\label{fig2}
\end{center}
\end{figure}

Fig.~\ref{fig2}a shows longitudinal resistivity ($\rho_\mathrm{xx}$) as a function of top and bottom gate voltages, plotted as a function of $n$ and $D$ at temperature $T=50$\,mK. Lighter colored regions of higher resistance correspond to the displacement-field-opened gap along the charge neutrality line. This gap is increasing by increasing $|D|$ as expected in bilayer graphene. 

\begin{figure*}[!tb]
\begin{center}
\includegraphics[width=1\linewidth]{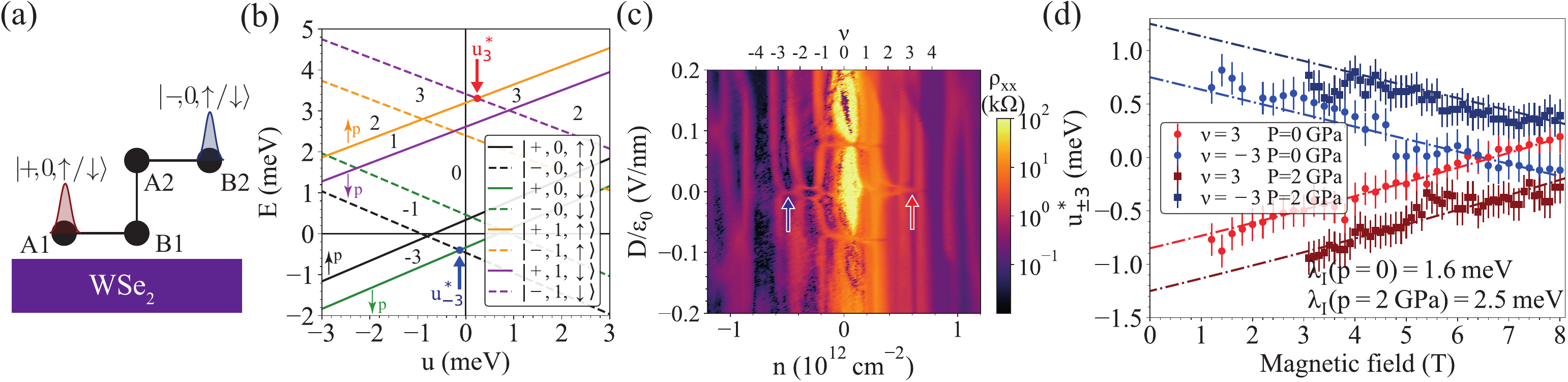}
\caption{Acquiring the Ising SOC strength from quantum Hall measurements. (a) Schematic representation of the real-space localization of the LL orbits. They can be represented with 3 quantum numbers: the valley ($+/-$), the orbital number ($n=0,1...$) and the spin ($\uparrow/\downarrow$). The $+$ orbits are localized at the bottom and the $-$ are localized at the top layer of the BLG. As the WSe$_2$ is at the bottom, the SOC only modifies the $+$ orbit energies. The $n=1$ LLs with the same valley index are localized on the same layer as the $n=0$ LLs (solid lines in b). (b) Calculated LL energies as a function of the interlayer potential difference ($u$) with $\lambda_I=1.6$\,meV at $B=8$\,T. The hopping parameters are given in the Supplemental Material\cite{supmat}. The $\nu=\pm3$ crossing points marked with a red and a blue dot and are found at $u=u_{\pm3}^*$. The arrows with letter p-s next to them show how the LL energies change by increasing pressure i.e. increasing the $\lambda_I$. The LL filling factors are shown with black numbers between LLs. (c) Four-probe resistance of the device at $B=8$\,T as a function of  $n$ and $D$ at $p=0$ and at $T=50$\,mK. On the top the corresponding Landau level (LL) filling factors ($\nu=nh/eB$) are shown. LLs correspond to extended areas of large resistance and crossings correspond to sharp features of large resistance. We use the $\nu=\pm3$ crossing points to extract $\lambda_I$. These are shown with red and blue arrows. (d) The position of the $\nu=\pm3$ crossing points extracted from the measurements. The dashed lines represent the fits of the model on the measurement data using $\lambda_I(p=0)=1.6$\,meV and $\lambda_I(p=2\,\mathrm{GPa})=2.5$\,meV.}
\label{fig3}
\end{center}
\end{figure*}

To confirm the presence of induced SOC in our heterostructure, we performed magnetoconductance measurements at low magnetic field $B$. In pristine BLG weak localization is expected\cite{Gorbachev2007}. However, the presence of SOC leads to weak antilocalization (WAL)\cite{Wang2015, Wang2016, Yang2016, Voelkl2017, Afzal2018, Wakamura2018, Zihlmann2018, Amann2022}. In our measurements, weak antilocalization (WAL) was observed as shown in Fig.\,\ref{fig2}b, which is a clear evidence for the presence of SOC in BLG. Further WAL measurements, at different pressures and at a series of $D$ fields, can be found in the Supplemental Material\cite{supmat}.

The SOC also modifies the Landau-level (LL) spectrum, which allows the extraction of the Ising term\cite{Island2019, Wang2019, Zhang2023}. The LLs in BLG can be parameterized with the quantum numbers of $\ket{\xi,n,s_z}$, where $n=0,1,\dots$ is the orbital number. Since for the low-energy bands, only $A1$ and $B2$ atoms (non-dimer sites) play a role, the layer and the pseudo-spin (sublattice) degree of freedom becomes identical, which is illustrated in Fig.\,\ref{fig3}a.  
Moreover, the lowest LLs from the $K$ valley (+) are localized on the bottom layer, while the lowest LLs with $K'$ index (-) are localized on the top layer. As a result, the LL energies of the different valleys depend oppositely on the application of a displacement field. As the SOC have effect only at the bottom layer, it affects only the LLs with $+$ index. It was shown that the $\lambda_R$ is negligible, only the Ising-type SOC has a measurable effect on the lowest LLs\cite{Khoo2018}. Based on the above considerations, the LL energies can be approximated by $E(\xi,n, s_z)\approx-E_Zs_z+n\Delta_{10}+\frac{u}{2}\alpha_{\xi,n,s_z}+\xi\frac{\lambda_I}{2}\zeta_{1,\xi,n}$\cite{Khoo2018}, where $E_Z$ is the Zeeman splitting, $\alpha_{\xi,n,s_z}$ is some layer polarization, $\zeta_{1,\xi,n}$ is some spin polarization on layer 1, both in the order of unity, and $\Delta_{10}$ is the orbital LL splitting. This approximation qualitatively agrees with the exact solution. However, to quantitatively assess $\lambda_I$, we performed numerical calculations of LL energies of BLG without taking into the effect of electron--electron interactions. The calculated LL energies are shown in Fig.\,\ref{fig3}b at $B=8$\,T as a function of interlayer potential difference $u$, which is related to the displacement field by $u =\frac{d}{\epsilon_{\text{BLG}}^\perp}D$, where $d$ is the BLG interlayer separation, $\epsilon_{\text{BLG}}^\perp$ is the out-of-plane dielectric constant of the BLG \cite{Khoo2018}. 
Since the Ising-type SOC behaves as an effective $B$-field, as we increase $\lambda_I$, LL energies of the bottom layer (indicated by the solid lines in Fig.\,\ref{fig3}b) shift according to their spin polarization, as indicated by the arrows with letter p-s next to them in the figure. We show the LL filling factors $\nu$ with numbers between the LLs. There are several crossings observable between the LLs, from which the LL crossing points at $\nu=\pm3$ are marked with red and blue dots. We study these crossings, since their position is unaffected by the electronic interactions \cite{Hunt2017} and their position is strongly affected by the VZ interaction. 

\begin{figure}[!tb]
\begin{center}
\includegraphics[width=1\linewidth]{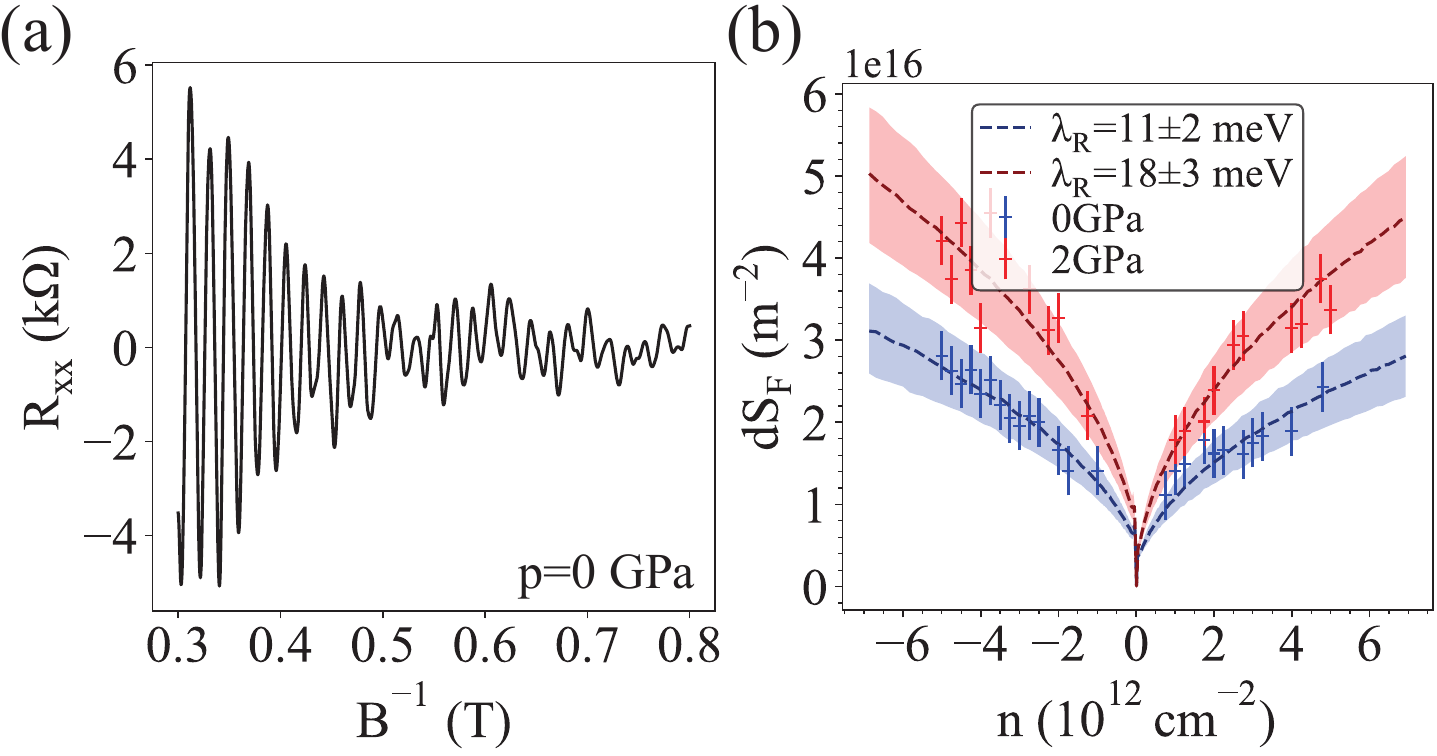}
\caption{Shubnikov\,--\,de Haas oscillations. (a) A typical four-probe resistance measurement as a function of $1/B$ after removing a quadratic background at $n=5\cdot10^{12}$\,cm$^{-2}$, $D=0$, $p=0$ and at $T=1.4$\,K. The beating is clearly visible due to the spin-split bands. (b) The difference of the Fermi surfaces as a function of $n$. We perform fast Fourier transform on the SdH data, where we saw two peaks. From the Fourier analysis we extract the frequency difference of the two peaks. The frequencies ($f_B$) were converted to Fermi-surfaces ($S_F=2\pi e f_B/\hbar$). The blue points were measured at $p=0$\,GPa and the red points were measured at $p=2$\,GPa. The dashed lines are the fitting of the model with the corresponding $\lambda_R$ values. The colored ranges show the confidence interval of the fits. }
\label{fig4}
\end{center}
\end{figure}

To experimentally  extract the LL crossings positions, we measure the longitudinal magnetoresistance as a function of $n$ and $D$ at a fixed magnetic field. In Fig.\,\ref{fig3}c we show $\rho_{xx}$ as a function of $n$ and $D$ at $B=8$\,T out-of-plane magnetic field. On the top we also show the corresponding Landau level (LL) filling factors $\nu=\frac{nh}{eB}$, where $e$ is the elementary charge and $h$ is the Planck-constant. The dark (bright) regions correspond to situations where the Fermi level is situated between (on) LLs, with (half-) integer $\nu$. Crosses between the LLs appear as bright spots along certain $\nu$s at particular $D$ values.
The $\nu=\pm3$ crossing points are highlighted with red and blue arrows in panel (b) and (c). We measure these crossing points as a function of the magnetic field by measuring a $D$ trace at a fixed filling factor at different magnetic fields (therefore also changing the density at every magnetic field to keep the filling factor constant). We extracted the position of the crossings as a function of $B$, which is shown with red and blue symbols for $\nu=\pm3$ in Fig.\,\ref{fig3}d. To determine the Ising-type SOC strength, we calculated the LL spectrum shown in panel (b) for different magnetic fields and $\lambda_I$ values. We fit the magnetic field dependence of crossings and the best fit is shown with dashed lines. From the fits we obtain $\lambda_I(p=0)=1.6\pm0.2$\,meV, which is in agreement with the experiments \cite{Zihlmann2018, Wang2019, Island2019, Sun2023} and the theoretical values in the literature \cite{David2019, Li2019, Naimer2021}.

To investigate the changes in the SOC coupling strength as the layer distance is decreased, we have performed measurement also at $p=2$\,GPa. Measurements under pressure were done using a pressure cell with a special sample holder, that allows wire-bonding our sample onto a printed circuit board and measurements in the range of $0-2$\,GPa. Details of the pressurization procedure is described in Ref.\cite{Fueloep2021a}. After repeating the procedure described for ambient conditions, the extracted crossing point values are shown in Fig.\,\ref{fig3}d with black and yellow markers for $\nu=\pm3$, respectively. Fitting similarly, we obtain $\lambda_I(p=2$\,GPa$)=2.5\pm0.2$\,meV, which is a significant enhancement. 

To extract the strength of the Rashba SOC coupling we also performed Shubnikov\,--\,de Haas (SdH) oscillation measurements. As visible in Fig.\,\ref{fig1}d, the Fermi surface (FS) is split due to the SOC coupling, and the splitting is dominated at high doping by the Rashba term. The SdH oscillations give us the opportunity to extract the splitting of the FS. Here in Fig.\,\ref{fig4}a, we subtract a quadratic background from the longitudinal resistance signal. The Fermi-surface is related to the frequency of the SdH oscillations by $S_F=2\pi e f_B/\hbar$, where $f_B$ is the oscillation frequency \cite{Soule1964}. In the figure a beating pattern is visible which is a sign of having two Fermi surfaces close to each other. We performed a fast Fourier transformation (FFT) on the signal to obtain the Fermi-surfaces, which is detailed in the Supplemental Material\cite{supmat}. 
We take the difference of the obtained Fermi-surfaces ($dS_F$) at a fixed $n$ as shown in Fig.\,\ref{fig4}b by blue symbols for a large range of charge density. To extract the Rashba SOC strength, we first calculate the Fermi-surfaces at a fixed $n$ using the previously obtained $\lambda_I$. We fit $dS_F$ from the model to our data using $\lambda_R$ as the only fitting parameter and find $\lambda_R(p=0)=11\pm2$\,meV, which is shown in Fig.\,\ref{fig4}b, where the dashed line and the shaded area shows the calculated FS splitting and the error, respectively. This is similar to the experimentally found values in the literature with similar techniques \cite{Wang2016, Wang2019, Sun2023}, but the theoretically predicted values are an order of magnitude smaller \cite{Gmitra2016,Alsharari2018,Li2019,David2019}. The reason for this discrepancy is not clear.

We have performed the same analysis at 2\,GPa. The extracted $dS_F$ are shown with red markers. We fit the model similarly, and find $\lambda_R(p=2$\,GPa$)=18\pm3$\,meV, which is shown with a red dashed line and the shaded area shows the error in the Figure. The relative increase is similar to the increase of $\lambda_I$.

The strength of SOC strongly depends on the rotation angle between WSe$_2$ and graphene\cite{David2019, Li2019, Naimer2021}. Though we did not control the rotation angle between the flakes, the large VZ splitting suggests rotation angles close to $\vartheta\sim11$° or $\vartheta\sim22$° according to Ref.\,\citep{Li2019} and according to Ref.\,\citep{Naimer2021} it is close to 0°. From the optical image of the flakes the twist angle is likely around $\vartheta\sim16$°. In the Supplemental Material we show similar measurements on two other devices\cite{supmat}. We found similar enhancement of the SOC strengths with pressure on these devices as well. On one of the samples, we find negative SOC at similar value ($\lambda_I=-1.7\pm1$\,meV) from Quantum Hall studies, which imply that $\vartheta$ is negative\cite{Naimer2021}. As the pressure doesn't change the twist angle\cite{Yankowitz2018, Szentpeteri2021}, our measurements provide clear evidence of increasing the proximity induced SOC strength by decreasing the interlayer distance.

The extracted $\lambda_I$ and its enhancement with pressure leads to such a large spin splitting, that could significantly modify and tune the correlated states in twisted graphene heterostructures\cite{Arora2020, Chou2024}.

\section{Conclusion}
In conclusion, we have observed from WAL measurements the presence of proximity SOC in graphene. To determine the strength of the Ising-type SOC term in our BLG/WSe$_2$ heterostructures, we performed quantum Hall measurements as a function of the displacement field and studied the LL crossings at $\nu = \pm 3$. We performed magnetoconductance study and from the Shubnikov-de Haas oscillation we extracted $\lambda_R$. We found that both are enhanced by more than 50\,\% by applying hydrostatic pressure of 2\,GPa on the device. Our study can be important in realizing novel phases in graphene/TMD heterostructures for which large SOC is required.

\section{Data availability}
Source data of the measurements and the python code for the band structure calculation are publicly available at \url{https://doi.org/10.5281/zenodo.13741039}.

\section{Author contribution}
B.Sz. fabricated the device. Measurements were performed by B.Sz. with the help of B.F., E.T., A.M. and M.K. B.Sz. did the data analysis and the theoretical calculations. B.Sz. and P.M. wrote the paper and all authors discussed the results and worked on the manuscript. K.W. and T.T. grew the hBN crystals. I.K. and A.M. was involved in AFM characterization of the samples during fabrication. The project was guided by Sz.Cs. and P.M.

\begin{acknowledgments}
This work acknowledges support from the MultiSpin and 2DSOTECH FlagERA networks, the OTKA K138433 and K134437 and PD134758 grants and the EIC Pathfinder Challenge grant QuKiT. This research was supported by the Ministry of Culture and Innovation and the National Research, Development and Innovation Office within the Quantum Information National Laboratory of Hungary (Grant No. 2022-2.1.1-NL-2022-00004 and ÚNKP-23-4-I-BME-36), by SuperGate networks and by the European Research Council ERC project Twistrain. We acknowledge COST Action CA 21144 superQUMAP. 
K.W. and T.T. acknowledge support from the JSPS KAKENHI (Grant Numbers 20H00354 and 23H02052) and World Premier International Research Center Initiative (WPI), MEXT, Japan.
\end{acknowledgments}

\bibliography{Szentpeteri_et_al_2024.bib}

\begin{thebibliography}{67}%
\makeatletter
\providecommand \@ifxundefined [1]{%
 \@ifx{#1\undefined}
}%
\providecommand \@ifnum [1]{%
 \ifnum #1\expandafter \@firstoftwo
 \else \expandafter \@secondoftwo
 \fi
}%
\providecommand \@ifx [1]{%
 \ifx #1\expandafter \@firstoftwo
 \else \expandafter \@secondoftwo
 \fi
}%
\providecommand \natexlab [1]{#1}%
\providecommand \enquote  [1]{``#1''}%
\providecommand \bibnamefont  [1]{#1}%
\providecommand \bibfnamefont [1]{#1}%
\providecommand \citenamefont [1]{#1}%
\providecommand \href@noop [0]{\@secondoftwo}%
\providecommand \href [0]{\begingroup \@sanitize@url \@href}%
\providecommand \@href[1]{\@@startlink{#1}\@@href}%
\providecommand \@@href[1]{\endgroup#1\@@endlink}%
\providecommand \@sanitize@url [0]{\catcode `\\12\catcode `\$12\catcode
  `\&12\catcode `\#12\catcode `\^12\catcode `\_12\catcode `\%12\relax}%
\providecommand \@@startlink[1]{}%
\providecommand \@@endlink[0]{}%
\providecommand \url  [0]{\begingroup\@sanitize@url \@url }%
\providecommand \@url [1]{\endgroup\@href {#1}{\urlprefix }}%
\providecommand \urlprefix  [0]{URL }%
\providecommand \Eprint [0]{\href }%
\providecommand \doibase [0]{https://doi.org/}%
\providecommand \selectlanguage [0]{\@gobble}%
\providecommand \bibinfo  [0]{\@secondoftwo}%
\providecommand \bibfield  [0]{\@secondoftwo}%
\providecommand \translation [1]{[#1]}%
\providecommand \BibitemOpen [0]{}%
\providecommand \bibitemStop [0]{}%
\providecommand \bibitemNoStop [0]{.\EOS\space}%
\providecommand \EOS [0]{\spacefactor3000\relax}%
\providecommand \BibitemShut  [1]{\csname bibitem#1\endcsname}%
\let\auto@bib@innerbib\@empty
\bibitem [{\citenamefont {Han}\ \emph {et~al.}(2014)\citenamefont {Han},
  \citenamefont {Kawakami}, \citenamefont {Gmitra},\ and\ \citenamefont
  {Fabian}}]{Han2014}%
  \BibitemOpen
  \bibfield  {author} {\bibinfo {author} {\bibfnamefont {W.}~\bibnamefont
  {Han}}, \bibinfo {author} {\bibfnamefont {R.~K.}\ \bibnamefont {Kawakami}},
  \bibinfo {author} {\bibfnamefont {M.}~\bibnamefont {Gmitra}},\ and\ \bibinfo
  {author} {\bibfnamefont {J.}~\bibnamefont {Fabian}},\ }\bibfield  {title}
  {\bibinfo {title} {Graphene spintronics},\ }\href
  {https://doi.org/10.1038/nnano.2014.214} {\bibfield  {journal} {\bibinfo
  {journal} {Nature Nanotechnology}\ }\textbf {\bibinfo {volume} {9}},\
  \bibinfo {pages} {794} (\bibinfo {year} {2014})}\BibitemShut {NoStop}%
\bibitem [{\citenamefont {Roche}\ \emph {et~al.}(2015)\citenamefont {Roche},
  \citenamefont {Åkerman}, \citenamefont {Beschoten}, \citenamefont
  {Charlier}, \citenamefont {Chshiev}, \citenamefont {Prasad~Dash},
  \citenamefont {Dlubak}, \citenamefont {Fabian}, \citenamefont {Fert},
  \citenamefont {Guimarães}, \citenamefont {Guinea}, \citenamefont
  {Grigorieva}, \citenamefont {Schönenberger}, \citenamefont {Seneor},
  \citenamefont {Stampfer}, \citenamefont {Valenzuela}, \citenamefont
  {Waintal},\ and\ \citenamefont {van Wees}}]{Roche2015}%
  \BibitemOpen
  \bibfield  {author} {\bibinfo {author} {\bibfnamefont {S.}~\bibnamefont
  {Roche}}, \bibinfo {author} {\bibfnamefont {J.}~\bibnamefont {Åkerman}},
  \bibinfo {author} {\bibfnamefont {B.}~\bibnamefont {Beschoten}}, \bibinfo
  {author} {\bibfnamefont {J.-C.}\ \bibnamefont {Charlier}}, \bibinfo {author}
  {\bibfnamefont {M.}~\bibnamefont {Chshiev}}, \bibinfo {author} {\bibfnamefont
  {S.}~\bibnamefont {Prasad~Dash}}, \bibinfo {author} {\bibfnamefont
  {B.}~\bibnamefont {Dlubak}}, \bibinfo {author} {\bibfnamefont
  {J.}~\bibnamefont {Fabian}}, \bibinfo {author} {\bibfnamefont
  {A.}~\bibnamefont {Fert}}, \bibinfo {author} {\bibfnamefont {M.}~\bibnamefont
  {Guimarães}}, \bibinfo {author} {\bibfnamefont {F.}~\bibnamefont {Guinea}},
  \bibinfo {author} {\bibfnamefont {I.}~\bibnamefont {Grigorieva}}, \bibinfo
  {author} {\bibfnamefont {C.}~\bibnamefont {Schönenberger}}, \bibinfo
  {author} {\bibfnamefont {P.}~\bibnamefont {Seneor}}, \bibinfo {author}
  {\bibfnamefont {C.}~\bibnamefont {Stampfer}}, \bibinfo {author}
  {\bibfnamefont {S.~O.}\ \bibnamefont {Valenzuela}}, \bibinfo {author}
  {\bibfnamefont {X.}~\bibnamefont {Waintal}},\ and\ \bibinfo {author}
  {\bibfnamefont {B.}~\bibnamefont {van Wees}},\ }\bibfield  {title} {\bibinfo
  {title} {Graphene spintronics: the european flagship perspective},\ }\href
  {https://doi.org/10.1088/2053-1583/2/3/030202} {\bibfield  {journal}
  {\bibinfo  {journal} {2D Materials}\ }\textbf {\bibinfo {volume} {2}},\
  \bibinfo {pages} {030202} (\bibinfo {year} {2015})}\BibitemShut {NoStop}%
\bibitem [{\citenamefont {Drögeler}\ \emph {et~al.}(2016)\citenamefont
  {Drögeler}, \citenamefont {Franzen}, \citenamefont {Volmer}, \citenamefont
  {Pohlmann}, \citenamefont {Banszerus}, \citenamefont {Wolter}, \citenamefont
  {Watanabe}, \citenamefont {Taniguchi}, \citenamefont {Stampfer},\ and\
  \citenamefont {Beschoten}}]{Droegeler2016}%
  \BibitemOpen
  \bibfield  {author} {\bibinfo {author} {\bibfnamefont {M.}~\bibnamefont
  {Drögeler}}, \bibinfo {author} {\bibfnamefont {C.}~\bibnamefont {Franzen}},
  \bibinfo {author} {\bibfnamefont {F.}~\bibnamefont {Volmer}}, \bibinfo
  {author} {\bibfnamefont {T.}~\bibnamefont {Pohlmann}}, \bibinfo {author}
  {\bibfnamefont {L.}~\bibnamefont {Banszerus}}, \bibinfo {author}
  {\bibfnamefont {M.}~\bibnamefont {Wolter}}, \bibinfo {author} {\bibfnamefont
  {K.}~\bibnamefont {Watanabe}}, \bibinfo {author} {\bibfnamefont
  {T.}~\bibnamefont {Taniguchi}}, \bibinfo {author} {\bibfnamefont
  {C.}~\bibnamefont {Stampfer}},\ and\ \bibinfo {author} {\bibfnamefont
  {B.}~\bibnamefont {Beschoten}},\ }\bibfield  {title} {\bibinfo {title} {Spin
  lifetimes exceeding 12 ns in graphene nonlocal spin valve devices},\ }\href
  {https://doi.org/10.1021/acs.nanolett.6b00497} {\bibfield  {journal}
  {\bibinfo  {journal} {Nano Letters}\ }\textbf {\bibinfo {volume} {16}},\
  \bibinfo {pages} {3533} (\bibinfo {year} {2016})}\BibitemShut {NoStop}%
\bibitem [{\citenamefont {Bisswanger}\ \emph {et~al.}(2022)\citenamefont
  {Bisswanger}, \citenamefont {Winter}, \citenamefont {Schmidt}, \citenamefont
  {Volmer}, \citenamefont {Watanabe}, \citenamefont {Taniguchi}, \citenamefont
  {Stampfer},\ and\ \citenamefont {Beschoten}}]{Bisswanger2022}%
  \BibitemOpen
  \bibfield  {author} {\bibinfo {author} {\bibfnamefont {T.}~\bibnamefont
  {Bisswanger}}, \bibinfo {author} {\bibfnamefont {Z.}~\bibnamefont {Winter}},
  \bibinfo {author} {\bibfnamefont {A.}~\bibnamefont {Schmidt}}, \bibinfo
  {author} {\bibfnamefont {F.}~\bibnamefont {Volmer}}, \bibinfo {author}
  {\bibfnamefont {K.}~\bibnamefont {Watanabe}}, \bibinfo {author}
  {\bibfnamefont {T.}~\bibnamefont {Taniguchi}}, \bibinfo {author}
  {\bibfnamefont {C.}~\bibnamefont {Stampfer}},\ and\ \bibinfo {author}
  {\bibfnamefont {B.}~\bibnamefont {Beschoten}},\ }\bibfield  {title} {\bibinfo
  {title} {Cvd bilayer graphene spin valves with 26 um spin diffusion length at
  room temperature},\ }\href {https://doi.org/10.1021/acs.nanolett.2c01119}
  {\bibfield  {journal} {\bibinfo  {journal} {Nano Letters}\ }\textbf {\bibinfo
  {volume} {22}},\ \bibinfo {pages} {4949} (\bibinfo {year}
  {2022})}\BibitemShut {NoStop}%
\bibitem [{\citenamefont {Eich}\ \emph
  {et~al.}(2018{\natexlab{a}})\citenamefont {Eich}, \citenamefont {Herman},
  \citenamefont {Pisoni}, \citenamefont {Overweg}, \citenamefont {Kurzmann},
  \citenamefont {Lee}, \citenamefont {Rickhaus}, \citenamefont {Watanabe},
  \citenamefont {Taniguchi}, \citenamefont {Sigrist}, \citenamefont {Ihn},\
  and\ \citenamefont {Ensslin}}]{Eich2018}%
  \BibitemOpen
  \bibfield  {author} {\bibinfo {author} {\bibfnamefont {M.}~\bibnamefont
  {Eich}}, \bibinfo {author} {\bibfnamefont {F.}~\bibnamefont {Herman}},
  \bibinfo {author} {\bibfnamefont {R.}~\bibnamefont {Pisoni}}, \bibinfo
  {author} {\bibfnamefont {H.}~\bibnamefont {Overweg}}, \bibinfo {author}
  {\bibfnamefont {A.}~\bibnamefont {Kurzmann}}, \bibinfo {author}
  {\bibfnamefont {Y.}~\bibnamefont {Lee}}, \bibinfo {author} {\bibfnamefont
  {P.}~\bibnamefont {Rickhaus}}, \bibinfo {author} {\bibfnamefont
  {K.}~\bibnamefont {Watanabe}}, \bibinfo {author} {\bibfnamefont
  {T.}~\bibnamefont {Taniguchi}}, \bibinfo {author} {\bibfnamefont
  {M.}~\bibnamefont {Sigrist}}, \bibinfo {author} {\bibfnamefont
  {T.}~\bibnamefont {Ihn}},\ and\ \bibinfo {author} {\bibfnamefont
  {K.}~\bibnamefont {Ensslin}},\ }\bibfield  {title} {\bibinfo {title} {Spin
  and valley states in gate-defined bilayer graphene quantum dots},\ }\href
  {https://doi.org/10.1103/physrevx.8.031023} {\bibfield  {journal} {\bibinfo
  {journal} {Physical Review X}\ }\textbf {\bibinfo {volume} {8}},\ \bibinfo
  {pages} {031023} (\bibinfo {year} {2018}{\natexlab{a}})}\BibitemShut
  {NoStop}%
\bibitem [{\citenamefont {Banszerus}\ \emph {et~al.}(2018)\citenamefont
  {Banszerus}, \citenamefont {Frohn}, \citenamefont {Epping}, \citenamefont
  {Neumaier}, \citenamefont {Watanabe}, \citenamefont {Taniguchi},\ and\
  \citenamefont {Stampfer}}]{Banszerus2018}%
  \BibitemOpen
  \bibfield  {author} {\bibinfo {author} {\bibfnamefont {L.}~\bibnamefont
  {Banszerus}}, \bibinfo {author} {\bibfnamefont {B.}~\bibnamefont {Frohn}},
  \bibinfo {author} {\bibfnamefont {A.}~\bibnamefont {Epping}}, \bibinfo
  {author} {\bibfnamefont {D.}~\bibnamefont {Neumaier}}, \bibinfo {author}
  {\bibfnamefont {K.}~\bibnamefont {Watanabe}}, \bibinfo {author}
  {\bibfnamefont {T.}~\bibnamefont {Taniguchi}},\ and\ \bibinfo {author}
  {\bibfnamefont {C.}~\bibnamefont {Stampfer}},\ }\bibfield  {title} {\bibinfo
  {title} {Gate-defined electron–hole double dots in bilayer graphene},\
  }\href {https://doi.org/10.1021/acs.nanolett.8b01303} {\bibfield  {journal}
  {\bibinfo  {journal} {Nano Letters}\ }\textbf {\bibinfo {volume} {18}},\
  \bibinfo {pages} {4785} (\bibinfo {year} {2018})}\BibitemShut {NoStop}%
\bibitem [{\citenamefont {Eich}\ \emph
  {et~al.}(2018{\natexlab{b}})\citenamefont {Eich}, \citenamefont {Pisoni},
  \citenamefont {Pally}, \citenamefont {Overweg}, \citenamefont {Kurzmann},
  \citenamefont {Lee}, \citenamefont {Rickhaus}, \citenamefont {Watanabe},
  \citenamefont {Taniguchi}, \citenamefont {Ensslin},\ and\ \citenamefont
  {Ihn}}]{Eich2018a}%
  \BibitemOpen
  \bibfield  {author} {\bibinfo {author} {\bibfnamefont {M.}~\bibnamefont
  {Eich}}, \bibinfo {author} {\bibfnamefont {R.}~\bibnamefont {Pisoni}},
  \bibinfo {author} {\bibfnamefont {A.}~\bibnamefont {Pally}}, \bibinfo
  {author} {\bibfnamefont {H.}~\bibnamefont {Overweg}}, \bibinfo {author}
  {\bibfnamefont {A.}~\bibnamefont {Kurzmann}}, \bibinfo {author}
  {\bibfnamefont {Y.}~\bibnamefont {Lee}}, \bibinfo {author} {\bibfnamefont
  {P.}~\bibnamefont {Rickhaus}}, \bibinfo {author} {\bibfnamefont
  {K.}~\bibnamefont {Watanabe}}, \bibinfo {author} {\bibfnamefont
  {T.}~\bibnamefont {Taniguchi}}, \bibinfo {author} {\bibfnamefont
  {K.}~\bibnamefont {Ensslin}},\ and\ \bibinfo {author} {\bibfnamefont
  {T.}~\bibnamefont {Ihn}},\ }\bibfield  {title} {\bibinfo {title} {Coupled
  quantum dots in bilayer graphene},\ }\href
  {https://doi.org/10.1021/acs.nanolett.8b01859} {\bibfield  {journal}
  {\bibinfo  {journal} {Nano Letters}\ }\textbf {\bibinfo {volume} {18}},\
  \bibinfo {pages} {5042} (\bibinfo {year} {2018}{\natexlab{b}})}\BibitemShut
  {NoStop}%
\bibitem [{\citenamefont {Banszerus}\ \emph {et~al.}(2020)\citenamefont
  {Banszerus}, \citenamefont {Möller}, \citenamefont {Icking}, \citenamefont
  {Watanabe}, \citenamefont {Taniguchi}, \citenamefont {Volk},\ and\
  \citenamefont {Stampfer}}]{Banszerus2020}%
  \BibitemOpen
  \bibfield  {author} {\bibinfo {author} {\bibfnamefont {L.}~\bibnamefont
  {Banszerus}}, \bibinfo {author} {\bibfnamefont {S.}~\bibnamefont {Möller}},
  \bibinfo {author} {\bibfnamefont {E.}~\bibnamefont {Icking}}, \bibinfo
  {author} {\bibfnamefont {K.}~\bibnamefont {Watanabe}}, \bibinfo {author}
  {\bibfnamefont {T.}~\bibnamefont {Taniguchi}}, \bibinfo {author}
  {\bibfnamefont {C.}~\bibnamefont {Volk}},\ and\ \bibinfo {author}
  {\bibfnamefont {C.}~\bibnamefont {Stampfer}},\ }\bibfield  {title} {\bibinfo
  {title} {Single-electron double quantum dots in bilayer graphene},\ }\href
  {https://doi.org/10.1021/acs.nanolett.9b05295} {\bibfield  {journal}
  {\bibinfo  {journal} {Nano Letters}\ }\textbf {\bibinfo {volume} {20}},\
  \bibinfo {pages} {2005} (\bibinfo {year} {2020})}\BibitemShut {NoStop}%
\bibitem [{\citenamefont {Eich}\ \emph {et~al.}(2020)\citenamefont {Eich},
  \citenamefont {Pisoni}, \citenamefont {Tong}, \citenamefont {Garreis},
  \citenamefont {Rickhaus}, \citenamefont {Watanabe}, \citenamefont
  {Taniguchi}, \citenamefont {Ihn}, \citenamefont {Ensslin},\ and\
  \citenamefont {Kurzmann}}]{Eich2020}%
  \BibitemOpen
  \bibfield  {author} {\bibinfo {author} {\bibfnamefont {M.}~\bibnamefont
  {Eich}}, \bibinfo {author} {\bibfnamefont {R.}~\bibnamefont {Pisoni}},
  \bibinfo {author} {\bibfnamefont {C.}~\bibnamefont {Tong}}, \bibinfo {author}
  {\bibfnamefont {R.}~\bibnamefont {Garreis}}, \bibinfo {author} {\bibfnamefont
  {P.}~\bibnamefont {Rickhaus}}, \bibinfo {author} {\bibfnamefont
  {K.}~\bibnamefont {Watanabe}}, \bibinfo {author} {\bibfnamefont
  {T.}~\bibnamefont {Taniguchi}}, \bibinfo {author} {\bibfnamefont
  {T.}~\bibnamefont {Ihn}}, \bibinfo {author} {\bibfnamefont {K.}~\bibnamefont
  {Ensslin}},\ and\ \bibinfo {author} {\bibfnamefont {A.}~\bibnamefont
  {Kurzmann}},\ }\bibfield  {title} {\bibinfo {title} {Coulomb dominated
  cavities in bilayer graphene},\ }\href
  {https://doi.org/10.1103/physrevresearch.2.022038} {\bibfield  {journal}
  {\bibinfo  {journal} {Physical Review Research}\ }\textbf {\bibinfo {volume}
  {2}},\ \bibinfo {pages} {022038(R)} (\bibinfo {year} {2020})}\BibitemShut
  {NoStop}%
\bibitem [{\citenamefont {Gachter}\ \emph {et~al.}(2022)\citenamefont
  {Gachter}, \citenamefont {Garreis}, \citenamefont {Gerber}, \citenamefont
  {Ruckriegel}, \citenamefont {Tong}, \citenamefont {Kratochwil}, \citenamefont
  {de~Vries}, \citenamefont {Kurzmann}, \citenamefont {Watanabe}, \citenamefont
  {Taniguchi}, \citenamefont {Ihn}, \citenamefont {Ensslin},\ and\
  \citenamefont {Huang}}]{Gaechter2022}%
  \BibitemOpen
  \bibfield  {author} {\bibinfo {author} {\bibfnamefont {L.~M.}\ \bibnamefont
  {Gachter}}, \bibinfo {author} {\bibfnamefont {R.}~\bibnamefont {Garreis}},
  \bibinfo {author} {\bibfnamefont {J.~D.}\ \bibnamefont {Gerber}}, \bibinfo
  {author} {\bibfnamefont {M.~J.}\ \bibnamefont {Ruckriegel}}, \bibinfo
  {author} {\bibfnamefont {C.}~\bibnamefont {Tong}}, \bibinfo {author}
  {\bibfnamefont {B.}~\bibnamefont {Kratochwil}}, \bibinfo {author}
  {\bibfnamefont {F.~K.}\ \bibnamefont {de~Vries}}, \bibinfo {author}
  {\bibfnamefont {A.}~\bibnamefont {Kurzmann}}, \bibinfo {author}
  {\bibfnamefont {K.}~\bibnamefont {Watanabe}}, \bibinfo {author}
  {\bibfnamefont {T.}~\bibnamefont {Taniguchi}}, \bibinfo {author}
  {\bibfnamefont {T.}~\bibnamefont {Ihn}}, \bibinfo {author} {\bibfnamefont
  {K.}~\bibnamefont {Ensslin}},\ and\ \bibinfo {author} {\bibfnamefont {W.~W.}\
  \bibnamefont {Huang}},\ }\bibfield  {title} {\bibinfo {title} {Single-shot
  spin readout in graphene quantum dots},\ }\href
  {https://doi.org/10.1103/prxquantum.3.020343} {\bibfield  {journal} {\bibinfo
   {journal} {PRX Quantum}\ }\textbf {\bibinfo {volume} {3}},\ \bibinfo {pages}
  {020343} (\bibinfo {year} {2022})}\BibitemShut {NoStop}%
\bibitem [{\citenamefont {Banszerus}\ \emph {et~al.}(2022)\citenamefont
  {Banszerus}, \citenamefont {Hecker}, \citenamefont {Möller}, \citenamefont
  {Icking}, \citenamefont {Watanabe}, \citenamefont {Taniguchi}, \citenamefont
  {Volk},\ and\ \citenamefont {Stampfer}}]{Banszerus2022}%
  \BibitemOpen
  \bibfield  {author} {\bibinfo {author} {\bibfnamefont {L.}~\bibnamefont
  {Banszerus}}, \bibinfo {author} {\bibfnamefont {K.}~\bibnamefont {Hecker}},
  \bibinfo {author} {\bibfnamefont {S.}~\bibnamefont {Möller}}, \bibinfo
  {author} {\bibfnamefont {E.}~\bibnamefont {Icking}}, \bibinfo {author}
  {\bibfnamefont {K.}~\bibnamefont {Watanabe}}, \bibinfo {author}
  {\bibfnamefont {T.}~\bibnamefont {Taniguchi}}, \bibinfo {author}
  {\bibfnamefont {C.}~\bibnamefont {Volk}},\ and\ \bibinfo {author}
  {\bibfnamefont {C.}~\bibnamefont {Stampfer}},\ }\bibfield  {title} {\bibinfo
  {title} {Spin relaxation in a single-electron graphene quantum dot},\ }\href
  {https://doi.org/10.1038/s41467-022-31231-5} {\bibfield  {journal} {\bibinfo
  {journal} {Nature Communications}\ }\textbf {\bibinfo {volume} {13}},\
  \bibinfo {pages} {3637} (\bibinfo {year} {2022})}\BibitemShut {NoStop}%
\bibitem [{\citenamefont {Garreis}\ \emph {et~al.}(2024)\citenamefont
  {Garreis}, \citenamefont {Tong}, \citenamefont {Terle}, \citenamefont
  {Ruckriegel}, \citenamefont {Gerber}, \citenamefont {Gächter}, \citenamefont
  {Watanabe}, \citenamefont {Taniguchi}, \citenamefont {Ihn}, \citenamefont
  {Ensslin},\ and\ \citenamefont {Huang}}]{Garreis2024}%
  \BibitemOpen
  \bibfield  {author} {\bibinfo {author} {\bibfnamefont {R.}~\bibnamefont
  {Garreis}}, \bibinfo {author} {\bibfnamefont {C.}~\bibnamefont {Tong}},
  \bibinfo {author} {\bibfnamefont {J.}~\bibnamefont {Terle}}, \bibinfo
  {author} {\bibfnamefont {M.~J.}\ \bibnamefont {Ruckriegel}}, \bibinfo
  {author} {\bibfnamefont {J.~D.}\ \bibnamefont {Gerber}}, \bibinfo {author}
  {\bibfnamefont {L.~M.}\ \bibnamefont {Gächter}}, \bibinfo {author}
  {\bibfnamefont {K.}~\bibnamefont {Watanabe}}, \bibinfo {author}
  {\bibfnamefont {T.}~\bibnamefont {Taniguchi}}, \bibinfo {author}
  {\bibfnamefont {T.}~\bibnamefont {Ihn}}, \bibinfo {author} {\bibfnamefont
  {K.}~\bibnamefont {Ensslin}},\ and\ \bibinfo {author} {\bibfnamefont {W.~W.}\
  \bibnamefont {Huang}},\ }\bibfield  {title} {\bibinfo {title} {Long-lived
  valley states in bilayer graphene quantum dots},\ }\href
  {https://doi.org/10.1038/s41567-023-02334-7} {\bibfield  {journal} {\bibinfo
  {journal} {Nature Physics}\ }\textbf {\bibinfo {volume} {20}},\ \bibinfo
  {pages} {428} (\bibinfo {year} {2024})}\BibitemShut {NoStop}%
\bibitem [{\citenamefont {Denisov}\ \emph {et~al.}(2024)\citenamefont
  {Denisov}, \citenamefont {Reckova}, \citenamefont {Cances}, \citenamefont
  {Ruckriegel}, \citenamefont {Masseroni}, \citenamefont {Adam}, \citenamefont
  {Tong}, \citenamefont {Gerber}, \citenamefont {Huang}, \citenamefont
  {Watanabe}, \citenamefont {Taniguchi}, \citenamefont {Ihn}, \citenamefont
  {Ensslin},\ and\ \citenamefont {Duprez}}]{Denisov2024}%
  \BibitemOpen
  \bibfield  {author} {\bibinfo {author} {\bibfnamefont {A.~O.}\ \bibnamefont
  {Denisov}}, \bibinfo {author} {\bibfnamefont {V.}~\bibnamefont {Reckova}},
  \bibinfo {author} {\bibfnamefont {S.}~\bibnamefont {Cances}}, \bibinfo
  {author} {\bibfnamefont {M.~J.}\ \bibnamefont {Ruckriegel}}, \bibinfo
  {author} {\bibfnamefont {M.}~\bibnamefont {Masseroni}}, \bibinfo {author}
  {\bibfnamefont {C.}~\bibnamefont {Adam}}, \bibinfo {author} {\bibfnamefont
  {C.}~\bibnamefont {Tong}}, \bibinfo {author} {\bibfnamefont {J.~D.}\
  \bibnamefont {Gerber}}, \bibinfo {author} {\bibfnamefont {W.~W.}\
  \bibnamefont {Huang}}, \bibinfo {author} {\bibfnamefont {K.}~\bibnamefont
  {Watanabe}}, \bibinfo {author} {\bibfnamefont {T.}~\bibnamefont {Taniguchi}},
  \bibinfo {author} {\bibfnamefont {T.}~\bibnamefont {Ihn}}, \bibinfo {author}
  {\bibfnamefont {K.}~\bibnamefont {Ensslin}},\ and\ \bibinfo {author}
  {\bibfnamefont {H.}~\bibnamefont {Duprez}},\ }\href
  {https://doi.org/10.48550/ARXIV.2403.08143} {\bibinfo {title} {Ultra-long
  relaxation of a kramers qubit formed in a bilayer graphene quantum dot}}
  (\bibinfo {year} {2024})\BibitemShut {NoStop}%
\bibitem [{\citenamefont {Gmitra}\ and\ \citenamefont
  {Fabian}(2015)}]{Gmitra2015b}%
  \BibitemOpen
  \bibfield  {author} {\bibinfo {author} {\bibfnamefont {M.}~\bibnamefont
  {Gmitra}}\ and\ \bibinfo {author} {\bibfnamefont {J.}~\bibnamefont
  {Fabian}},\ }\bibfield  {title} {\bibinfo {title} {Graphene on
  transition-metal dichalcogenides: A platform for proximity spin-orbit physics
  and optospintronics},\ }\href {https://doi.org/10.1103/physrevb.92.155403}
  {\bibfield  {journal} {\bibinfo  {journal} {Physical Review B}\ }\textbf
  {\bibinfo {volume} {92}},\ \bibinfo {pages} {155403} (\bibinfo {year}
  {2015})}\BibitemShut {NoStop}%
\bibitem [{\citenamefont {Gmitra}\ and\ \citenamefont
  {Fabian}(2017)}]{Gmitra2017}%
  \BibitemOpen
  \bibfield  {author} {\bibinfo {author} {\bibfnamefont {M.}~\bibnamefont
  {Gmitra}}\ and\ \bibinfo {author} {\bibfnamefont {J.}~\bibnamefont
  {Fabian}},\ }\bibfield  {title} {\bibinfo {title} {Proximity effects in
  bilayer graphene on monolayer {WSe}2 : Field-effect spin valley locking,
  spin-orbit valve, and spin transistor},\ }\href
  {https://doi.org/10.1103/physrevlett.119.146401} {\bibfield  {journal}
  {\bibinfo  {journal} {Physical Review Letters}\ }\textbf {\bibinfo {volume}
  {119}},\ \bibinfo {pages} {146401} (\bibinfo {year} {2017})}\BibitemShut
  {NoStop}%
\bibitem [{\citenamefont {Wang}\ \emph {et~al.}(2015)\citenamefont {Wang},
  \citenamefont {Ki}, \citenamefont {Chen}, \citenamefont {Berger},
  \citenamefont {MacDonald},\ and\ \citenamefont {Morpurgo}}]{Wang2015}%
  \BibitemOpen
  \bibfield  {author} {\bibinfo {author} {\bibfnamefont {Z.}~\bibnamefont
  {Wang}}, \bibinfo {author} {\bibfnamefont {D.}~\bibnamefont {Ki}}, \bibinfo
  {author} {\bibfnamefont {H.}~\bibnamefont {Chen}}, \bibinfo {author}
  {\bibfnamefont {H.}~\bibnamefont {Berger}}, \bibinfo {author} {\bibfnamefont
  {A.~H.}\ \bibnamefont {MacDonald}},\ and\ \bibinfo {author} {\bibfnamefont
  {A.~F.}\ \bibnamefont {Morpurgo}},\ }\bibfield  {title} {\bibinfo {title}
  {Strong interface-induced spin{\textendash}orbit interaction in graphene on
  {WS}2},\ }\href {https://doi.org/10.1038/ncomms9339} {\bibfield  {journal}
  {\bibinfo  {journal} {Nature Communications}\ }\textbf {\bibinfo {volume}
  {6}},\ \bibinfo {pages} {8339} (\bibinfo {year} {2015})}\BibitemShut
  {NoStop}%
\bibitem [{\citenamefont {Wang}\ \emph {et~al.}(2016)\citenamefont {Wang},
  \citenamefont {Ki}, \citenamefont {Khoo}, \citenamefont {Mauro},
  \citenamefont {Berger}, \citenamefont {Levitov},\ and\ \citenamefont
  {Morpurgo}}]{Wang2016}%
  \BibitemOpen
  \bibfield  {author} {\bibinfo {author} {\bibfnamefont {Z.}~\bibnamefont
  {Wang}}, \bibinfo {author} {\bibfnamefont {D.-K.}\ \bibnamefont {Ki}},
  \bibinfo {author} {\bibfnamefont {J.~Y.}\ \bibnamefont {Khoo}}, \bibinfo
  {author} {\bibfnamefont {D.}~\bibnamefont {Mauro}}, \bibinfo {author}
  {\bibfnamefont {H.}~\bibnamefont {Berger}}, \bibinfo {author} {\bibfnamefont
  {L.~S.}\ \bibnamefont {Levitov}},\ and\ \bibinfo {author} {\bibfnamefont
  {A.~F.}\ \bibnamefont {Morpurgo}},\ }\bibfield  {title} {\bibinfo {title}
  {Origin and magnitude of `designer' spin-orbit interaction in graphene on
  semiconducting transition metal dichalcogenides},\ }\href
  {https://doi.org/10.1103/physrevx.6.041020} {\bibfield  {journal} {\bibinfo
  {journal} {Physical Review X}\ }\textbf {\bibinfo {volume} {6}},\ \bibinfo
  {pages} {041020} (\bibinfo {year} {2016})}\BibitemShut {NoStop}%
\bibitem [{\citenamefont {Yang}\ \emph {et~al.}(2016)\citenamefont {Yang},
  \citenamefont {Tu}, \citenamefont {Kim}, \citenamefont {Wu}, \citenamefont
  {Wang}, \citenamefont {Alicea}, \citenamefont {Wu}, \citenamefont
  {Bockrath},\ and\ \citenamefont {Shi}}]{Yang2016}%
  \BibitemOpen
  \bibfield  {author} {\bibinfo {author} {\bibfnamefont {B.}~\bibnamefont
  {Yang}}, \bibinfo {author} {\bibfnamefont {M.-F.}\ \bibnamefont {Tu}},
  \bibinfo {author} {\bibfnamefont {J.}~\bibnamefont {Kim}}, \bibinfo {author}
  {\bibfnamefont {Y.}~\bibnamefont {Wu}}, \bibinfo {author} {\bibfnamefont
  {H.}~\bibnamefont {Wang}}, \bibinfo {author} {\bibfnamefont {J.}~\bibnamefont
  {Alicea}}, \bibinfo {author} {\bibfnamefont {R.}~\bibnamefont {Wu}}, \bibinfo
  {author} {\bibfnamefont {M.}~\bibnamefont {Bockrath}},\ and\ \bibinfo
  {author} {\bibfnamefont {J.}~\bibnamefont {Shi}},\ }\bibfield  {title}
  {\bibinfo {title} {Tunable spin{\textendash}orbit coupling and
  symmetry-protected edge states in graphene/{WS}2},\ }\href
  {https://doi.org/10.1088/2053-1583/3/3/031012} {\bibfield  {journal}
  {\bibinfo  {journal} {2D Materials}\ }\textbf {\bibinfo {volume} {3}},\
  \bibinfo {pages} {031012} (\bibinfo {year} {2016})}\BibitemShut {NoStop}%
\bibitem [{\citenamefont {Yang}\ \emph {et~al.}(2017)\citenamefont {Yang},
  \citenamefont {Lohmann}, \citenamefont {Barroso}, \citenamefont {Liao},
  \citenamefont {Lin}, \citenamefont {Liu}, \citenamefont {Bartels},
  \citenamefont {Watanabe}, \citenamefont {Taniguchi},\ and\ \citenamefont
  {Shi}}]{Yang2017}%
  \BibitemOpen
  \bibfield  {author} {\bibinfo {author} {\bibfnamefont {B.}~\bibnamefont
  {Yang}}, \bibinfo {author} {\bibfnamefont {M.}~\bibnamefont {Lohmann}},
  \bibinfo {author} {\bibfnamefont {D.}~\bibnamefont {Barroso}}, \bibinfo
  {author} {\bibfnamefont {I.}~\bibnamefont {Liao}}, \bibinfo {author}
  {\bibfnamefont {Z.}~\bibnamefont {Lin}}, \bibinfo {author} {\bibfnamefont
  {Y.}~\bibnamefont {Liu}}, \bibinfo {author} {\bibfnamefont {L.}~\bibnamefont
  {Bartels}}, \bibinfo {author} {\bibfnamefont {K.}~\bibnamefont {Watanabe}},
  \bibinfo {author} {\bibfnamefont {T.}~\bibnamefont {Taniguchi}},\ and\
  \bibinfo {author} {\bibfnamefont {J.}~\bibnamefont {Shi}},\ }\bibfield
  {title} {\bibinfo {title} {Strong electron-hole symmetric rashba spin-orbit
  coupling in graphene/monolayer transition metal dichalcogenide
  heterostructures},\ }\href {https://doi.org/10.1103/physrevb.96.041409}
  {\bibfield  {journal} {\bibinfo  {journal} {Physical Review B}\ }\textbf
  {\bibinfo {volume} {96}},\ \bibinfo {pages} {041409(R)} (\bibinfo {year}
  {2017})}\BibitemShut {NoStop}%
\bibitem [{\citenamefont {Wakamura}\ \emph {et~al.}(2018)\citenamefont
  {Wakamura}, \citenamefont {Reale}, \citenamefont {Palczynski}, \citenamefont
  {Gueron}, \citenamefont {Mattevi},\ and\ \citenamefont
  {Bouchiat}}]{Wakamura2018}%
  \BibitemOpen
  \bibfield  {author} {\bibinfo {author} {\bibfnamefont {T.}~\bibnamefont
  {Wakamura}}, \bibinfo {author} {\bibfnamefont {F.}~\bibnamefont {Reale}},
  \bibinfo {author} {\bibfnamefont {P.}~\bibnamefont {Palczynski}}, \bibinfo
  {author} {\bibfnamefont {S.}~\bibnamefont {Gueron}}, \bibinfo {author}
  {\bibfnamefont {C.}~\bibnamefont {Mattevi}},\ and\ \bibinfo {author}
  {\bibfnamefont {H.}~\bibnamefont {Bouchiat}},\ }\bibfield  {title} {\bibinfo
  {title} {Strong anisotropic spin-orbit interaction induced in graphene by
  monolayer ws2},\ }\href {https://doi.org/10.1103/physrevlett.120.106802}
  {\bibfield  {journal} {\bibinfo  {journal} {Physical Review Letters}\
  }\textbf {\bibinfo {volume} {120}},\ \bibinfo {pages} {106802} (\bibinfo
  {year} {2018})}\BibitemShut {NoStop}%
\bibitem [{\citenamefont {Zihlmann}\ \emph {et~al.}(2018)\citenamefont
  {Zihlmann}, \citenamefont {Cummings}, \citenamefont {Garcia}, \citenamefont
  {Kedves}, \citenamefont {Watanabe}, \citenamefont {Taniguchi}, \citenamefont
  {Schonenberger},\ and\ \citenamefont {Makk}}]{Zihlmann2018}%
  \BibitemOpen
  \bibfield  {author} {\bibinfo {author} {\bibfnamefont {S.}~\bibnamefont
  {Zihlmann}}, \bibinfo {author} {\bibfnamefont {A.~W.}\ \bibnamefont
  {Cummings}}, \bibinfo {author} {\bibfnamefont {J.~H.}\ \bibnamefont
  {Garcia}}, \bibinfo {author} {\bibfnamefont {M.}~\bibnamefont {Kedves}},
  \bibinfo {author} {\bibfnamefont {K.}~\bibnamefont {Watanabe}}, \bibinfo
  {author} {\bibfnamefont {T.}~\bibnamefont {Taniguchi}}, \bibinfo {author}
  {\bibfnamefont {C.}~\bibnamefont {Schonenberger}},\ and\ \bibinfo {author}
  {\bibfnamefont {P.}~\bibnamefont {Makk}},\ }\bibfield  {title} {\bibinfo
  {title} {Large spin relaxation anisotropy and valley-zeeman spin-orbit
  coupling in wse2/graphene/h-bn heterostructures},\ }\href
  {https://doi.org/10.1103/physrevb.97.075434} {\bibfield  {journal} {\bibinfo
  {journal} {Physical Review B}\ }\textbf {\bibinfo {volume} {97}},\ \bibinfo
  {pages} {075434} (\bibinfo {year} {2018})}\BibitemShut {NoStop}%
\bibitem [{\citenamefont {Amann}\ \emph {et~al.}(2022)\citenamefont {Amann},
  \citenamefont {Volkl}, \citenamefont {Rockinger}, \citenamefont {Kochan},
  \citenamefont {Watanabe}, \citenamefont {Taniguchi}, \citenamefont {Fabian},
  \citenamefont {Weiss},\ and\ \citenamefont {Eroms}}]{Amann2022}%
  \BibitemOpen
  \bibfield  {author} {\bibinfo {author} {\bibfnamefont {J.}~\bibnamefont
  {Amann}}, \bibinfo {author} {\bibfnamefont {T.}~\bibnamefont {Volkl}},
  \bibinfo {author} {\bibfnamefont {T.}~\bibnamefont {Rockinger}}, \bibinfo
  {author} {\bibfnamefont {D.}~\bibnamefont {Kochan}}, \bibinfo {author}
  {\bibfnamefont {K.}~\bibnamefont {Watanabe}}, \bibinfo {author}
  {\bibfnamefont {T.}~\bibnamefont {Taniguchi}}, \bibinfo {author}
  {\bibfnamefont {J.}~\bibnamefont {Fabian}}, \bibinfo {author} {\bibfnamefont
  {D.}~\bibnamefont {Weiss}},\ and\ \bibinfo {author} {\bibfnamefont
  {J.}~\bibnamefont {Eroms}},\ }\bibfield  {title} {\bibinfo {title}
  {Counterintuitive gate dependence of weak antilocalization in bilayer
  graphene/wse2 heterostructures},\ }\href
  {https://doi.org/10.1103/physrevb.105.115425} {\bibfield  {journal} {\bibinfo
   {journal} {Physical Review B}\ }\textbf {\bibinfo {volume} {105}},\ \bibinfo
  {pages} {115425} (\bibinfo {year} {2022})}\BibitemShut {NoStop}%
\bibitem [{\citenamefont {Afzal}\ \emph {et~al.}(2018)\citenamefont {Afzal},
  \citenamefont {Khan}, \citenamefont {Nazir}, \citenamefont {Dastgeer},
  \citenamefont {Aftab}, \citenamefont {Akhtar}, \citenamefont {Seo},\ and\
  \citenamefont {Eom}}]{Afzal2018}%
  \BibitemOpen
  \bibfield  {author} {\bibinfo {author} {\bibfnamefont {A.~M.}\ \bibnamefont
  {Afzal}}, \bibinfo {author} {\bibfnamefont {M.~F.}\ \bibnamefont {Khan}},
  \bibinfo {author} {\bibfnamefont {G.}~\bibnamefont {Nazir}}, \bibinfo
  {author} {\bibfnamefont {G.}~\bibnamefont {Dastgeer}}, \bibinfo {author}
  {\bibfnamefont {S.}~\bibnamefont {Aftab}}, \bibinfo {author} {\bibfnamefont
  {I.}~\bibnamefont {Akhtar}}, \bibinfo {author} {\bibfnamefont
  {Y.}~\bibnamefont {Seo}},\ and\ \bibinfo {author} {\bibfnamefont
  {J.}~\bibnamefont {Eom}},\ }\bibfield  {title} {\bibinfo {title} {Gate
  modulation of the spin-orbit interaction in bilayer graphene encapsulated by
  ws2 films},\ }\href {https://doi.org/10.1038/s41598-018-21787-y} {\bibfield
  {journal} {\bibinfo  {journal} {Scientific Reports}\ }\textbf {\bibinfo
  {volume} {8}},\ \bibinfo {pages} {3412} (\bibinfo {year} {2018})}\BibitemShut
  {NoStop}%
\bibitem [{\citenamefont {Tiwari}\ \emph {et~al.}(2022)\citenamefont {Tiwari},
  \citenamefont {Jat}, \citenamefont {Udupa}, \citenamefont {Narang},
  \citenamefont {Watanabe}, \citenamefont {Taniguchi}, \citenamefont {Sen},\
  and\ \citenamefont {Bid}}]{Tiwari2022}%
  \BibitemOpen
  \bibfield  {author} {\bibinfo {author} {\bibfnamefont {P.}~\bibnamefont
  {Tiwari}}, \bibinfo {author} {\bibfnamefont {M.~K.}\ \bibnamefont {Jat}},
  \bibinfo {author} {\bibfnamefont {A.}~\bibnamefont {Udupa}}, \bibinfo
  {author} {\bibfnamefont {D.~S.}\ \bibnamefont {Narang}}, \bibinfo {author}
  {\bibfnamefont {K.}~\bibnamefont {Watanabe}}, \bibinfo {author}
  {\bibfnamefont {T.}~\bibnamefont {Taniguchi}}, \bibinfo {author}
  {\bibfnamefont {D.}~\bibnamefont {Sen}},\ and\ \bibinfo {author}
  {\bibfnamefont {A.}~\bibnamefont {Bid}},\ }\bibfield  {title} {\bibinfo
  {title} {Experimental observation of spin-split energy dispersion in
  high-mobility single-layer graphene/wse2 heterostructures},\ }\href
  {https://doi.org/10.1038/s41699-022-00348-y} {\bibfield  {journal} {\bibinfo
  {journal} {npj 2D Materials and Applications}\ }\textbf {\bibinfo {volume}
  {6}},\ \bibinfo {pages} {68} (\bibinfo {year} {2022})}\BibitemShut {NoStop}%
\bibitem [{\citenamefont {Rao}\ \emph {et~al.}(2023)\citenamefont {Rao},
  \citenamefont {Kang}, \citenamefont {Xue}, \citenamefont {Ye}, \citenamefont
  {Feng}, \citenamefont {Watanabe}, \citenamefont {Taniguchi}, \citenamefont
  {Wang}, \citenamefont {Liu},\ and\ \citenamefont {Ki}}]{Rao2023}%
  \BibitemOpen
  \bibfield  {author} {\bibinfo {author} {\bibfnamefont {Q.}~\bibnamefont
  {Rao}}, \bibinfo {author} {\bibfnamefont {W.-H.}\ \bibnamefont {Kang}},
  \bibinfo {author} {\bibfnamefont {H.}~\bibnamefont {Xue}}, \bibinfo {author}
  {\bibfnamefont {Z.}~\bibnamefont {Ye}}, \bibinfo {author} {\bibfnamefont
  {X.}~\bibnamefont {Feng}}, \bibinfo {author} {\bibfnamefont {K.}~\bibnamefont
  {Watanabe}}, \bibinfo {author} {\bibfnamefont {T.}~\bibnamefont {Taniguchi}},
  \bibinfo {author} {\bibfnamefont {N.}~\bibnamefont {Wang}}, \bibinfo {author}
  {\bibfnamefont {M.-H.}\ \bibnamefont {Liu}},\ and\ \bibinfo {author}
  {\bibfnamefont {D.-K.}\ \bibnamefont {Ki}},\ }\bibfield  {title} {\bibinfo
  {title} {Ballistic transport spectroscopy of spin-orbit-coupled bands in
  monolayer graphene on wse2},\ }\href
  {https://doi.org/10.1038/s41467-023-41826-1} {\bibfield  {journal} {\bibinfo
  {journal} {Nature Communications}\ }\textbf {\bibinfo {volume} {14}},\
  \bibinfo {pages} {6124} (\bibinfo {year} {2023})}\BibitemShut {NoStop}%
\bibitem [{\citenamefont {Omar}\ and\ \citenamefont {van
  Wees}(2017)}]{Omar2017}%
  \BibitemOpen
  \bibfield  {author} {\bibinfo {author} {\bibfnamefont {S.}~\bibnamefont
  {Omar}}\ and\ \bibinfo {author} {\bibfnamefont {B.~J.}\ \bibnamefont {van
  Wees}},\ }\bibfield  {title} {\bibinfo {title} {Graphene-ws2 heterostructures
  for tunable spin injection and spin transport},\ }\href
  {https://doi.org/10.1103/physrevb.95.081404} {\bibfield  {journal} {\bibinfo
  {journal} {Physical Review B}\ }\textbf {\bibinfo {volume} {95}},\ \bibinfo
  {pages} {081404(R)} (\bibinfo {year} {2017})}\BibitemShut {NoStop}%
\bibitem [{\citenamefont {Avsar}\ \emph {et~al.}(2014)\citenamefont {Avsar},
  \citenamefont {Tan}, \citenamefont {Taychatanapat}, \citenamefont
  {Balakrishnan}, \citenamefont {Koon}, \citenamefont {Yeo}, \citenamefont
  {Lahiri}, \citenamefont {Carvalho}, \citenamefont {Rodin}, \citenamefont
  {O'Farrell}, \citenamefont {Eda}, \citenamefont {Neto},\ and\ \citenamefont
  {Özyilmaz}}]{Avsar2014}%
  \BibitemOpen
  \bibfield  {author} {\bibinfo {author} {\bibfnamefont {A.}~\bibnamefont
  {Avsar}}, \bibinfo {author} {\bibfnamefont {J.~Y.}\ \bibnamefont {Tan}},
  \bibinfo {author} {\bibfnamefont {T.}~\bibnamefont {Taychatanapat}}, \bibinfo
  {author} {\bibfnamefont {J.}~\bibnamefont {Balakrishnan}}, \bibinfo {author}
  {\bibfnamefont {G.}~\bibnamefont {Koon}}, \bibinfo {author} {\bibfnamefont
  {Y.}~\bibnamefont {Yeo}}, \bibinfo {author} {\bibfnamefont {J.}~\bibnamefont
  {Lahiri}}, \bibinfo {author} {\bibfnamefont {A.}~\bibnamefont {Carvalho}},
  \bibinfo {author} {\bibfnamefont {A.~S.}\ \bibnamefont {Rodin}}, \bibinfo
  {author} {\bibfnamefont {E.}~\bibnamefont {O'Farrell}}, \bibinfo {author}
  {\bibfnamefont {G.}~\bibnamefont {Eda}}, \bibinfo {author} {\bibfnamefont
  {A.~H.~C.}\ \bibnamefont {Neto}},\ and\ \bibinfo {author} {\bibfnamefont
  {B.}~\bibnamefont {Özyilmaz}},\ }\bibfield  {title} {\bibinfo {title}
  {Spin{\textendash}orbit proximity effect in graphene},\ }\href
  {https://doi.org/10.1038/ncomms5875} {\bibfield  {journal} {\bibinfo
  {journal} {Nature Communications}\ }\textbf {\bibinfo {volume} {5}},\
  \bibinfo {pages} {4875} (\bibinfo {year} {2014})}\BibitemShut {NoStop}%
\bibitem [{\citenamefont {Ben{\'{\i}}tez}\ \emph {et~al.}(2017)\citenamefont
  {Ben{\'{\i}}tez}, \citenamefont {Sierra}, \citenamefont {Torres},
  \citenamefont {Arrighi}, \citenamefont {Bonell}, \citenamefont {Costache},\
  and\ \citenamefont {Valenzuela}}]{Benitez2017}%
  \BibitemOpen
  \bibfield  {author} {\bibinfo {author} {\bibfnamefont {L.~A.}\ \bibnamefont
  {Ben{\'{\i}}tez}}, \bibinfo {author} {\bibfnamefont {J.~F.}\ \bibnamefont
  {Sierra}}, \bibinfo {author} {\bibfnamefont {W.~S.}\ \bibnamefont {Torres}},
  \bibinfo {author} {\bibfnamefont {A.}~\bibnamefont {Arrighi}}, \bibinfo
  {author} {\bibfnamefont {F.}~\bibnamefont {Bonell}}, \bibinfo {author}
  {\bibfnamefont {M.~V.}\ \bibnamefont {Costache}},\ and\ \bibinfo {author}
  {\bibfnamefont {S.~O.}\ \bibnamefont {Valenzuela}},\ }\bibfield  {title}
  {\bibinfo {title} {Strongly anisotropic spin relaxation in
  graphene{\textendash}transition metal dichalcogenide heterostructures at room
  temperature},\ }\href {https://doi.org/10.1038/s41567-017-0019-2} {\bibfield
  {journal} {\bibinfo  {journal} {Nature Physics}\ }\textbf {\bibinfo {volume}
  {14}},\ \bibinfo {pages} {303} (\bibinfo {year} {2017})}\BibitemShut
  {NoStop}%
\bibitem [{\citenamefont {Ghiasi}\ \emph {et~al.}(2017)\citenamefont {Ghiasi},
  \citenamefont {Ingla-Ayn{\'{e}}s}, \citenamefont {Kaverzin},\ and\
  \citenamefont {van Wees}}]{Ghiasi2017}%
  \BibitemOpen
  \bibfield  {author} {\bibinfo {author} {\bibfnamefont {T.~S.}\ \bibnamefont
  {Ghiasi}}, \bibinfo {author} {\bibfnamefont {J.}~\bibnamefont
  {Ingla-Ayn{\'{e}}s}}, \bibinfo {author} {\bibfnamefont {A.~A.}\ \bibnamefont
  {Kaverzin}},\ and\ \bibinfo {author} {\bibfnamefont {B.~J.}\ \bibnamefont
  {van Wees}},\ }\bibfield  {title} {\bibinfo {title} {Large proximity-induced
  spin lifetime anisotropy in transition-metal dichalcogenide/graphene
  heterostructures},\ }\href {https://doi.org/10.1021/acs.nanolett.7b03460}
  {\bibfield  {journal} {\bibinfo  {journal} {Nano Letters}\ }\textbf {\bibinfo
  {volume} {17}},\ \bibinfo {pages} {7528} (\bibinfo {year}
  {2017})}\BibitemShut {NoStop}%
\bibitem [{\citenamefont {Yan}\ \emph {et~al.}(2016)\citenamefont {Yan},
  \citenamefont {Txoperena}, \citenamefont {Llopis}, \citenamefont {Dery},
  \citenamefont {Hueso},\ and\ \citenamefont {Casanova}}]{Yan2016}%
  \BibitemOpen
  \bibfield  {author} {\bibinfo {author} {\bibfnamefont {W.}~\bibnamefont
  {Yan}}, \bibinfo {author} {\bibfnamefont {O.}~\bibnamefont {Txoperena}},
  \bibinfo {author} {\bibfnamefont {R.}~\bibnamefont {Llopis}}, \bibinfo
  {author} {\bibfnamefont {H.}~\bibnamefont {Dery}}, \bibinfo {author}
  {\bibfnamefont {L.~E.}\ \bibnamefont {Hueso}},\ and\ \bibinfo {author}
  {\bibfnamefont {F.}~\bibnamefont {Casanova}},\ }\bibfield  {title} {\bibinfo
  {title} {A two-dimensional spin field-effect switch},\ }\href
  {https://doi.org/10.1038/ncomms13372} {\bibfield  {journal} {\bibinfo
  {journal} {Nature Communications}\ }\textbf {\bibinfo {volume} {7}},\
  \bibinfo {pages} {13372} (\bibinfo {year} {2016})}\BibitemShut {NoStop}%
\bibitem [{\citenamefont {Dankert}\ and\ \citenamefont
  {Dash}(2017)}]{Dankert2017}%
  \BibitemOpen
  \bibfield  {author} {\bibinfo {author} {\bibfnamefont {A.}~\bibnamefont
  {Dankert}}\ and\ \bibinfo {author} {\bibfnamefont {S.~P.}\ \bibnamefont
  {Dash}},\ }\bibfield  {title} {\bibinfo {title} {Electrical gate control of
  spin current in van der waals heterostructures at room temperature},\ }\href
  {https://doi.org/10.1038/ncomms16093} {\bibfield  {journal} {\bibinfo
  {journal} {Nature Communications}\ }\textbf {\bibinfo {volume} {8}},\
  \bibinfo {pages} {16093} (\bibinfo {year} {2017})}\BibitemShut {NoStop}%
\bibitem [{\citenamefont {Sun}\ \emph {et~al.}(2023)\citenamefont {Sun},
  \citenamefont {Rademaker}, \citenamefont {Mauro}, \citenamefont {Scarfato},
  \citenamefont {P{\'a}sztor}, \citenamefont {Gutiérrez-Lezama}, \citenamefont
  {Wang}, \citenamefont {Martinez-Castro}, \citenamefont {Morpurgo},\ and\
  \citenamefont {Renner}}]{Sun2023}%
  \BibitemOpen
  \bibfield  {author} {\bibinfo {author} {\bibfnamefont {L.}~\bibnamefont
  {Sun}}, \bibinfo {author} {\bibfnamefont {L.}~\bibnamefont {Rademaker}},
  \bibinfo {author} {\bibfnamefont {D.}~\bibnamefont {Mauro}}, \bibinfo
  {author} {\bibfnamefont {A.}~\bibnamefont {Scarfato}}, \bibinfo {author}
  {\bibfnamefont {{\'A}.}~\bibnamefont {P{\'a}sztor}}, \bibinfo {author}
  {\bibfnamefont {I.}~\bibnamefont {Gutiérrez-Lezama}}, \bibinfo {author}
  {\bibfnamefont {Z.}~\bibnamefont {Wang}}, \bibinfo {author} {\bibfnamefont
  {J.}~\bibnamefont {Martinez-Castro}}, \bibinfo {author} {\bibfnamefont
  {A.~F.}\ \bibnamefont {Morpurgo}},\ and\ \bibinfo {author} {\bibfnamefont
  {C.}~\bibnamefont {Renner}},\ }\bibfield  {title} {\bibinfo {title}
  {Determining spin-orbit coupling in graphene by quasiparticle interference
  imaging},\ }\href {https://doi.org/10.1038/s41467-023-39453-x} {\bibfield
  {journal} {\bibinfo  {journal} {Nature Communications}\ }\textbf {\bibinfo
  {volume} {14}},\ \bibinfo {pages} {3771} (\bibinfo {year}
  {2023})}\BibitemShut {NoStop}%
\bibitem [{\citenamefont {Omar}\ and\ \citenamefont {van
  Wees}(2018)}]{Omar2018}%
  \BibitemOpen
  \bibfield  {author} {\bibinfo {author} {\bibfnamefont {S.}~\bibnamefont
  {Omar}}\ and\ \bibinfo {author} {\bibfnamefont {B.~J.}\ \bibnamefont {van
  Wees}},\ }\bibfield  {title} {\bibinfo {title} {Spin transport in
  high-mobility graphene on {WS}2 substrate with electric-field tunable
  proximity spin-orbit interaction},\ }\href
  {https://doi.org/10.1103/physrevb.97.045414} {\bibfield  {journal} {\bibinfo
  {journal} {Physical Review B}\ }\textbf {\bibinfo {volume} {97}},\ \bibinfo
  {pages} {045414} (\bibinfo {year} {2018})}\BibitemShut {NoStop}%
\bibitem [{\citenamefont {Benitez}\ \emph {et~al.}(2020)\citenamefont
  {Benitez}, \citenamefont {Torres}, \citenamefont {Sierra}, \citenamefont
  {Timmermans}, \citenamefont {Garcia}, \citenamefont {Roche}, \citenamefont
  {Costache},\ and\ \citenamefont {Valenzuela}}]{Benitez2019}%
  \BibitemOpen
  \bibfield  {author} {\bibinfo {author} {\bibfnamefont {L.~A.}\ \bibnamefont
  {Benitez}}, \bibinfo {author} {\bibfnamefont {W.~S.}\ \bibnamefont {Torres}},
  \bibinfo {author} {\bibfnamefont {J.~F.}\ \bibnamefont {Sierra}}, \bibinfo
  {author} {\bibfnamefont {M.}~\bibnamefont {Timmermans}}, \bibinfo {author}
  {\bibfnamefont {J.~H.}\ \bibnamefont {Garcia}}, \bibinfo {author}
  {\bibfnamefont {S.}~\bibnamefont {Roche}}, \bibinfo {author} {\bibfnamefont
  {M.~V.}\ \bibnamefont {Costache}},\ and\ \bibinfo {author} {\bibfnamefont
  {S.~O.}\ \bibnamefont {Valenzuela}},\ }\bibfield  {title} {\bibinfo {title}
  {Tunable room-temperature spin galvanic and spin hall effects in van der
  waals heterostructures},\ }\href {https://doi.org/10.1038/s41563-019-0575-1}
  {\bibfield  {journal} {\bibinfo  {journal} {Nature Materials}\ }\textbf
  {\bibinfo {volume} {19}},\ \bibinfo {pages} {170} (\bibinfo {year}
  {2020})}\BibitemShut {NoStop}%
\bibitem [{\citenamefont {Herling}\ \emph {et~al.}(2020)\citenamefont
  {Herling}, \citenamefont {Safeer}, \citenamefont {Ingla-Aynés},
  \citenamefont {Ontoso}, \citenamefont {Hueso},\ and\ \citenamefont
  {Casanova}}]{Herling2020}%
  \BibitemOpen
  \bibfield  {author} {\bibinfo {author} {\bibfnamefont {F.}~\bibnamefont
  {Herling}}, \bibinfo {author} {\bibfnamefont {C.~K.}\ \bibnamefont {Safeer}},
  \bibinfo {author} {\bibfnamefont {J.}~\bibnamefont {Ingla-Aynés}}, \bibinfo
  {author} {\bibfnamefont {N.}~\bibnamefont {Ontoso}}, \bibinfo {author}
  {\bibfnamefont {L.~E.}\ \bibnamefont {Hueso}},\ and\ \bibinfo {author}
  {\bibfnamefont {F.}~\bibnamefont {Casanova}},\ }\bibfield  {title} {\bibinfo
  {title} {Gate tunability of highly efficient spin-to-charge conversion by
  spin hall effect in graphene proximitized with wse2},\ }\href
  {https://doi.org/10.1063/5.0006101} {\bibfield  {journal} {\bibinfo
  {journal} {APL Materials}\ }\textbf {\bibinfo {volume} {8}},\ \bibinfo
  {pages} {071103} (\bibinfo {year} {2020})}\BibitemShut {NoStop}%
\bibitem [{\citenamefont {Safeer}\ \emph {et~al.}(2019)\citenamefont {Safeer},
  \citenamefont {Ingla-Ayn{\'{e}}s}, \citenamefont {Herling}, \citenamefont
  {Garcia}, \citenamefont {Vila}, \citenamefont {Ontoso}, \citenamefont
  {Calvo}, \citenamefont {Roche}, \citenamefont {Hueso},\ and\ \citenamefont
  {Casanova}}]{Safeer2019}%
  \BibitemOpen
  \bibfield  {author} {\bibinfo {author} {\bibfnamefont {C.~K.}\ \bibnamefont
  {Safeer}}, \bibinfo {author} {\bibfnamefont {J.}~\bibnamefont
  {Ingla-Ayn{\'{e}}s}}, \bibinfo {author} {\bibfnamefont {F.}~\bibnamefont
  {Herling}}, \bibinfo {author} {\bibfnamefont {J.~H.}\ \bibnamefont {Garcia}},
  \bibinfo {author} {\bibfnamefont {M.}~\bibnamefont {Vila}}, \bibinfo {author}
  {\bibfnamefont {N.}~\bibnamefont {Ontoso}}, \bibinfo {author} {\bibfnamefont
  {M.~R.}\ \bibnamefont {Calvo}}, \bibinfo {author} {\bibfnamefont
  {S.}~\bibnamefont {Roche}}, \bibinfo {author} {\bibfnamefont {L.~E.}\
  \bibnamefont {Hueso}},\ and\ \bibinfo {author} {\bibfnamefont
  {F.}~\bibnamefont {Casanova}},\ }\bibfield  {title} {\bibinfo {title}
  {Room-temperature spin hall effect in graphene/{MoS}2 van der waals
  heterostructures},\ }\href {https://doi.org/10.1021/acs.nanolett.8b04368}
  {\bibfield  {journal} {\bibinfo  {journal} {Nano Letters}\ }\textbf {\bibinfo
  {volume} {19}},\ \bibinfo {pages} {1074} (\bibinfo {year}
  {2019})}\BibitemShut {NoStop}%
\bibitem [{\citenamefont {Yang}\ \emph {et~al.}(2024)\citenamefont {Yang},
  \citenamefont {Martín-García}, \citenamefont {Kimák}, \citenamefont
  {Schmoranzerová}, \citenamefont {Dolan}, \citenamefont {Chi}, \citenamefont
  {Gobbi}, \citenamefont {Němec}, \citenamefont {Hueso},\ and\ \citenamefont
  {Casanova}}]{Yang2024}%
  \BibitemOpen
  \bibfield  {author} {\bibinfo {author} {\bibfnamefont {H.}~\bibnamefont
  {Yang}}, \bibinfo {author} {\bibfnamefont {B.}~\bibnamefont
  {Martín-García}}, \bibinfo {author} {\bibfnamefont {J.}~\bibnamefont
  {Kimák}}, \bibinfo {author} {\bibfnamefont {E.}~\bibnamefont
  {Schmoranzerová}}, \bibinfo {author} {\bibfnamefont {E.}~\bibnamefont
  {Dolan}}, \bibinfo {author} {\bibfnamefont {Z.}~\bibnamefont {Chi}}, \bibinfo
  {author} {\bibfnamefont {M.}~\bibnamefont {Gobbi}}, \bibinfo {author}
  {\bibfnamefont {P.}~\bibnamefont {Němec}}, \bibinfo {author} {\bibfnamefont
  {L.~E.}\ \bibnamefont {Hueso}},\ and\ \bibinfo {author} {\bibfnamefont
  {F.}~\bibnamefont {Casanova}},\ }\bibfield  {title} {\bibinfo {title}
  {Twist-angle-tunable spin texture in wse2/graphene van der waals
  heterostructures},\ }\bibfield  {journal} {\bibinfo  {journal} {Nature
  Materials}\ }\href {https://doi.org/10.1038/s41563-024-01985-y}
  {10.1038/s41563-024-01985-y} (\bibinfo {year} {2024})\BibitemShut {NoStop}%
\bibitem [{\citenamefont {Zollner}\ and\ \citenamefont
  {Fabian}(2021)}]{Zollner2021}%
  \BibitemOpen
  \bibfield  {author} {\bibinfo {author} {\bibfnamefont {K.}~\bibnamefont
  {Zollner}}\ and\ \bibinfo {author} {\bibfnamefont {J.}~\bibnamefont
  {Fabian}},\ }\bibfield  {title} {\bibinfo {title} {Bilayer graphene
  encapsulated within monolayers of {WS}2 or cr2ge2te6 : Tunable proximity
  spin-orbit or exchange coupling},\ }\href
  {https://doi.org/10.1103/physrevb.104.075126} {\bibfield  {journal} {\bibinfo
   {journal} {Physical Review B}\ }\textbf {\bibinfo {volume} {104}},\ \bibinfo
  {pages} {075126} (\bibinfo {year} {2021})}\BibitemShut {NoStop}%
\bibitem [{\citenamefont {Lin}\ \emph {et~al.}(2022)\citenamefont {Lin},
  \citenamefont {Zhang}, \citenamefont {Morissette}, \citenamefont {Wang},
  \citenamefont {Liu}, \citenamefont {Rhodes}, \citenamefont {Watanabe},
  \citenamefont {Taniguchi}, \citenamefont {Hone},\ and\ \citenamefont
  {Li}}]{Lin2022}%
  \BibitemOpen
  \bibfield  {author} {\bibinfo {author} {\bibfnamefont {J.-X.}\ \bibnamefont
  {Lin}}, \bibinfo {author} {\bibfnamefont {Y.-H.}\ \bibnamefont {Zhang}},
  \bibinfo {author} {\bibfnamefont {E.}~\bibnamefont {Morissette}}, \bibinfo
  {author} {\bibfnamefont {Z.}~\bibnamefont {Wang}}, \bibinfo {author}
  {\bibfnamefont {S.}~\bibnamefont {Liu}}, \bibinfo {author} {\bibfnamefont
  {D.}~\bibnamefont {Rhodes}}, \bibinfo {author} {\bibfnamefont
  {K.}~\bibnamefont {Watanabe}}, \bibinfo {author} {\bibfnamefont
  {T.}~\bibnamefont {Taniguchi}}, \bibinfo {author} {\bibfnamefont
  {J.}~\bibnamefont {Hone}},\ and\ \bibinfo {author} {\bibfnamefont {J.~I.~A.}\
  \bibnamefont {Li}},\ }\bibfield  {title} {\bibinfo {title}
  {Spin-orbit–driven ferromagnetism at half moiré filling in magic-angle
  twisted bilayer graphene},\ }\href {https://doi.org/10.1126/science.abh2889}
  {\bibfield  {journal} {\bibinfo  {journal} {Science}\ }\textbf {\bibinfo
  {volume} {375}},\ \bibinfo {pages} {437} (\bibinfo {year}
  {2022})}\BibitemShut {NoStop}%
\bibitem [{\citenamefont {Bhowmik}\ \emph {et~al.}(2023)\citenamefont
  {Bhowmik}, \citenamefont {Ghawri}, \citenamefont {Park}, \citenamefont {Lee},
  \citenamefont {Datta}, \citenamefont {Soni}, \citenamefont {Watanabe},
  \citenamefont {Taniguchi}, \citenamefont {Ghosh}, \citenamefont {Jung},\ and\
  \citenamefont {Chandni}}]{Bhowmik2023}%
  \BibitemOpen
  \bibfield  {author} {\bibinfo {author} {\bibfnamefont {S.}~\bibnamefont
  {Bhowmik}}, \bibinfo {author} {\bibfnamefont {B.}~\bibnamefont {Ghawri}},
  \bibinfo {author} {\bibfnamefont {Y.}~\bibnamefont {Park}}, \bibinfo {author}
  {\bibfnamefont {D.}~\bibnamefont {Lee}}, \bibinfo {author} {\bibfnamefont
  {S.}~\bibnamefont {Datta}}, \bibinfo {author} {\bibfnamefont
  {R.}~\bibnamefont {Soni}}, \bibinfo {author} {\bibfnamefont {K.}~\bibnamefont
  {Watanabe}}, \bibinfo {author} {\bibfnamefont {T.}~\bibnamefont {Taniguchi}},
  \bibinfo {author} {\bibfnamefont {A.}~\bibnamefont {Ghosh}}, \bibinfo
  {author} {\bibfnamefont {J.}~\bibnamefont {Jung}},\ and\ \bibinfo {author}
  {\bibfnamefont {U.}~\bibnamefont {Chandni}},\ }\bibfield  {title} {\bibinfo
  {title} {Spin-orbit coupling-enhanced valley ordering of malleable bands in
  twisted bilayer graphene on wse2},\ }\href
  {https://doi.org/10.1038/s41467-023-39855-x} {\bibfield  {journal} {\bibinfo
  {journal} {Nature Communications}\ }\textbf {\bibinfo {volume} {14}},\
  \bibinfo {pages} {4055} (\bibinfo {year} {2023})}\BibitemShut {NoStop}%
\bibitem [{\citenamefont {Chou}\ \emph {et~al.}(2024)\citenamefont {Chou},
  \citenamefont {Tan}, \citenamefont {Wu},\ and\ \citenamefont
  {Das~Sarma}}]{Chou2024}%
  \BibitemOpen
  \bibfield  {author} {\bibinfo {author} {\bibfnamefont {Y.-Z.}\ \bibnamefont
  {Chou}}, \bibinfo {author} {\bibfnamefont {Y.}~\bibnamefont {Tan}}, \bibinfo
  {author} {\bibfnamefont {F.}~\bibnamefont {Wu}},\ and\ \bibinfo {author}
  {\bibfnamefont {S.}~\bibnamefont {Das~Sarma}},\ }\bibfield  {title} {\bibinfo
  {title} {Topological flat bands, valley polarization, and interband
  superconductivity in magic-angle twisted bilayer graphene with proximitized
  spin-orbit couplings},\ }\href {https://doi.org/10.1103/physrevb.110.l041108}
  {\bibfield  {journal} {\bibinfo  {journal} {Physical Review B}\ }\textbf
  {\bibinfo {volume} {110}},\ \bibinfo {pages} {L041108} (\bibinfo {year}
  {2024})}\BibitemShut {NoStop}%
\bibitem [{\citenamefont {Zhang}\ \emph {et~al.}(2024)\citenamefont {Zhang},
  \citenamefont {Shavit}, \citenamefont {Ma}, \citenamefont {Han},
  \citenamefont {Watanabe}, \citenamefont {Taniguchi}, \citenamefont {Hsieh},
  \citenamefont {Lewandowski}, \citenamefont {von Oppen}, \citenamefont
  {Oreg},\ and\ \citenamefont {Nadj-Perge}}]{Zhang2024}%
  \BibitemOpen
  \bibfield  {author} {\bibinfo {author} {\bibfnamefont {Y.}~\bibnamefont
  {Zhang}}, \bibinfo {author} {\bibfnamefont {G.}~\bibnamefont {Shavit}},
  \bibinfo {author} {\bibfnamefont {H.}~\bibnamefont {Ma}}, \bibinfo {author}
  {\bibfnamefont {Y.}~\bibnamefont {Han}}, \bibinfo {author} {\bibfnamefont
  {K.}~\bibnamefont {Watanabe}}, \bibinfo {author} {\bibfnamefont
  {T.}~\bibnamefont {Taniguchi}}, \bibinfo {author} {\bibfnamefont
  {D.}~\bibnamefont {Hsieh}}, \bibinfo {author} {\bibfnamefont
  {C.}~\bibnamefont {Lewandowski}}, \bibinfo {author} {\bibfnamefont
  {F.}~\bibnamefont {von Oppen}}, \bibinfo {author} {\bibfnamefont
  {Y.}~\bibnamefont {Oreg}},\ and\ \bibinfo {author} {\bibfnamefont
  {S.}~\bibnamefont {Nadj-Perge}},\ }\href
  {https://doi.org/10.48550/ARXIV.2408.10335} {\bibinfo {title}
  {Twist-programmable superconductivity in spin-orbit coupled bilayer
  graphene}} (\bibinfo {year} {2024})\BibitemShut {NoStop}%
\bibitem [{\citenamefont {Fülöp}\ \emph
  {et~al.}(2021{\natexlab{a}})\citenamefont {Fülöp}, \citenamefont
  {M{\'{a}}rffy}, \citenamefont {Zihlmann}, \citenamefont {Gmitra},
  \citenamefont {T{\'{o}}v{\'{a}}ri}, \citenamefont {Szentp{\'{e}}teri},
  \citenamefont {Kedves}, \citenamefont {Watanabe}, \citenamefont {Taniguchi},
  \citenamefont {Fabian}, \citenamefont {Schönenberger}, \citenamefont
  {Makk},\ and\ \citenamefont {Csonka}}]{Fueloep2021}%
  \BibitemOpen
  \bibfield  {author} {\bibinfo {author} {\bibfnamefont {B.}~\bibnamefont
  {Fülöp}}, \bibinfo {author} {\bibfnamefont {A.}~\bibnamefont
  {M{\'{a}}rffy}}, \bibinfo {author} {\bibfnamefont {S.}~\bibnamefont
  {Zihlmann}}, \bibinfo {author} {\bibfnamefont {M.}~\bibnamefont {Gmitra}},
  \bibinfo {author} {\bibfnamefont {E.}~\bibnamefont {T{\'{o}}v{\'{a}}ri}},
  \bibinfo {author} {\bibfnamefont {B.}~\bibnamefont {Szentp{\'{e}}teri}},
  \bibinfo {author} {\bibfnamefont {M.}~\bibnamefont {Kedves}}, \bibinfo
  {author} {\bibfnamefont {K.}~\bibnamefont {Watanabe}}, \bibinfo {author}
  {\bibfnamefont {T.}~\bibnamefont {Taniguchi}}, \bibinfo {author}
  {\bibfnamefont {J.}~\bibnamefont {Fabian}}, \bibinfo {author} {\bibfnamefont
  {C.}~\bibnamefont {Schönenberger}}, \bibinfo {author} {\bibfnamefont
  {P.}~\bibnamefont {Makk}},\ and\ \bibinfo {author} {\bibfnamefont
  {S.}~\bibnamefont {Csonka}},\ }\bibfield  {title} {\bibinfo {title} {Boosting
  proximity spin{\textendash}orbit coupling in graphene/{WSe}2 heterostructures
  via hydrostatic pressure},\ }\href
  {https://doi.org/10.1038/s41699-021-00262-9} {\bibfield  {journal} {\bibinfo
  {journal} {npj 2D Materials and Applications}\ }\textbf {\bibinfo {volume}
  {5}},\ \bibinfo {pages} {82} (\bibinfo {year} {2021}{\natexlab{a}})},\
  \Eprint {https://arxiv.org/abs/2103.13325} {2103.13325} \BibitemShut
  {NoStop}%
\bibitem [{\citenamefont {Kedves}\ \emph {et~al.}(2023)\citenamefont {Kedves},
  \citenamefont {Szentpéteri}, \citenamefont {Márffy}, \citenamefont
  {Tóvári}, \citenamefont {Papadopoulos}, \citenamefont {Rout}, \citenamefont
  {Watanabe}, \citenamefont {Taniguchi}, \citenamefont {Goswami}, \citenamefont
  {Csonka},\ and\ \citenamefont {Makk}}]{Kedves2023}%
  \BibitemOpen
  \bibfield  {author} {\bibinfo {author} {\bibfnamefont {M.}~\bibnamefont
  {Kedves}}, \bibinfo {author} {\bibfnamefont {B.}~\bibnamefont
  {Szentpéteri}}, \bibinfo {author} {\bibfnamefont {A.}~\bibnamefont
  {Márffy}}, \bibinfo {author} {\bibfnamefont {E.}~\bibnamefont {Tóvári}},
  \bibinfo {author} {\bibfnamefont {N.}~\bibnamefont {Papadopoulos}}, \bibinfo
  {author} {\bibfnamefont {P.~K.}\ \bibnamefont {Rout}}, \bibinfo {author}
  {\bibfnamefont {K.}~\bibnamefont {Watanabe}}, \bibinfo {author}
  {\bibfnamefont {T.}~\bibnamefont {Taniguchi}}, \bibinfo {author}
  {\bibfnamefont {S.}~\bibnamefont {Goswami}}, \bibinfo {author} {\bibfnamefont
  {S.}~\bibnamefont {Csonka}},\ and\ \bibinfo {author} {\bibfnamefont
  {P.}~\bibnamefont {Makk}},\ }\bibfield  {title} {\bibinfo {title}
  {Stabilizing the inverted phase of a wse2/blg/wse2 heterostructure via
  hydrostatic pressure},\ }\href {https://doi.org/10.1021/acs.nanolett.3c03029}
  {\bibfield  {journal} {\bibinfo  {journal} {Nano Letters}\ }\textbf {\bibinfo
  {volume} {23}},\ \bibinfo {pages} {9508} (\bibinfo {year}
  {2023})}\BibitemShut {NoStop}%
\bibitem [{\citenamefont {Khoo}\ \emph {et~al.}(2017)\citenamefont {Khoo},
  \citenamefont {Morpurgo},\ and\ \citenamefont {Levitov}}]{Khoo2017}%
  \BibitemOpen
  \bibfield  {author} {\bibinfo {author} {\bibfnamefont {J.~Y.}\ \bibnamefont
  {Khoo}}, \bibinfo {author} {\bibfnamefont {A.~F.}\ \bibnamefont {Morpurgo}},\
  and\ \bibinfo {author} {\bibfnamefont {L.}~\bibnamefont {Levitov}},\
  }\bibfield  {title} {\bibinfo {title} {On-demand spin–orbit interaction
  from which-layer tunability in bilayer graphene},\ }\href
  {https://doi.org/10.1021/acs.nanolett.7b03604} {\bibfield  {journal}
  {\bibinfo  {journal} {Nano Letters}\ }\textbf {\bibinfo {volume} {17}},\
  \bibinfo {pages} {7003} (\bibinfo {year} {2017})}\BibitemShut {NoStop}%
\bibitem [{sup()}]{supmat}%
  \BibitemOpen
  \href@noop {} {}\bibinfo {note} {See Supplemental Material
  [\url{https://journals.aps.org/prb/abstract/10.1103/PhysRevB.111.205415}] for
  device fabrication and further transport measurements on the device presented
  in the main text and on two other devices; conversion of gate voltages to
  charge density and electric displacement field; details of the low-energy
  model in zero and large transverse magnetic fields; measurements of the
  $\nu=\pm3$ LL crossings; Measurements of SdH oscillations and the extraction
  of the Fermi-surfaces from them; more WL measurements at different
  displacement fields. The extraction of the Ising-type SOC strength in two
  other devices and the Rashba-type SOC strength in one device. The
  Supplemental Material also contains Refs.\cite{Kim2016a, Solozhenko1995,
  Yankowitz2018, Fueloep2021a, McCann2013, Jung2014, Island2019, Khoo2017,
  Amann2022, Naimer2021, Wang2019}.}\BibitemShut {Stop}%
\bibitem [{\citenamefont {McCann}\ and\ \citenamefont
  {Koshino}(2013)}]{McCann2013}%
  \BibitemOpen
  \bibfield  {author} {\bibinfo {author} {\bibfnamefont {E.}~\bibnamefont
  {McCann}}\ and\ \bibinfo {author} {\bibfnamefont {M.}~\bibnamefont
  {Koshino}},\ }\bibfield  {title} {\bibinfo {title} {The electronic properties
  of bilayer graphene},\ }\href {https://doi.org/10.1088/0034-4885/76/5/056503}
  {\bibfield  {journal} {\bibinfo  {journal} {Reports on Progress in Physics}\
  }\textbf {\bibinfo {volume} {76}},\ \bibinfo {pages} {056503} (\bibinfo
  {year} {2013})}\BibitemShut {NoStop}%
\bibitem [{\citenamefont {Gorbachev}\ \emph {et~al.}(2007)\citenamefont
  {Gorbachev}, \citenamefont {Tikhonenko}, \citenamefont {Mayorov},
  \citenamefont {Horsell},\ and\ \citenamefont {Savchenko}}]{Gorbachev2007}%
  \BibitemOpen
  \bibfield  {author} {\bibinfo {author} {\bibfnamefont {R.~V.}\ \bibnamefont
  {Gorbachev}}, \bibinfo {author} {\bibfnamefont {F.~V.}\ \bibnamefont
  {Tikhonenko}}, \bibinfo {author} {\bibfnamefont {A.~S.}\ \bibnamefont
  {Mayorov}}, \bibinfo {author} {\bibfnamefont {D.~W.}\ \bibnamefont
  {Horsell}},\ and\ \bibinfo {author} {\bibfnamefont {A.~K.}\ \bibnamefont
  {Savchenko}},\ }\bibfield  {title} {\bibinfo {title} {Weak localization in
  bilayer graphene},\ }\href {https://doi.org/10.1103/physrevlett.98.176805}
  {\bibfield  {journal} {\bibinfo  {journal} {Physical Review Letters}\
  }\textbf {\bibinfo {volume} {98}},\ \bibinfo {pages} {176805} (\bibinfo
  {year} {2007})}\BibitemShut {NoStop}%
\bibitem [{\citenamefont {Volkl}\ \emph {et~al.}(2017)\citenamefont {Volkl},
  \citenamefont {Rockinger}, \citenamefont {Drienovsky}, \citenamefont
  {Watanabe}, \citenamefont {Taniguchi}, \citenamefont {Weiss},\ and\
  \citenamefont {Eroms}}]{Voelkl2017}%
  \BibitemOpen
  \bibfield  {author} {\bibinfo {author} {\bibfnamefont {T.}~\bibnamefont
  {Volkl}}, \bibinfo {author} {\bibfnamefont {T.}~\bibnamefont {Rockinger}},
  \bibinfo {author} {\bibfnamefont {M.}~\bibnamefont {Drienovsky}}, \bibinfo
  {author} {\bibfnamefont {K.}~\bibnamefont {Watanabe}}, \bibinfo {author}
  {\bibfnamefont {T.}~\bibnamefont {Taniguchi}}, \bibinfo {author}
  {\bibfnamefont {D.}~\bibnamefont {Weiss}},\ and\ \bibinfo {author}
  {\bibfnamefont {J.}~\bibnamefont {Eroms}},\ }\bibfield  {title} {\bibinfo
  {title} {Magnetotransport in heterostructures of transition metal
  dichalcogenides and graphene},\ }\href
  {https://doi.org/10.1103/physrevb.96.125405} {\bibfield  {journal} {\bibinfo
  {journal} {Physical Review B}\ }\textbf {\bibinfo {volume} {96}},\ \bibinfo
  {pages} {125405} (\bibinfo {year} {2017})}\BibitemShut {NoStop}%
\bibitem [{\citenamefont {Island}\ \emph {et~al.}(2019)\citenamefont {Island},
  \citenamefont {Cui}, \citenamefont {Lewandowski}, \citenamefont {Khoo},
  \citenamefont {Spanton}, \citenamefont {Zhou}, \citenamefont {Rhodes},
  \citenamefont {Hone}, \citenamefont {Taniguchi}, \citenamefont {Watanabe},
  \citenamefont {Levitov}, \citenamefont {Zaletel},\ and\ \citenamefont
  {Young}}]{Island2019}%
  \BibitemOpen
  \bibfield  {author} {\bibinfo {author} {\bibfnamefont {J.~O.}\ \bibnamefont
  {Island}}, \bibinfo {author} {\bibfnamefont {X.}~\bibnamefont {Cui}},
  \bibinfo {author} {\bibfnamefont {C.}~\bibnamefont {Lewandowski}}, \bibinfo
  {author} {\bibfnamefont {J.~Y.}\ \bibnamefont {Khoo}}, \bibinfo {author}
  {\bibfnamefont {E.~M.}\ \bibnamefont {Spanton}}, \bibinfo {author}
  {\bibfnamefont {H.}~\bibnamefont {Zhou}}, \bibinfo {author} {\bibfnamefont
  {D.}~\bibnamefont {Rhodes}}, \bibinfo {author} {\bibfnamefont {J.~C.}\
  \bibnamefont {Hone}}, \bibinfo {author} {\bibfnamefont {T.}~\bibnamefont
  {Taniguchi}}, \bibinfo {author} {\bibfnamefont {K.}~\bibnamefont {Watanabe}},
  \bibinfo {author} {\bibfnamefont {L.~S.}\ \bibnamefont {Levitov}}, \bibinfo
  {author} {\bibfnamefont {M.~P.}\ \bibnamefont {Zaletel}},\ and\ \bibinfo
  {author} {\bibfnamefont {A.~F.}\ \bibnamefont {Young}},\ }\bibfield  {title}
  {\bibinfo {title} {Spin{\textendash}orbit-driven band inversion in bilayer
  graphene by the van der waals proximity effect},\ }\href
  {https://doi.org/10.1038/s41586-019-1304-2} {\bibfield  {journal} {\bibinfo
  {journal} {Nature}\ }\textbf {\bibinfo {volume} {571}},\ \bibinfo {pages}
  {85} (\bibinfo {year} {2019})}\BibitemShut {NoStop}%
\bibitem [{\citenamefont {Wang}\ \emph {et~al.}(2019)\citenamefont {Wang},
  \citenamefont {Che}, \citenamefont {Cao}, \citenamefont {Lyu}, \citenamefont
  {Watanabe}, \citenamefont {Taniguchi}, \citenamefont {Lau},\ and\
  \citenamefont {Bockrath}}]{Wang2019}%
  \BibitemOpen
  \bibfield  {author} {\bibinfo {author} {\bibfnamefont {D.}~\bibnamefont
  {Wang}}, \bibinfo {author} {\bibfnamefont {S.}~\bibnamefont {Che}}, \bibinfo
  {author} {\bibfnamefont {G.}~\bibnamefont {Cao}}, \bibinfo {author}
  {\bibfnamefont {R.}~\bibnamefont {Lyu}}, \bibinfo {author} {\bibfnamefont
  {K.}~\bibnamefont {Watanabe}}, \bibinfo {author} {\bibfnamefont
  {T.}~\bibnamefont {Taniguchi}}, \bibinfo {author} {\bibfnamefont {C.~N.}\
  \bibnamefont {Lau}},\ and\ \bibinfo {author} {\bibfnamefont {M.}~\bibnamefont
  {Bockrath}},\ }\bibfield  {title} {\bibinfo {title} {Quantum hall effect
  measurement of spin-orbit coupling strengths in ultraclean bilayer
  graphene/{WSe}2 heterostructures},\ }\href
  {https://doi.org/10.1021/acs.nanolett.9b02445} {\bibfield  {journal}
  {\bibinfo  {journal} {Nano Letters}\ }\textbf {\bibinfo {volume} {19}},\
  \bibinfo {pages} {7028} (\bibinfo {year} {2019})}\BibitemShut {NoStop}%
\bibitem [{\citenamefont {Zhang}\ \emph {et~al.}(2023)\citenamefont {Zhang},
  \citenamefont {Polski}, \citenamefont {Thomson}, \citenamefont
  {Lantagne-Hurtubise}, \citenamefont {Lewandowski}, \citenamefont {Zhou},
  \citenamefont {Watanabe}, \citenamefont {Taniguchi}, \citenamefont {Alicea},\
  and\ \citenamefont {Nadj-Perge}}]{Zhang2023}%
  \BibitemOpen
  \bibfield  {author} {\bibinfo {author} {\bibfnamefont {Y.}~\bibnamefont
  {Zhang}}, \bibinfo {author} {\bibfnamefont {R.}~\bibnamefont {Polski}},
  \bibinfo {author} {\bibfnamefont {A.}~\bibnamefont {Thomson}}, \bibinfo
  {author} {\bibfnamefont {{\'E}.}~\bibnamefont {Lantagne-Hurtubise}}, \bibinfo
  {author} {\bibfnamefont {C.}~\bibnamefont {Lewandowski}}, \bibinfo {author}
  {\bibfnamefont {H.}~\bibnamefont {Zhou}}, \bibinfo {author} {\bibfnamefont
  {K.}~\bibnamefont {Watanabe}}, \bibinfo {author} {\bibfnamefont
  {T.}~\bibnamefont {Taniguchi}}, \bibinfo {author} {\bibfnamefont
  {J.}~\bibnamefont {Alicea}},\ and\ \bibinfo {author} {\bibfnamefont
  {S.}~\bibnamefont {Nadj-Perge}},\ }\bibfield  {title} {\bibinfo {title}
  {Enhanced superconductivity in spin–orbit proximitized bilayer graphene},\
  }\href {https://doi.org/10.1038/s41586-022-05446-x} {\bibfield  {journal}
  {\bibinfo  {journal} {Nature}\ }\textbf {\bibinfo {volume} {613}},\ \bibinfo
  {pages} {268} (\bibinfo {year} {2023})}\BibitemShut {NoStop}%
\bibitem [{\citenamefont {Khoo}\ and\ \citenamefont
  {Levitov}(2018)}]{Khoo2018}%
  \BibitemOpen
  \bibfield  {author} {\bibinfo {author} {\bibfnamefont {J.~Y.}\ \bibnamefont
  {Khoo}}\ and\ \bibinfo {author} {\bibfnamefont {L.}~\bibnamefont {Levitov}},\
  }\bibfield  {title} {\bibinfo {title} {Tunable quantum hall edge conduction
  in bilayer graphene through spin-orbit interaction},\ }\href
  {https://doi.org/10.1103/physrevb.98.115307} {\bibfield  {journal} {\bibinfo
  {journal} {Physical Review B}\ }\textbf {\bibinfo {volume} {98}},\ \bibinfo
  {pages} {115307} (\bibinfo {year} {2018})}\BibitemShut {NoStop}%
\bibitem [{\citenamefont {Hunt}\ \emph {et~al.}(2017)\citenamefont {Hunt},
  \citenamefont {Li}, \citenamefont {Zibrov}, \citenamefont {Wang},
  \citenamefont {Taniguchi}, \citenamefont {Watanabe}, \citenamefont {Hone},
  \citenamefont {Dean}, \citenamefont {Zaletel}, \citenamefont {Ashoori},\ and\
  \citenamefont {Young}}]{Hunt2017}%
  \BibitemOpen
  \bibfield  {author} {\bibinfo {author} {\bibfnamefont {B.~M.}\ \bibnamefont
  {Hunt}}, \bibinfo {author} {\bibfnamefont {J.~I.~A.}\ \bibnamefont {Li}},
  \bibinfo {author} {\bibfnamefont {A.~A.}\ \bibnamefont {Zibrov}}, \bibinfo
  {author} {\bibfnamefont {L.}~\bibnamefont {Wang}}, \bibinfo {author}
  {\bibfnamefont {T.}~\bibnamefont {Taniguchi}}, \bibinfo {author}
  {\bibfnamefont {K.}~\bibnamefont {Watanabe}}, \bibinfo {author}
  {\bibfnamefont {J.}~\bibnamefont {Hone}}, \bibinfo {author} {\bibfnamefont
  {C.~R.}\ \bibnamefont {Dean}}, \bibinfo {author} {\bibfnamefont
  {M.}~\bibnamefont {Zaletel}}, \bibinfo {author} {\bibfnamefont {R.~C.}\
  \bibnamefont {Ashoori}},\ and\ \bibinfo {author} {\bibfnamefont {A.~F.}\
  \bibnamefont {Young}},\ }\bibfield  {title} {\bibinfo {title} {Direct
  measurement of discrete valley and orbital quantum numbers in bilayer
  graphene},\ }\href {https://doi.org/10.1038/s41467-017-00824-w} {\bibfield
  {journal} {\bibinfo  {journal} {Nature Communications}\ }\textbf {\bibinfo
  {volume} {8}},\ \bibinfo {pages} {948} (\bibinfo {year} {2017})}\BibitemShut
  {NoStop}%
\bibitem [{\citenamefont {David}\ \emph {et~al.}(2019)\citenamefont {David},
  \citenamefont {Rakyta}, \citenamefont {Korm{\'{a}}nyos},\ and\ \citenamefont
  {Burkard}}]{David2019}%
  \BibitemOpen
  \bibfield  {author} {\bibinfo {author} {\bibfnamefont {A.}~\bibnamefont
  {David}}, \bibinfo {author} {\bibfnamefont {P.}~\bibnamefont {Rakyta}},
  \bibinfo {author} {\bibfnamefont {A.}~\bibnamefont {Korm{\'{a}}nyos}},\ and\
  \bibinfo {author} {\bibfnamefont {G.}~\bibnamefont {Burkard}},\ }\bibfield
  {title} {\bibinfo {title} {Induced spin-orbit coupling in twisted
  graphene{\textendash}transition metal dichalcogenide heterobilayers:
  Twistronics meets spintronics},\ }\href
  {https://doi.org/10.1103/physrevb.100.085412} {\bibfield  {journal} {\bibinfo
   {journal} {Physical Review B}\ }\textbf {\bibinfo {volume} {100}},\ \bibinfo
  {pages} {085412} (\bibinfo {year} {2019})}\BibitemShut {NoStop}%
\bibitem [{\citenamefont {Li}\ and\ \citenamefont {Koshino}(2019)}]{Li2019}%
  \BibitemOpen
  \bibfield  {author} {\bibinfo {author} {\bibfnamefont {Y.}~\bibnamefont
  {Li}}\ and\ \bibinfo {author} {\bibfnamefont {M.}~\bibnamefont {Koshino}},\
  }\bibfield  {title} {\bibinfo {title} {Twist-angle dependence of the
  proximity spin-orbit coupling in graphene on transition-metal
  dichalcogenides},\ }\href {https://doi.org/10.1103/physrevb.99.075438}
  {\bibfield  {journal} {\bibinfo  {journal} {Physical Review B}\ }\textbf
  {\bibinfo {volume} {99}},\ \bibinfo {pages} {075438} (\bibinfo {year}
  {2019})}\BibitemShut {NoStop}%
\bibitem [{\citenamefont {Naimer}\ \emph {et~al.}(2021)\citenamefont {Naimer},
  \citenamefont {Zollner}, \citenamefont {Gmitra},\ and\ \citenamefont
  {Fabian}}]{Naimer2021}%
  \BibitemOpen
  \bibfield  {author} {\bibinfo {author} {\bibfnamefont {T.}~\bibnamefont
  {Naimer}}, \bibinfo {author} {\bibfnamefont {K.}~\bibnamefont {Zollner}},
  \bibinfo {author} {\bibfnamefont {M.}~\bibnamefont {Gmitra}},\ and\ \bibinfo
  {author} {\bibfnamefont {J.}~\bibnamefont {Fabian}},\ }\bibfield  {title}
  {\bibinfo {title} {Twist-angle dependent proximity induced spin-orbit
  coupling in graphene/transition metal dichalcogenide heterostructures},\
  }\href {https://doi.org/10.1103/physrevb.104.195156} {\bibfield  {journal}
  {\bibinfo  {journal} {Physical Review B}\ }\textbf {\bibinfo {volume}
  {104}},\ \bibinfo {pages} {195156} (\bibinfo {year} {2021})}\BibitemShut
  {NoStop}%
\bibitem [{\citenamefont {Fülöp}\ \emph
  {et~al.}(2021{\natexlab{b}})\citenamefont {Fülöp}, \citenamefont
  {M{\'{a}}rffy}, \citenamefont {T{\'{o}}v{\'{a}}ri}, \citenamefont {Kedves},
  \citenamefont {Zihlmann}, \citenamefont {Indolese}, \citenamefont
  {Kov{\'{a}}cs-Krausz}, \citenamefont {Watanabe}, \citenamefont {Taniguchi},
  \citenamefont {Schonenberger}, \citenamefont {K{\'{e}}zsm{\'{a}}rki},
  \citenamefont {Makk},\ and\ \citenamefont {Csonka}}]{Fueloep2021a}%
  \BibitemOpen
  \bibfield  {author} {\bibinfo {author} {\bibfnamefont {B.}~\bibnamefont
  {Fülöp}}, \bibinfo {author} {\bibfnamefont {A.}~\bibnamefont
  {M{\'{a}}rffy}}, \bibinfo {author} {\bibfnamefont {E.}~\bibnamefont
  {T{\'{o}}v{\'{a}}ri}}, \bibinfo {author} {\bibfnamefont {M.}~\bibnamefont
  {Kedves}}, \bibinfo {author} {\bibfnamefont {S.}~\bibnamefont {Zihlmann}},
  \bibinfo {author} {\bibfnamefont {D.}~\bibnamefont {Indolese}}, \bibinfo
  {author} {\bibfnamefont {Z.}~\bibnamefont {Kov{\'{a}}cs-Krausz}}, \bibinfo
  {author} {\bibfnamefont {K.}~\bibnamefont {Watanabe}}, \bibinfo {author}
  {\bibfnamefont {T.}~\bibnamefont {Taniguchi}}, \bibinfo {author}
  {\bibfnamefont {C.}~\bibnamefont {Schonenberger}}, \bibinfo {author}
  {\bibfnamefont {I.}~\bibnamefont {K{\'{e}}zsm{\'{a}}rki}}, \bibinfo {author}
  {\bibfnamefont {P.}~\bibnamefont {Makk}},\ and\ \bibinfo {author}
  {\bibfnamefont {S.}~\bibnamefont {Csonka}},\ }\bibfield  {title} {\bibinfo
  {title} {New method of transport measurements on van der waals
  heterostructures under pressure},\ }\href {https://doi.org/10.1063/5.0058583}
  {\bibfield  {journal} {\bibinfo  {journal} {Journal of Applied Physics}\
  }\textbf {\bibinfo {volume} {130}},\ \bibinfo {pages} {064303} (\bibinfo
  {year} {2021}{\natexlab{b}})}\BibitemShut {NoStop}%
\bibitem [{\citenamefont {Soule}\ \emph {et~al.}(1964)\citenamefont {Soule},
  \citenamefont {McClure},\ and\ \citenamefont {Smith}}]{Soule1964}%
  \BibitemOpen
  \bibfield  {author} {\bibinfo {author} {\bibfnamefont {D.~E.}\ \bibnamefont
  {Soule}}, \bibinfo {author} {\bibfnamefont {J.~W.}\ \bibnamefont {McClure}},\
  and\ \bibinfo {author} {\bibfnamefont {L.~B.}\ \bibnamefont {Smith}},\
  }\bibfield  {title} {\bibinfo {title} {Study of the shubnikov-de haas effect.
  determination of the fermi surfaces in graphite},\ }\href
  {https://doi.org/10.1103/physrev.134.a453} {\bibfield  {journal} {\bibinfo
  {journal} {Physical Review}\ }\textbf {\bibinfo {volume} {134}},\ \bibinfo
  {pages} {A453} (\bibinfo {year} {1964})}\BibitemShut {NoStop}%
\bibitem [{\citenamefont {Gmitra}\ \emph {et~al.}(2016)\citenamefont {Gmitra},
  \citenamefont {Kochan}, \citenamefont {Hogl},\ and\ \citenamefont
  {Fabian}}]{Gmitra2016}%
  \BibitemOpen
  \bibfield  {author} {\bibinfo {author} {\bibfnamefont {M.}~\bibnamefont
  {Gmitra}}, \bibinfo {author} {\bibfnamefont {D.}~\bibnamefont {Kochan}},
  \bibinfo {author} {\bibfnamefont {P.}~\bibnamefont {Hogl}},\ and\ \bibinfo
  {author} {\bibfnamefont {J.}~\bibnamefont {Fabian}},\ }\bibfield  {title}
  {\bibinfo {title} {Trivial and inverted dirac bands and the emergence of
  quantum spin hall states in graphene on transition-metal dichalcogenides},\
  }\href {https://doi.org/10.1103/physrevb.93.155104} {\bibfield  {journal}
  {\bibinfo  {journal} {Physical Review B}\ }\textbf {\bibinfo {volume} {93}},\
  \bibinfo {pages} {155104} (\bibinfo {year} {2016})}\BibitemShut {NoStop}%
\bibitem [{\citenamefont {Alsharari}\ \emph {et~al.}(2018)\citenamefont
  {Alsharari}, \citenamefont {Asmar},\ and\ \citenamefont
  {Ulloa}}]{Alsharari2018}%
  \BibitemOpen
  \bibfield  {author} {\bibinfo {author} {\bibfnamefont {A.~M.}\ \bibnamefont
  {Alsharari}}, \bibinfo {author} {\bibfnamefont {M.~M.}\ \bibnamefont
  {Asmar}},\ and\ \bibinfo {author} {\bibfnamefont {S.~E.}\ \bibnamefont
  {Ulloa}},\ }\bibfield  {title} {\bibinfo {title} {Topological phases and
  twisting of graphene on a dichalcogenide monolayer},\ }\href
  {https://doi.org/10.1103/physrevb.98.195129} {\bibfield  {journal} {\bibinfo
  {journal} {Physical Review B}\ }\textbf {\bibinfo {volume} {98}},\ \bibinfo
  {pages} {195129} (\bibinfo {year} {2018})}\BibitemShut {NoStop}%
\bibitem [{\citenamefont {Yankowitz}\ \emph {et~al.}(2018)\citenamefont
  {Yankowitz}, \citenamefont {Jung}, \citenamefont {Laksono}, \citenamefont
  {Leconte}, \citenamefont {Chittari}, \citenamefont {Watanabe}, \citenamefont
  {Taniguchi}, \citenamefont {Adam}, \citenamefont {Graf},\ and\ \citenamefont
  {Dean}}]{Yankowitz2018}%
  \BibitemOpen
  \bibfield  {author} {\bibinfo {author} {\bibfnamefont {M.}~\bibnamefont
  {Yankowitz}}, \bibinfo {author} {\bibfnamefont {J.}~\bibnamefont {Jung}},
  \bibinfo {author} {\bibfnamefont {E.}~\bibnamefont {Laksono}}, \bibinfo
  {author} {\bibfnamefont {N.}~\bibnamefont {Leconte}}, \bibinfo {author}
  {\bibfnamefont {B.~L.}\ \bibnamefont {Chittari}}, \bibinfo {author}
  {\bibfnamefont {K.}~\bibnamefont {Watanabe}}, \bibinfo {author}
  {\bibfnamefont {T.}~\bibnamefont {Taniguchi}}, \bibinfo {author}
  {\bibfnamefont {S.}~\bibnamefont {Adam}}, \bibinfo {author} {\bibfnamefont
  {D.}~\bibnamefont {Graf}},\ and\ \bibinfo {author} {\bibfnamefont {C.~R.}\
  \bibnamefont {Dean}},\ }\bibfield  {title} {\bibinfo {title} {Dynamic
  band-structure tuning of graphene moiré superlattices with pressure},\
  }\href {https://doi.org/10.1038/s41586-018-0107-1} {\bibfield  {journal}
  {\bibinfo  {journal} {Nature}\ }\textbf {\bibinfo {volume} {557}},\ \bibinfo
  {pages} {404} (\bibinfo {year} {2018})}\BibitemShut {NoStop}%
\bibitem [{\citenamefont {Szentp{\'{e}}teri}\ \emph {et~al.}(2021)\citenamefont
  {Szentp{\'{e}}teri}, \citenamefont {Rickhaus}, \citenamefont {de~Vries},
  \citenamefont {M{\'{a}}rffy}, \citenamefont {Fülöp}, \citenamefont
  {T{\'{o}}v{\'{a}}ri}, \citenamefont {Watanabe}, \citenamefont {Taniguchi},
  \citenamefont {Korm{\'{a}}nyos}, \citenamefont {Csonka},\ and\ \citenamefont
  {Makk}}]{Szentpeteri2021}%
  \BibitemOpen
  \bibfield  {author} {\bibinfo {author} {\bibfnamefont {B.}~\bibnamefont
  {Szentp{\'{e}}teri}}, \bibinfo {author} {\bibfnamefont {P.}~\bibnamefont
  {Rickhaus}}, \bibinfo {author} {\bibfnamefont {F.~K.}\ \bibnamefont
  {de~Vries}}, \bibinfo {author} {\bibfnamefont {A.}~\bibnamefont
  {M{\'{a}}rffy}}, \bibinfo {author} {\bibfnamefont {B.}~\bibnamefont
  {Fülöp}}, \bibinfo {author} {\bibfnamefont {E.}~\bibnamefont
  {T{\'{o}}v{\'{a}}ri}}, \bibinfo {author} {\bibfnamefont {K.}~\bibnamefont
  {Watanabe}}, \bibinfo {author} {\bibfnamefont {T.}~\bibnamefont {Taniguchi}},
  \bibinfo {author} {\bibfnamefont {A.}~\bibnamefont {Korm{\'{a}}nyos}},
  \bibinfo {author} {\bibfnamefont {S.}~\bibnamefont {Csonka}},\ and\ \bibinfo
  {author} {\bibfnamefont {P.}~\bibnamefont {Makk}},\ }\bibfield  {title}
  {\bibinfo {title} {Tailoring the band structure of twisted double bilayer
  graphene with pressure},\ }\href
  {https://doi.org/10.1021/acs.nanolett.1c03066} {\bibfield  {journal}
  {\bibinfo  {journal} {Nano Letters}\ }\textbf {\bibinfo {volume} {21}},\
  \bibinfo {pages} {8777} (\bibinfo {year} {2021})}\BibitemShut {NoStop}%
\bibitem [{\citenamefont {Arora}\ \emph {et~al.}(2020)\citenamefont {Arora},
  \citenamefont {Polski}, \citenamefont {Zhang}, \citenamefont {Thomson},
  \citenamefont {Choi}, \citenamefont {Kim}, \citenamefont {Lin}, \citenamefont
  {Wilson}, \citenamefont {Xu}, \citenamefont {Chu}, \citenamefont {Watanabe},
  \citenamefont {Taniguchi}, \citenamefont {Alicea},\ and\ \citenamefont
  {Nadj-Perge}}]{Arora2020}%
  \BibitemOpen
  \bibfield  {author} {\bibinfo {author} {\bibfnamefont {H.~S.}\ \bibnamefont
  {Arora}}, \bibinfo {author} {\bibfnamefont {R.}~\bibnamefont {Polski}},
  \bibinfo {author} {\bibfnamefont {Y.}~\bibnamefont {Zhang}}, \bibinfo
  {author} {\bibfnamefont {A.}~\bibnamefont {Thomson}}, \bibinfo {author}
  {\bibfnamefont {Y.}~\bibnamefont {Choi}}, \bibinfo {author} {\bibfnamefont
  {H.}~\bibnamefont {Kim}}, \bibinfo {author} {\bibfnamefont {Z.}~\bibnamefont
  {Lin}}, \bibinfo {author} {\bibfnamefont {I.~Z.}\ \bibnamefont {Wilson}},
  \bibinfo {author} {\bibfnamefont {X.}~\bibnamefont {Xu}}, \bibinfo {author}
  {\bibfnamefont {J.-H.}\ \bibnamefont {Chu}}, \bibinfo {author} {\bibfnamefont
  {K.}~\bibnamefont {Watanabe}}, \bibinfo {author} {\bibfnamefont
  {T.}~\bibnamefont {Taniguchi}}, \bibinfo {author} {\bibfnamefont
  {J.}~\bibnamefont {Alicea}},\ and\ \bibinfo {author} {\bibfnamefont
  {S.}~\bibnamefont {Nadj-Perge}},\ }\bibfield  {title} {\bibinfo {title}
  {Superconductivity in metallic twisted bilayer graphene stabilized by wse2},\
  }\href {https://doi.org/10.1038/s41586-020-2473-8} {\bibfield  {journal}
  {\bibinfo  {journal} {Nature}\ }\textbf {\bibinfo {volume} {583}},\ \bibinfo
  {pages} {379} (\bibinfo {year} {2020})}\BibitemShut {NoStop}%
\bibitem [{\citenamefont {Kim}\ \emph {et~al.}(2016)\citenamefont {Kim},
  \citenamefont {Yankowitz}, \citenamefont {Fallahazad}, \citenamefont {Kang},
  \citenamefont {Movva}, \citenamefont {Huang}, \citenamefont {Larentis},
  \citenamefont {Corbet}, \citenamefont {Taniguchi}, \citenamefont {Watanabe},
  \citenamefont {Banerjee}, \citenamefont {LeRoy},\ and\ \citenamefont
  {Tutuc}}]{Kim2016a}%
  \BibitemOpen
  \bibfield  {author} {\bibinfo {author} {\bibfnamefont {K.}~\bibnamefont
  {Kim}}, \bibinfo {author} {\bibfnamefont {M.}~\bibnamefont {Yankowitz}},
  \bibinfo {author} {\bibfnamefont {B.}~\bibnamefont {Fallahazad}}, \bibinfo
  {author} {\bibfnamefont {S.}~\bibnamefont {Kang}}, \bibinfo {author}
  {\bibfnamefont {H.~C.~P.}\ \bibnamefont {Movva}}, \bibinfo {author}
  {\bibfnamefont {S.}~\bibnamefont {Huang}}, \bibinfo {author} {\bibfnamefont
  {S.}~\bibnamefont {Larentis}}, \bibinfo {author} {\bibfnamefont {C.~M.}\
  \bibnamefont {Corbet}}, \bibinfo {author} {\bibfnamefont {T.}~\bibnamefont
  {Taniguchi}}, \bibinfo {author} {\bibfnamefont {K.}~\bibnamefont {Watanabe}},
  \bibinfo {author} {\bibfnamefont {S.~K.}\ \bibnamefont {Banerjee}}, \bibinfo
  {author} {\bibfnamefont {B.~J.}\ \bibnamefont {LeRoy}},\ and\ \bibinfo
  {author} {\bibfnamefont {E.}~\bibnamefont {Tutuc}},\ }\bibfield  {title}
  {\bibinfo {title} {van der waals heterostructures with high accuracy
  rotational alignment},\ }\href {https://doi.org/10.1021/acs.nanolett.5b05263}
  {\bibfield  {journal} {\bibinfo  {journal} {Nano Letters}\ }\textbf {\bibinfo
  {volume} {16}},\ \bibinfo {pages} {1989} (\bibinfo {year}
  {2016})}\BibitemShut {NoStop}%
\bibitem [{\citenamefont {Solozhenko}\ \emph {et~al.}(1995)\citenamefont
  {Solozhenko}, \citenamefont {Will},\ and\ \citenamefont
  {Elf}}]{Solozhenko1995}%
  \BibitemOpen
  \bibfield  {author} {\bibinfo {author} {\bibfnamefont {V.}~\bibnamefont
  {Solozhenko}}, \bibinfo {author} {\bibfnamefont {G.}~\bibnamefont {Will}},\
  and\ \bibinfo {author} {\bibfnamefont {F.}~\bibnamefont {Elf}},\ }\bibfield
  {title} {\bibinfo {title} {Isothermal compression of hexagonal graphite-like
  boron nitride up to 12 gpa},\ }\href
  {https://doi.org/10.1016/0038-1098(95)00381-9} {\bibfield  {journal}
  {\bibinfo  {journal} {Solid State Communications}\ }\textbf {\bibinfo
  {volume} {96}},\ \bibinfo {pages} {1} (\bibinfo {year} {1995})}\BibitemShut
  {NoStop}%
\bibitem [{\citenamefont {Jung}\ and\ \citenamefont
  {MacDonald}(2014)}]{Jung2014}%
  \BibitemOpen
  \bibfield  {author} {\bibinfo {author} {\bibfnamefont {J.}~\bibnamefont
  {Jung}}\ and\ \bibinfo {author} {\bibfnamefont {A.~H.}\ \bibnamefont
  {MacDonald}},\ }\bibfield  {title} {\bibinfo {title} {Accurate tight-binding
  models for the$\pi$bands of bilayer graphene},\ }\href
  {https://doi.org/10.1103/physrevb.89.035405} {\bibfield  {journal} {\bibinfo
  {journal} {Physical Review B}\ }\textbf {\bibinfo {volume} {89}},\ \bibinfo
  {pages} {035405} (\bibinfo {year} {2014})}\BibitemShut {NoStop}%
\end{thebibliography}%


\begin{thebibliography}{11}%
\makeatletter
\providecommand \@ifxundefined [1]{%
 \@ifx{#1\undefined}
}%
\providecommand \@ifnum [1]{%
 \ifnum #1\expandafter \@firstoftwo
 \else \expandafter \@secondoftwo
 \fi
}%
\providecommand \@ifx [1]{%
 \ifx #1\expandafter \@firstoftwo
 \else \expandafter \@secondoftwo
 \fi
}%
\providecommand \natexlab [1]{#1}%
\providecommand \enquote  [1]{``#1''}%
\providecommand \bibnamefont  [1]{#1}%
\providecommand \bibfnamefont [1]{#1}%
\providecommand \citenamefont [1]{#1}%
\providecommand \href@noop [0]{\@secondoftwo}%
\providecommand \href [0]{\begingroup \@sanitize@url \@href}%
\providecommand \@href[1]{\@@startlink{#1}\@@href}%
\providecommand \@@href[1]{\endgroup#1\@@endlink}%
\providecommand \@sanitize@url [0]{\catcode `\\12\catcode `\$12\catcode
  `\&12\catcode `\#12\catcode `\^12\catcode `\_12\catcode `\%12\relax}%
\providecommand \@@startlink[1]{}%
\providecommand \@@endlink[0]{}%
\providecommand \url  [0]{\begingroup\@sanitize@url \@url }%
\providecommand \@url [1]{\endgroup\@href {#1}{\urlprefix }}%
\providecommand \urlprefix  [0]{URL }%
\providecommand \Eprint [0]{\href }%
\providecommand \doibase [0]{https://doi.org/}%
\providecommand \selectlanguage [0]{\@gobble}%
\providecommand \bibinfo  [0]{\@secondoftwo}%
\providecommand \bibfield  [0]{\@secondoftwo}%
\providecommand \translation [1]{[#1]}%
\providecommand \BibitemOpen [0]{}%
\providecommand \bibitemStop [0]{}%
\providecommand \bibitemNoStop [0]{.\EOS\space}%
\providecommand \EOS [0]{\spacefactor3000\relax}%
\providecommand \BibitemShut  [1]{\csname bibitem#1\endcsname}%
\let\auto@bib@innerbib\@empty
\bibitem [{\citenamefont {Kim}\ \emph {et~al.}(2016)\citenamefont {Kim},
  \citenamefont {Yankowitz}, \citenamefont {Fallahazad}, \citenamefont {Kang},
  \citenamefont {Movva}, \citenamefont {Huang}, \citenamefont {Larentis},
  \citenamefont {Corbet}, \citenamefont {Taniguchi}, \citenamefont {Watanabe},
  \citenamefont {Banerjee}, \citenamefont {LeRoy},\ and\ \citenamefont
  {Tutuc}}]{Kim2016a}%
  \BibitemOpen
  \bibfield  {author} {\bibinfo {author} {\bibfnamefont {K.}~\bibnamefont
  {Kim}}, \bibinfo {author} {\bibfnamefont {M.}~\bibnamefont {Yankowitz}},
  \bibinfo {author} {\bibfnamefont {B.}~\bibnamefont {Fallahazad}}, \bibinfo
  {author} {\bibfnamefont {S.}~\bibnamefont {Kang}}, \bibinfo {author}
  {\bibfnamefont {H.~C.~P.}\ \bibnamefont {Movva}}, \bibinfo {author}
  {\bibfnamefont {S.}~\bibnamefont {Huang}}, \bibinfo {author} {\bibfnamefont
  {S.}~\bibnamefont {Larentis}}, \bibinfo {author} {\bibfnamefont {C.~M.}\
  \bibnamefont {Corbet}}, \bibinfo {author} {\bibfnamefont {T.}~\bibnamefont
  {Taniguchi}}, \bibinfo {author} {\bibfnamefont {K.}~\bibnamefont {Watanabe}},
  \bibinfo {author} {\bibfnamefont {S.~K.}\ \bibnamefont {Banerjee}}, \bibinfo
  {author} {\bibfnamefont {B.~J.}\ \bibnamefont {LeRoy}},\ and\ \bibinfo
  {author} {\bibfnamefont {E.}~\bibnamefont {Tutuc}},\ }\bibfield  {title}
  {\bibinfo {title} {van der waals heterostructures with high accuracy
  rotational alignment},\ }\href {https://doi.org/10.1021/acs.nanolett.5b05263}
  {\bibfield  {journal} {\bibinfo  {journal} {Nano Letters}\ }\textbf {\bibinfo
  {volume} {16}},\ \bibinfo {pages} {1989} (\bibinfo {year}
  {2016})}\BibitemShut {NoStop}%
\bibitem [{\citenamefont {Solozhenko}\ \emph {et~al.}(1995)\citenamefont
  {Solozhenko}, \citenamefont {Will},\ and\ \citenamefont
  {Elf}}]{Solozhenko1995}%
  \BibitemOpen
  \bibfield  {author} {\bibinfo {author} {\bibfnamefont {V.}~\bibnamefont
  {Solozhenko}}, \bibinfo {author} {\bibfnamefont {G.}~\bibnamefont {Will}},\
  and\ \bibinfo {author} {\bibfnamefont {F.}~\bibnamefont {Elf}},\ }\bibfield
  {title} {\bibinfo {title} {Isothermal compression of hexagonal graphite-like
  boron nitride up to 12 gpa},\ }\href
  {https://doi.org/10.1016/0038-1098(95)00381-9} {\bibfield  {journal}
  {\bibinfo  {journal} {Solid State Communications}\ }\textbf {\bibinfo
  {volume} {96}},\ \bibinfo {pages} {1} (\bibinfo {year} {1995})}\BibitemShut
  {NoStop}%
\bibitem [{\citenamefont {Yankowitz}\ \emph {et~al.}(2018)\citenamefont
  {Yankowitz}, \citenamefont {Jung}, \citenamefont {Laksono}, \citenamefont
  {Leconte}, \citenamefont {Chittari}, \citenamefont {Watanabe}, \citenamefont
  {Taniguchi}, \citenamefont {Adam}, \citenamefont {Graf},\ and\ \citenamefont
  {Dean}}]{Yankowitz2018}%
  \BibitemOpen
  \bibfield  {author} {\bibinfo {author} {\bibfnamefont {M.}~\bibnamefont
  {Yankowitz}}, \bibinfo {author} {\bibfnamefont {J.}~\bibnamefont {Jung}},
  \bibinfo {author} {\bibfnamefont {E.}~\bibnamefont {Laksono}}, \bibinfo
  {author} {\bibfnamefont {N.}~\bibnamefont {Leconte}}, \bibinfo {author}
  {\bibfnamefont {B.~L.}\ \bibnamefont {Chittari}}, \bibinfo {author}
  {\bibfnamefont {K.}~\bibnamefont {Watanabe}}, \bibinfo {author}
  {\bibfnamefont {T.}~\bibnamefont {Taniguchi}}, \bibinfo {author}
  {\bibfnamefont {S.}~\bibnamefont {Adam}}, \bibinfo {author} {\bibfnamefont
  {D.}~\bibnamefont {Graf}},\ and\ \bibinfo {author} {\bibfnamefont {C.~R.}\
  \bibnamefont {Dean}},\ }\bibfield  {title} {\bibinfo {title} {Dynamic
  band-structure tuning of graphene moiré superlattices with pressure},\
  }\href {https://doi.org/10.1038/s41586-018-0107-1} {\bibfield  {journal}
  {\bibinfo  {journal} {Nature}\ }\textbf {\bibinfo {volume} {557}},\ \bibinfo
  {pages} {404} (\bibinfo {year} {2018})}\BibitemShut {NoStop}%
\bibitem [{\citenamefont {Fülöp}\ \emph {et~al.}(2021)\citenamefont
  {Fülöp}, \citenamefont {M{\'{a}}rffy}, \citenamefont {T{\'{o}}v{\'{a}}ri},
  \citenamefont {Kedves}, \citenamefont {Zihlmann}, \citenamefont {Indolese},
  \citenamefont {Kov{\'{a}}cs-Krausz}, \citenamefont {Watanabe}, \citenamefont
  {Taniguchi}, \citenamefont {Schönenberger}, \citenamefont
  {K{\'{e}}zsm{\'{a}}rki}, \citenamefont {Makk},\ and\ \citenamefont
  {Csonka}}]{Fueloep2021a}%
  \BibitemOpen
  \bibfield  {author} {\bibinfo {author} {\bibfnamefont {B.}~\bibnamefont
  {Fülöp}}, \bibinfo {author} {\bibfnamefont {A.}~\bibnamefont
  {M{\'{a}}rffy}}, \bibinfo {author} {\bibfnamefont {E.}~\bibnamefont
  {T{\'{o}}v{\'{a}}ri}}, \bibinfo {author} {\bibfnamefont {M.}~\bibnamefont
  {Kedves}}, \bibinfo {author} {\bibfnamefont {S.}~\bibnamefont {Zihlmann}},
  \bibinfo {author} {\bibfnamefont {D.}~\bibnamefont {Indolese}}, \bibinfo
  {author} {\bibfnamefont {Z.}~\bibnamefont {Kov{\'{a}}cs-Krausz}}, \bibinfo
  {author} {\bibfnamefont {K.}~\bibnamefont {Watanabe}}, \bibinfo {author}
  {\bibfnamefont {T.}~\bibnamefont {Taniguchi}}, \bibinfo {author}
  {\bibfnamefont {C.}~\bibnamefont {Schönenberger}}, \bibinfo {author}
  {\bibfnamefont {I.}~\bibnamefont {K{\'{e}}zsm{\'{a}}rki}}, \bibinfo {author}
  {\bibfnamefont {P.}~\bibnamefont {Makk}},\ and\ \bibinfo {author}
  {\bibfnamefont {S.}~\bibnamefont {Csonka}},\ }\bibfield  {title} {\bibinfo
  {title} {New method of transport measurements on van der waals
  heterostructures under pressure},\ }\href {https://doi.org/10.1063/5.0058583}
  {\bibfield  {journal} {\bibinfo  {journal} {Journal of Applied Physics}\
  }\textbf {\bibinfo {volume} {130}},\ \bibinfo {pages} {064303} (\bibinfo
  {year} {2021})}\BibitemShut {NoStop}%
\bibitem [{\citenamefont {McCann}\ and\ \citenamefont
  {Koshino}(2013)}]{McCann2013}%
  \BibitemOpen
  \bibfield  {author} {\bibinfo {author} {\bibfnamefont {E.}~\bibnamefont
  {McCann}}\ and\ \bibinfo {author} {\bibfnamefont {M.}~\bibnamefont
  {Koshino}},\ }\bibfield  {title} {\bibinfo {title} {The electronic properties
  of bilayer graphene},\ }\href {https://doi.org/10.1088/0034-4885/76/5/056503}
  {\bibfield  {journal} {\bibinfo  {journal} {Reports on Progress in Physics}\
  }\textbf {\bibinfo {volume} {76}},\ \bibinfo {pages} {056503} (\bibinfo
  {year} {2013})}\BibitemShut {NoStop}%
\bibitem [{\citenamefont {Jung}\ and\ \citenamefont
  {MacDonald}(2014)}]{Jung2014}%
  \BibitemOpen
  \bibfield  {author} {\bibinfo {author} {\bibfnamefont {J.}~\bibnamefont
  {Jung}}\ and\ \bibinfo {author} {\bibfnamefont {A.~H.}\ \bibnamefont
  {MacDonald}},\ }\bibfield  {title} {\bibinfo {title} {Accurate tight-binding
  models for the$\pi$bands of bilayer graphene},\ }\href
  {https://doi.org/10.1103/physrevb.89.035405} {\bibfield  {journal} {\bibinfo
  {journal} {Physical Review B}\ }\textbf {\bibinfo {volume} {89}},\ \bibinfo
  {pages} {035405} (\bibinfo {year} {2014})}\BibitemShut {NoStop}%
\bibitem [{\citenamefont {Island}\ \emph {et~al.}(2019)\citenamefont {Island},
  \citenamefont {Cui}, \citenamefont {Lewandowski}, \citenamefont {Khoo},
  \citenamefont {Spanton}, \citenamefont {Zhou}, \citenamefont {Rhodes},
  \citenamefont {Hone}, \citenamefont {Taniguchi}, \citenamefont {Watanabe},
  \citenamefont {Levitov}, \citenamefont {Zaletel},\ and\ \citenamefont
  {Young}}]{Island2019}%
  \BibitemOpen
  \bibfield  {author} {\bibinfo {author} {\bibfnamefont {J.~O.}\ \bibnamefont
  {Island}}, \bibinfo {author} {\bibfnamefont {X.}~\bibnamefont {Cui}},
  \bibinfo {author} {\bibfnamefont {C.}~\bibnamefont {Lewandowski}}, \bibinfo
  {author} {\bibfnamefont {J.~Y.}\ \bibnamefont {Khoo}}, \bibinfo {author}
  {\bibfnamefont {E.~M.}\ \bibnamefont {Spanton}}, \bibinfo {author}
  {\bibfnamefont {H.}~\bibnamefont {Zhou}}, \bibinfo {author} {\bibfnamefont
  {D.}~\bibnamefont {Rhodes}}, \bibinfo {author} {\bibfnamefont {J.~C.}\
  \bibnamefont {Hone}}, \bibinfo {author} {\bibfnamefont {T.}~\bibnamefont
  {Taniguchi}}, \bibinfo {author} {\bibfnamefont {K.}~\bibnamefont {Watanabe}},
  \bibinfo {author} {\bibfnamefont {L.~S.}\ \bibnamefont {Levitov}}, \bibinfo
  {author} {\bibfnamefont {M.~P.}\ \bibnamefont {Zaletel}},\ and\ \bibinfo
  {author} {\bibfnamefont {A.~F.}\ \bibnamefont {Young}},\ }\bibfield  {title}
  {\bibinfo {title} {Spin{\textendash}orbit-driven band inversion in bilayer
  graphene by the van der waals proximity effect},\ }\href
  {https://doi.org/10.1038/s41586-019-1304-2} {\bibfield  {journal} {\bibinfo
  {journal} {Nature}\ }\textbf {\bibinfo {volume} {571}},\ \bibinfo {pages}
  {85} (\bibinfo {year} {2019})}\BibitemShut {NoStop}%
\bibitem [{\citenamefont {Khoo}\ \emph {et~al.}(2017)\citenamefont {Khoo},
  \citenamefont {Morpurgo},\ and\ \citenamefont {Levitov}}]{Khoo2017}%
  \BibitemOpen
  \bibfield  {author} {\bibinfo {author} {\bibfnamefont {J.~Y.}\ \bibnamefont
  {Khoo}}, \bibinfo {author} {\bibfnamefont {A.~F.}\ \bibnamefont {Morpurgo}},\
  and\ \bibinfo {author} {\bibfnamefont {L.}~\bibnamefont {Levitov}},\
  }\bibfield  {title} {\bibinfo {title} {On-demand spin–orbit interaction
  from which-layer tunability in bilayer graphene},\ }\href
  {https://doi.org/10.1021/acs.nanolett.7b03604} {\bibfield  {journal}
  {\bibinfo  {journal} {Nano Letters}\ }\textbf {\bibinfo {volume} {17}},\
  \bibinfo {pages} {7003} (\bibinfo {year} {2017})}\BibitemShut {NoStop}%
\bibitem [{\citenamefont {Amann}\ \emph {et~al.}(2022)\citenamefont {Amann},
  \citenamefont {Völkl}, \citenamefont {Rockinger}, \citenamefont {Kochan},
  \citenamefont {Watanabe}, \citenamefont {Taniguchi}, \citenamefont {Fabian},
  \citenamefont {Weiss},\ and\ \citenamefont {Eroms}}]{Amann2022}%
  \BibitemOpen
  \bibfield  {author} {\bibinfo {author} {\bibfnamefont {J.}~\bibnamefont
  {Amann}}, \bibinfo {author} {\bibfnamefont {T.}~\bibnamefont {Völkl}},
  \bibinfo {author} {\bibfnamefont {T.}~\bibnamefont {Rockinger}}, \bibinfo
  {author} {\bibfnamefont {D.}~\bibnamefont {Kochan}}, \bibinfo {author}
  {\bibfnamefont {K.}~\bibnamefont {Watanabe}}, \bibinfo {author}
  {\bibfnamefont {T.}~\bibnamefont {Taniguchi}}, \bibinfo {author}
  {\bibfnamefont {J.}~\bibnamefont {Fabian}}, \bibinfo {author} {\bibfnamefont
  {D.}~\bibnamefont {Weiss}},\ and\ \bibinfo {author} {\bibfnamefont
  {J.}~\bibnamefont {Eroms}},\ }\bibfield  {title} {\bibinfo {title}
  {Counterintuitive gate dependence of weak antilocalization in bilayer
  graphene/wse2 heterostructures},\ }\href
  {https://doi.org/10.1103/physrevb.105.115425} {\bibfield  {journal} {\bibinfo
   {journal} {Physical Review B}\ }\textbf {\bibinfo {volume} {105}},\ \bibinfo
  {pages} {115425} (\bibinfo {year} {2022})}\BibitemShut {NoStop}%
\bibitem [{\citenamefont {Naimer}\ \emph {et~al.}(2021)\citenamefont {Naimer},
  \citenamefont {Zollner}, \citenamefont {Gmitra},\ and\ \citenamefont
  {Fabian}}]{Naimer2021}%
  \BibitemOpen
  \bibfield  {author} {\bibinfo {author} {\bibfnamefont {T.}~\bibnamefont
  {Naimer}}, \bibinfo {author} {\bibfnamefont {K.}~\bibnamefont {Zollner}},
  \bibinfo {author} {\bibfnamefont {M.}~\bibnamefont {Gmitra}},\ and\ \bibinfo
  {author} {\bibfnamefont {J.}~\bibnamefont {Fabian}},\ }\bibfield  {title}
  {\bibinfo {title} {Twist-angle dependent proximity induced spin-orbit
  coupling in graphene/transition metal dichalcogenide heterostructures},\
  }\href {https://doi.org/10.1103/physrevb.104.195156} {\bibfield  {journal}
  {\bibinfo  {journal} {Physical Review B}\ }\textbf {\bibinfo {volume}
  {104}},\ \bibinfo {pages} {195156} (\bibinfo {year} {2021})}\BibitemShut
  {NoStop}%
\bibitem [{\citenamefont {Wang}\ \emph {et~al.}(2019)\citenamefont {Wang},
  \citenamefont {Che}, \citenamefont {Cao}, \citenamefont {Lyu}, \citenamefont
  {Watanabe}, \citenamefont {Taniguchi}, \citenamefont {Lau},\ and\
  \citenamefont {Bockrath}}]{Wang2019}%
  \BibitemOpen
  \bibfield  {author} {\bibinfo {author} {\bibfnamefont {D.}~\bibnamefont
  {Wang}}, \bibinfo {author} {\bibfnamefont {S.}~\bibnamefont {Che}}, \bibinfo
  {author} {\bibfnamefont {G.}~\bibnamefont {Cao}}, \bibinfo {author}
  {\bibfnamefont {R.}~\bibnamefont {Lyu}}, \bibinfo {author} {\bibfnamefont
  {K.}~\bibnamefont {Watanabe}}, \bibinfo {author} {\bibfnamefont
  {T.}~\bibnamefont {Taniguchi}}, \bibinfo {author} {\bibfnamefont {C.~N.}\
  \bibnamefont {Lau}},\ and\ \bibinfo {author} {\bibfnamefont {M.}~\bibnamefont
  {Bockrath}},\ }\bibfield  {title} {\bibinfo {title} {Quantum hall effect
  measurement of spin-orbit coupling strengths in ultraclean bilayer
  graphene/{WSe}2 heterostructures},\ }\href
  {https://doi.org/10.1021/acs.nanolett.9b02445} {\bibfield  {journal}
  {\bibinfo  {journal} {Nano Letters}\ }\textbf {\bibinfo {volume} {19}},\
  \bibinfo {pages} {7028} (\bibinfo {year} {2019})}\BibitemShut {NoStop}%
\end{thebibliography}%
\end{document}


\setcounter{equation}{0}
\setcounter{figure}{0}
\setcounter{table}{0}
\setcounter{page}{1}
\makeatletter
\renewcommand{\theequation}{S\arabic{equation}}
\renewcommand{\thefigure}{S\arabic{figure}}
\renewcommand{\bibnumfmt}[1]{[S#1]}
\renewcommand{\citenumfont}[1]{S#1}
\renewcommand{\thetable}{S\arabic{table}}
\setcounter{section}{0}
\renewcommand{\thesection}{S\arabic{section}}

\title{Supplemental Material for Increasing the proximity induced spin\,–\,orbit coupling in bilayer graphene/WSe$_2$ heterostructures with pressure}
\author{Bálint Szentpéteri$^{1,2}$, Albin Márffy$^{1,2}$, Máté Kedves$^{1,2}$, Endre Tóvári$^{1,2}$,  Bálint Fülöp$^{1,2}$, István Kükemezey$^{4}$, András Magyarkuti$^{4}$, Kenji Watanabe$^5$, Takashi Taniguchi$^5$, Szabolcs Csonka$^{1,3,6}$, Péter Makk$^{1,2}$}
\affiliation{$^1$Department of Physics, Budapest University of Technology and Economics and
Nanoelectronics Momentum Research Group of the Hungarian Academy of Sciences,
Budafoki ut 8, 1111 Budapest, Hungary\\
$^2$MTA-BME Correlated van der Waals Structures Momentum Research Group, Műegyetem rkp. 3., H-1111 Budapest, Hungary\\
$^3$MTA-BME Superconducting Nanoelectronics Momentum Research Group, Műegyetem rkp. 3., H-1111 Budapest, Hungary\\
$^4$Semilab Co. Ltd. Prielle Kornélia u. 2, 1117 Budapest, Hungary\\
$^5$Research Center for Functional Materials, National Institute for Materials Science, 1-1 Namiki, Tsukuba 305-0044, Japan\\
$^6$HUN-REN Centre for Energy Research, Institute of Technical Physics and Materials Science, Konkoly Thege Miklós út 29-33., H-1121 Budapest, Hungary
}
\maketitle

\section{Device fabrication}
\begin{figure*}[!ht]
\begin{center}
\includegraphics[width=.9\linewidth]{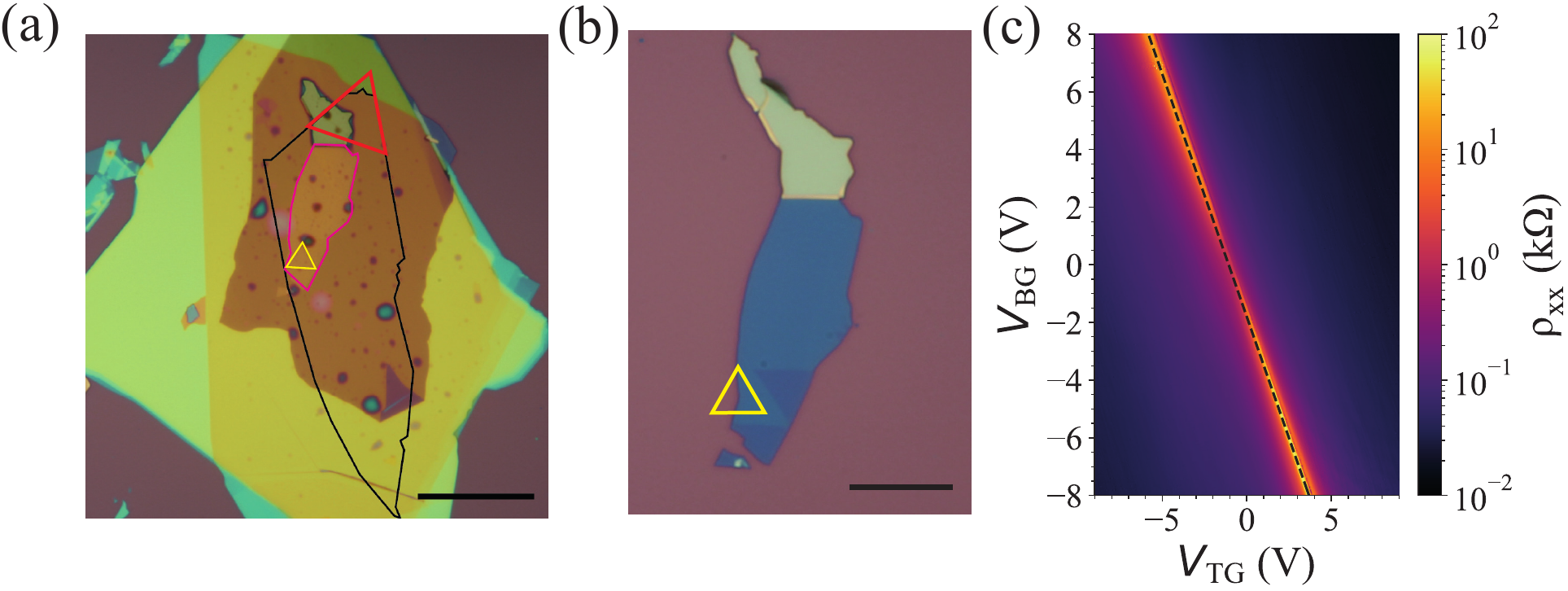}
\caption{(a) Optical microscope image of the measured stack. The scale bar is \SI{20}{\micro\metre}. The black and magenta lines show the perimeter of the graphene and the WSe$_2$, respectively. The red triangle is an estimate of the crystal orientation of the graphene from its edges. The yellow triangle is an estimate of the crystal orientation of the WSe$_2$. (b) Optical microscope image of the used WSe$_2$ showing the same yellow triangle as in (a). The scale bar is \SI{10}{\micro\metre}. (c) $\rho_{xx}$ versus the top and bottom gate voltage. The slope of the black dashed line is used to get the ratio between the lever arms.}
\label{supfig_sample}
\end{center}
\end{figure*}
The van der Waals heterostructure is fabricated using the dry-transfer technique\cite{Kim2016a} at ambient conditions. With the top hBN the BLG is picked up which is followed with the WSe$_2$, the bottom hBN and finally the bottom graphite, that serves as a global bottom gate. After the assembly of the stack the device is fabricated in a cleanroom facility, using standard electron beam lithography techniques to define the different layers. Side contacts are created with reactive ion etching and evaporation of 10\,nm of chromium (Cr) and 80\,nm of gold (Au). After the side contacts are made, the heterostructure is etched into a Hall-bar shape by reactive ion etching using SF6/CHF3/O2 gases. After this, atomic layer deposition is used to create a 30\,nm aluminium oxide dielectric layer, which isolates the top gate layer of 10/90\,nm Cr/Au from the device. The thickness of top hBN is\,22 nm, the bottom hBN is 58\,nm thick and the thickness of the WSe$_2$ layer is 7\,nm. The surface topology of the stack and the thicknesses of the flakes were measured with an AFM made by Semilab. The twist angle of the graphene and the WSe$_2$ is estimated from the optical images shown in Fig.\,\ref{supfig_sample}. The angle between the two triangles, which is used to estimate the orientation of the crystal lattices, is $\theta\sim16$°.

\section{Transport measurements}
Transport measurements were carried out in a Hall-geometry with typical AC voltage excitation of 1 mV using a standard lock-in technique at 1167.3\,Hz.
\subsection{Lever arms}
\begin{figure}[!ht]
\begin{center}
\includegraphics[width=1\linewidth]{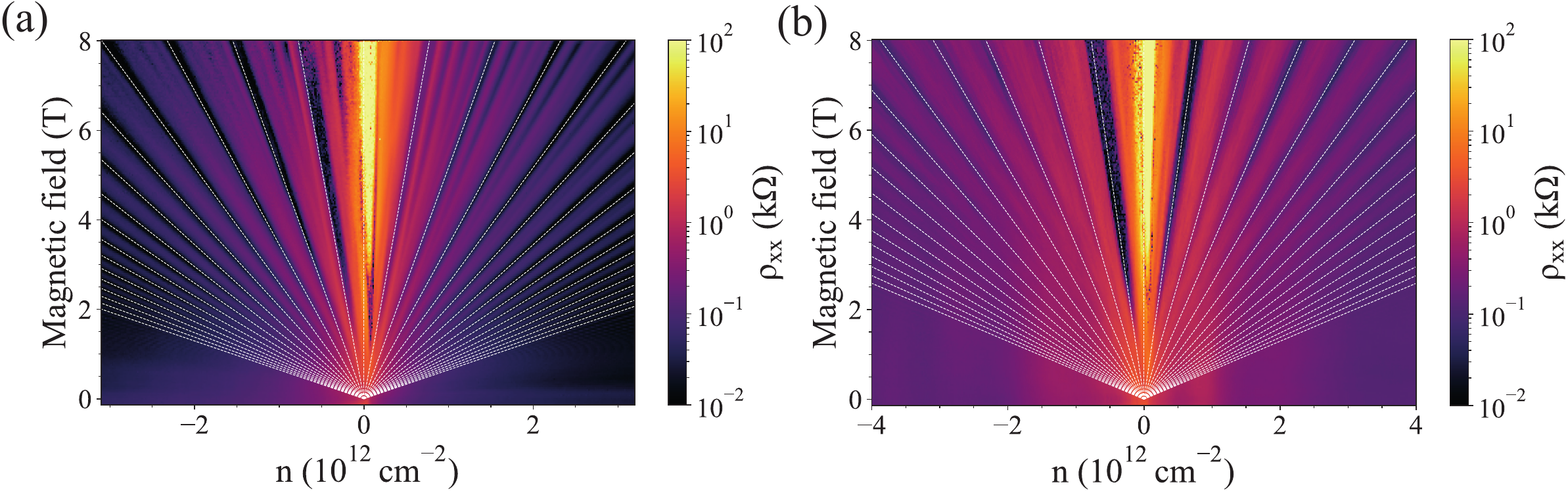}
\caption{Magnetic oscillation measurements of $\rho_{xx}$ (a) at $p=0$ and (b) at $p=2$\,GPa respectively at $T=50$\,mK. The white dashed lines shows the Landau levels at different fillings ($\nu=0,\pm4,\pm8\dots$). }
\label{supfig_fanplot}
\end{center}
\end{figure}
To obtain the lever arms first we get the ratio between the top ($\alpha_{\text{TG}}$) and bottom gate lever arms ($\alpha_{\text{TG}}$)  which is equal to the slope of the charge neutrality line, shown with a dashed line in Fig.~\ref{supfig_sample}c. The charge density ($n$) and electric displacement field ($D$) is related to the top and bottom gate voltage by
\begin{eqnarray}
n&=&\alpha_{\text{TG}}V_{\text{TG}}+\alpha_{\text{BG}}V_{\text{BG}}+n_0,\\
\frac{D}{\epsilon_0}&=&\frac{e}{2\epsilon_0}\left(\alpha_{\text{TG}}V_{\text{TG}}-\alpha_{\text{BG}}V_{\text{BG}}\right)+\frac{D_0}{\epsilon_0},\nonumber
\end{eqnarray}
 where  $\epsilon_0$ is the vacuum permittivity, $V_{\text{TG}}$ and $V_{\text{BG}}$ are the top and bottom gate voltage, respectively, $n_0$ is the zero-carrier density, whereas ${D_0}$ is a built-in offset electric field. $n_0$ and $D_0$ are obtained from the charge density and displacement field of the minimum resistance along the charge neutrality line. Finally, the lever arms are obtained from the quantum oscillations in magnetoconductance measurements (Fig.~\ref{supfig_fanplot}) using the full filling of the Landau levels (LL) being at $n=\nu eB/h$, where $e$ is the elementary charge, $h$ is the Planck's constant, $B$ is the transverse magnetic field and $\nu=0,\pm4,\pm8\dots$ is the filling factor of the LLs. The extracted lever arms are shown in Table \ref{1tabl}. The pressure dependence of the lever arms originates from the compression of the dielectrics and pressure-dependent dielectric constant of the hBN as already reported in Refs.\,\cite{Solozhenko1995, Yankowitz2018}.
\begin{table}[!ht]
\centering
\begin{tabular}{|ccc|}
\hline 
P (GPa)&0&2\\
\hline 
$\alpha_{TG}$ $(\frac{10^{15}}{\mathrm{Vm}^2})$&4.09(4)    &4.45(4)\\ 
$\alpha_{BG}$ $(\frac{10^{15}}{\mathrm{Vm}^2})$&2.47(3)    &2.80(3)\\ 
$n_0$ ($10^{15}/$m$^2$)   &4.20(5)  &4.79(5)\\
$D_0/\epsilon_0$ (V/nm) &0.047(2)    &0.069(2)\\  
\hline 
\end{tabular}
\caption{{The extracted parameters for the calculation of $n$ and $D$ at $p=0$ and $p=2$\,GPa with the uncertainty of the last digit in the bracket originating from the measurement resolution, the read out precision and from the fitting.}}
\label{1tabl}
\end{table}

\subsection{Pressurization}
The device is measured using a special high-pressure sample holder equipped with a circuit board suitable for transport measurements at cryogenic temperature. In the cell the pressure was mediated with kerosene and it was applied with a hydraulic press at room temperature. Our pressure cell is described in more detail in Ref.\,\cite{Fueloep2021a}.
\section{Model}
The effective low-energy Hamiltonian of the pristine BLG near the Dirac points can be written in the $\ket{A1,B1,A2,B2}$ basis\cite{McCann2013,Jung2014} as
\begin{equation}
\label{BLGHamiltonian}
H_0=\left(\begin{array}{cccc}
\frac{u}{2}&v_0\pi^\dagger&-v_4\pi^\dagger&v_3\pi\\
v_0\pi&\frac{u}{2}+\Delta'&\gamma_1&-v_4\pi^\dagger\\
-v_4\pi&\gamma_1&-\frac{u}{2}+\Delta'&v_0\pi^\dagger\\
v_3\pi^\dagger&-v_4\pi&v_0\pi&-\frac{u}{2}
\end{array}\right),
\end{equation}
where $\pi=\hbar(\xi k_x+ik_y)$, $v_{i}=\frac{\sqrt{3}a}{2\hbar}\gamma_{i}$ and $\hbar$ is the reduced Planck constant. $\gamma_i$-s in $v_i$-s are the hopping parameters: $\gamma_0=2.61$ eV is the intralayer nearest neighbour hopping, $\gamma_1= 0.361$ eV is the interlayer coupling between orbitals on the dimer sites B1 and A2, and $\gamma_4= 0.138$ eV is the interlayer coupling between dimer and non-dimer orbitals A1 and A2 or B1 and B2. The parameter $\Delta' = 0.015$ eV describes the energy difference between dimer and non-dimer sites and $\xi$ is the valley index. The interlayer potential difference is related to the displacements field $D$ by $u =\frac{d}{\epsilon_{\text{BLG}}^\perp}D$, where $d=0.33$ nm is the BLG interlayer separation, $\epsilon_{\text{BLG}}^\perp=4.3$ is the out-of-plane dielectric constant of the BLG \cite{Island2019}. $\gamma_3=0.1$ eV is the interlayer coupling of A1 and B2, which is responsible for the trigonal warping.

The effect of WSe$_2$ is taken into account by adding a spin orbit term as $H=H_0+H_{SOC}$, where $H_{SOC}=H_I+H_R$ has two terms, the Ising-type or valley-Zeeman SOC $H_I=\xi\frac{\lambda_I}{2} \sigma_z$ and the Rashba-type SOC term given as $H_R=\frac{\lambda_R}{2} (\xi\sigma_x s_y - \sigma_y s_x)$, where $s_i$ and $\sigma_i$ are Pauli-matrices acting on the spin and pseudo spin (lattice) degree of freedom \cite{Khoo2017}. We suppose that the SOC acts on only the layer which is close to the WSe$_2$.

\begin{figure}[!ht]
\begin{center}
\includegraphics[width=.6\linewidth]{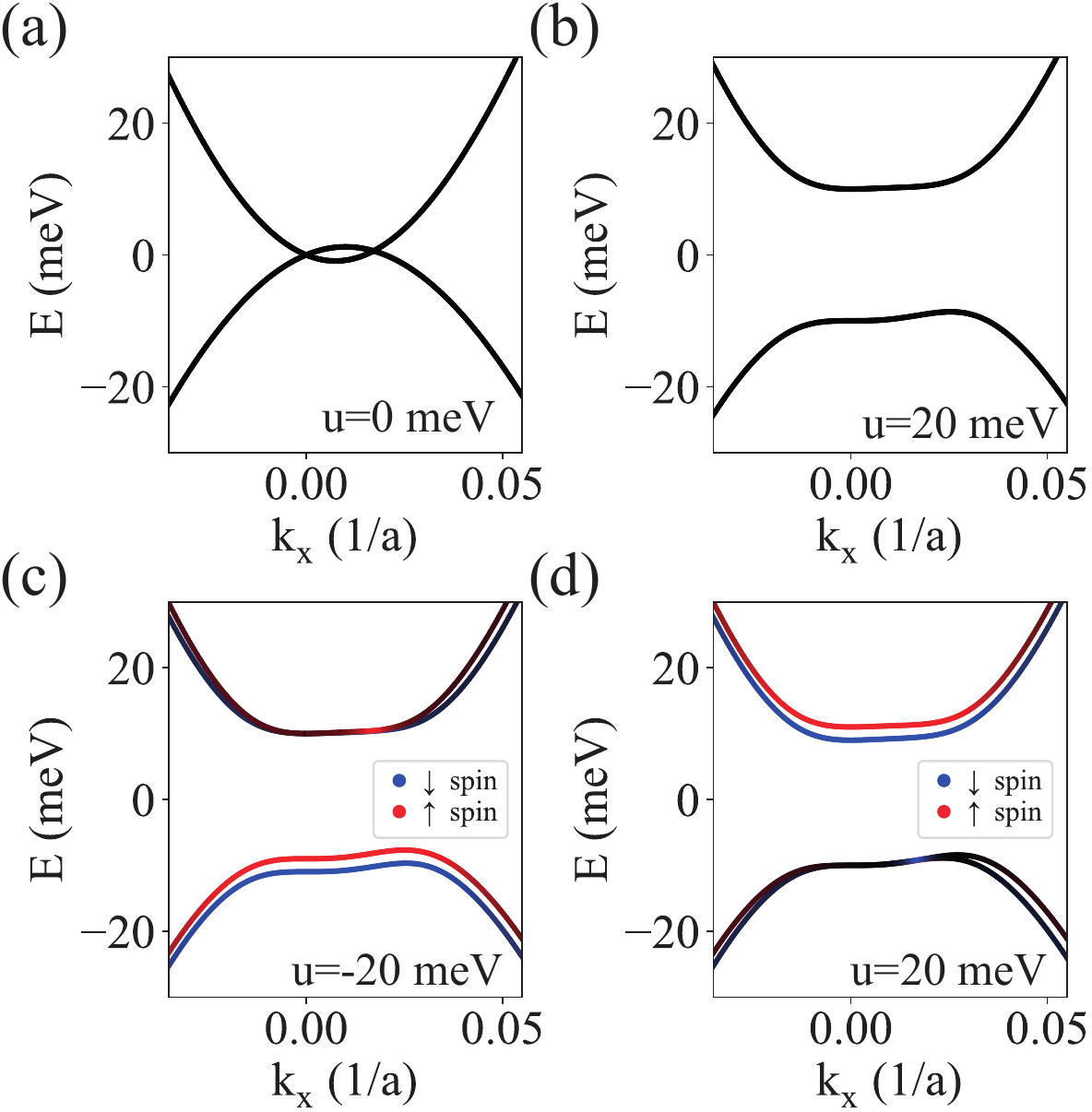}
\caption{Electric field dependence of the low-energy spectrum of bilayer graphene (a,b) without SOC at u=0 and u=20\,meV, respectively, and (c,d) including the SOC using the parameters of $\lambda_I=2$\,meV, $\lambda_R=10$\,meV and $u=\pm20$\,meV. When $u>0$ the conduction band is spin-split, and when $u<0$ the valence band is spin-split.}
\label{supfig_spectrumDdep}
\end{center}
\end{figure}

The spectrum without spin-orbit coupling is shown in Fig.~\ref{supfig_spectrumDdep}a around the $K$ point. There are two doubly degenerate bands which intersect at the $K$ point. Including a finite $u$ (Fig.~\ref{supfig_spectrumDdep}b), a gap opens and layer-polarizes the conduction and the valence bands. By also including the SOC, the bands are spin-split (shown by the colors in Fig.\,\ref{supfig_spectrumDdep}). In case of $u>0$ only the conduction band is spin-split (panel d) and if $u<0$ only the valence band is spin-split (shown by the colors in \ref{supfig_spectrumDdep}c).

To calculate the Landau level spectrum in large transverse magnetic field, we follow the footsteps of Ref.\,\cite{Khoo2017}. We neglect the $\gamma_3$ hopping and introduce the ladder operators as $\hat{a}=\frac{l_B}{2} (q_x + i q_y)$, $\hat{a}^\dagger=\frac{l_B}{2} (q_x - i q_y)$, which we write back to the original Hamiltonian. Here, $l_B$ is the magnetic length and $q_i=k_i -e/\hbar A_i$ with $A_i$ the vector potential. 

\begin{figure}[!ht]
\begin{center}
\includegraphics[width=.9\linewidth]{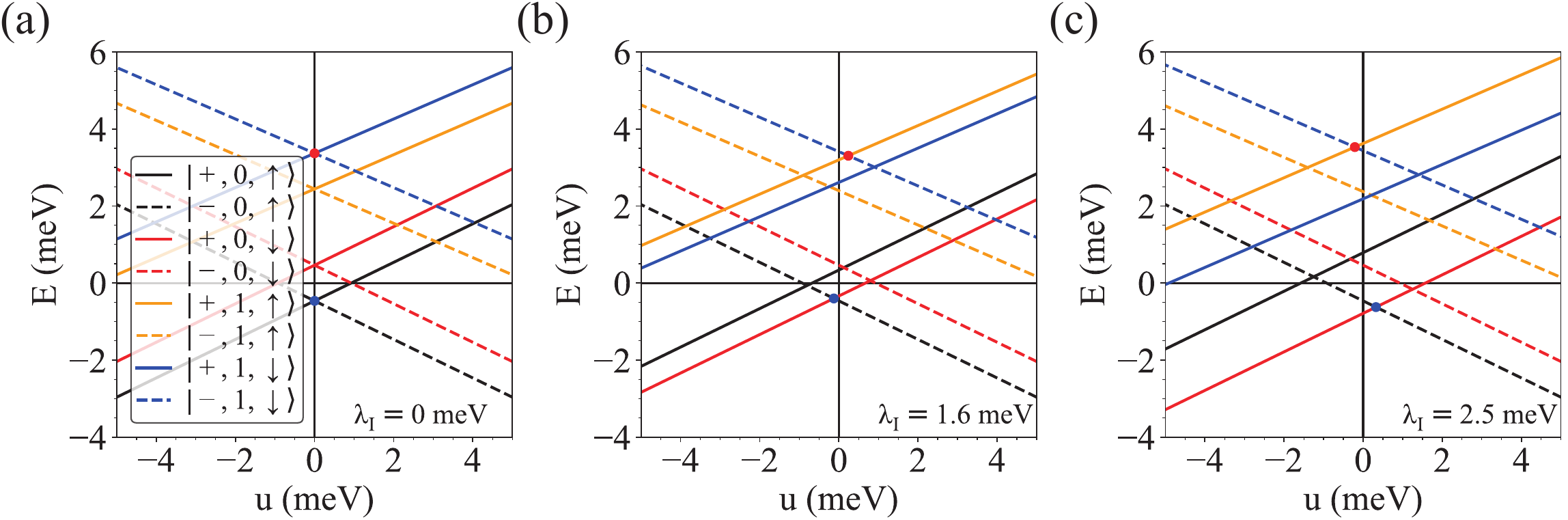}
\caption{The Landau level spectrum of BLG at $B=8$\,T with (a) $\lambda_I=0$, (b) $\lambda_I=1.6$\,meV and (c) $\lambda_I=2.5$\,meV. Only the Landau levels with index $+$ ($K$ valley, solid lines) depend on $\lambda_I$.}
\label{supfig_QHEspectrumB8T}
\end{center}
\end{figure}

The Landau levels' (LL) $u$ dependence is shown in Fig.~\ref{supfig_QHEspectrumB8T}. at fixed $B=8$\,T magnetic field. In case of $\lambda_I=0$ (Fig.~\ref{supfig_QHEspectrumB8T}a), the LLs are symmetric to $u$. The splittings at $u=0$ originates from $\gamma_4$, $\Delta'$ and from the Zeeman-splitting $E_Z=-g\mu_BB/2$, where $\mu_B$ is the Bohr magneton and $g=2$ is the Landé g-factor. The $\nu=\pm3$ crossing points, denoted with a red and a blue dot, are at zero. The SOC change their positions as the SOC acts only on the LLs in the $K$ valley ($+$) and these are localized to the bottom layer where we included the SOC. At fixed $B$ field the crossing points are non-monotonically depend on $\lambda_I$: it increases till the two LLs get degenerate eg. blue and yellow, then it decreases. 

\begin{figure}[!ht]
\begin{center}
\includegraphics[width=1\linewidth]{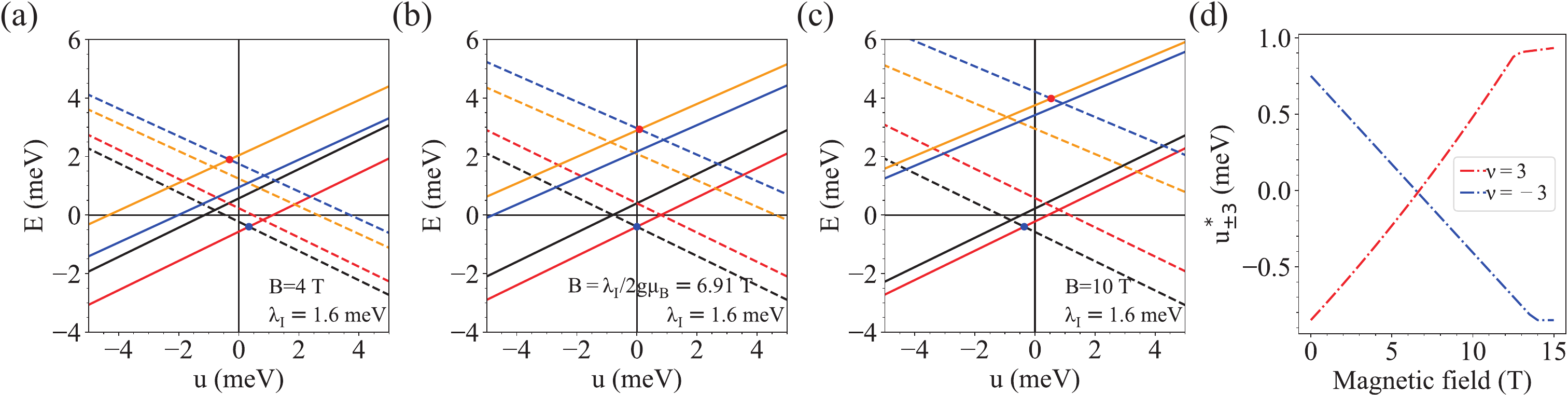}
\caption{The Landau level spectrum of BLG with $\lambda_I=1.6$\,meV at (a) $B=4$\,T (b) $B=6.91$\,T where $2g\mu_BB=\lambda_I$ and (c) B=10\,T. Here $\mu_B$ is the Bohr magneton and $g=2$ is the Landé g-factor. (d) $u_{\pm3}^*$ magnetic field dependence with $\lambda_I=1.6$\,meV.}
\label{supfig_QHEspectrum_Bdep}
\end{center}
\end{figure}

The evolution of the LLs with the magnetic field is illustrated in Fig.~\ref{supfig_QHEspectrum_Bdep}. When the Zeeman energy equals the Ising-type SOC $2g\mu_BB=\lambda_I/2$, the Zeeman splitting cancels the effective Zeeman splitting from the SOC. At lower magnetic fields the sign of the position of the $\nu=3$ crossing point $u_3^*$ is negative and at higher magnetic fields $u_3^*$ is positive.

\section{Measurement of $\nu=\pm3$ crossings}
\begin{figure}[!ht]
\begin{center}
\includegraphics[width=.7\linewidth]{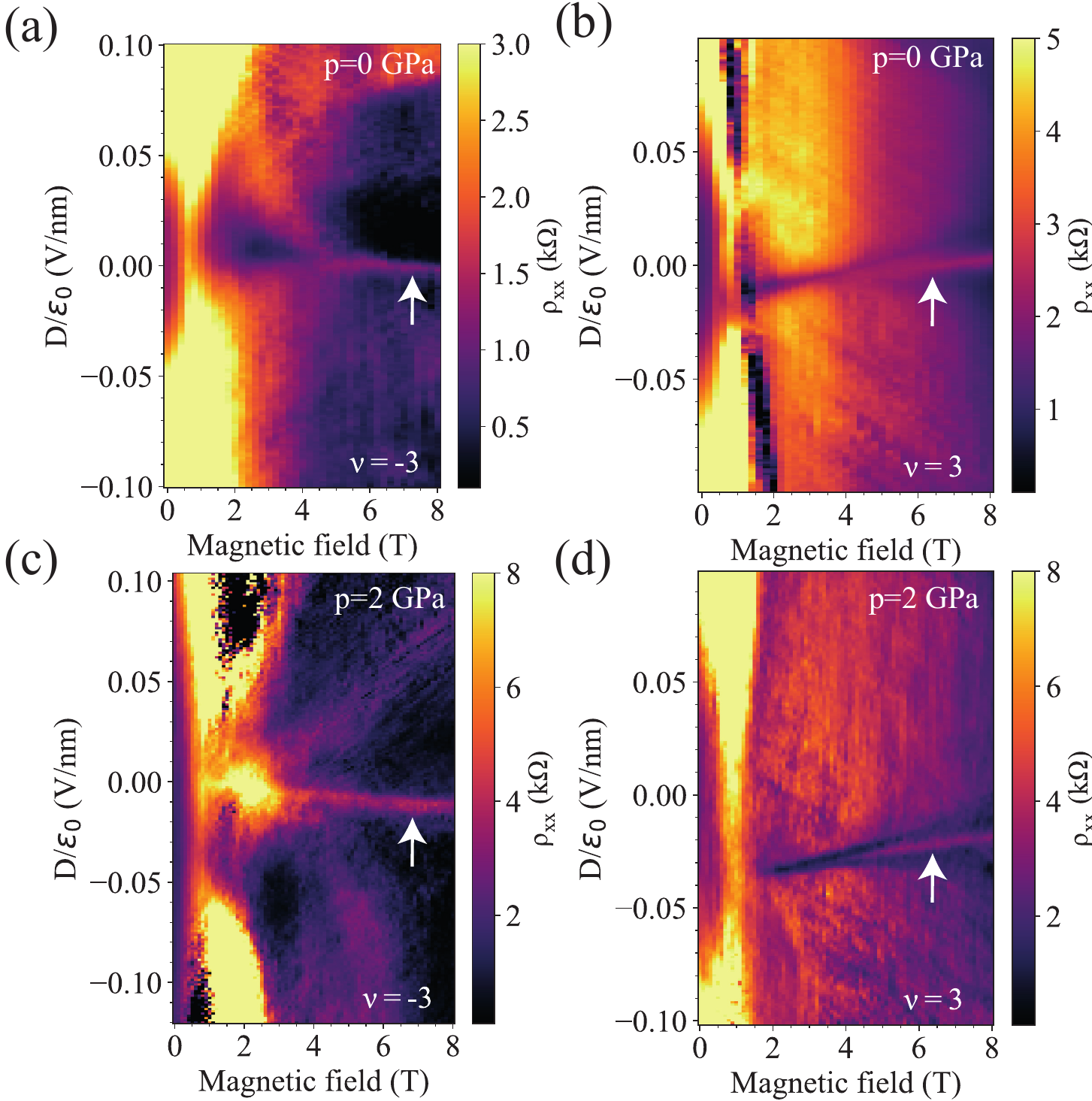}
\caption{Measurements of the $\nu=\pm3$ crossing points. $\rho_{xx}$ as a function of $B$ and $D$ while setting $n=\nu e/h|B|$ as $\nu$ fixed at $T=50$\,mK at $p=0$ (a-b) and at $p=2$\,GPa (c-d). The evolution of $\nu=-3$ is shown in (a) and (c) and the evolution of $\nu=3$ is shown in (b) and (d). The white arrows are guide to the eye to show the peak of the crossings.}
\label{supfig_nupm3}
\end{center}
\end{figure}

\begin{figure}[!ht]
\begin{center}
\includegraphics[width=.9\linewidth]{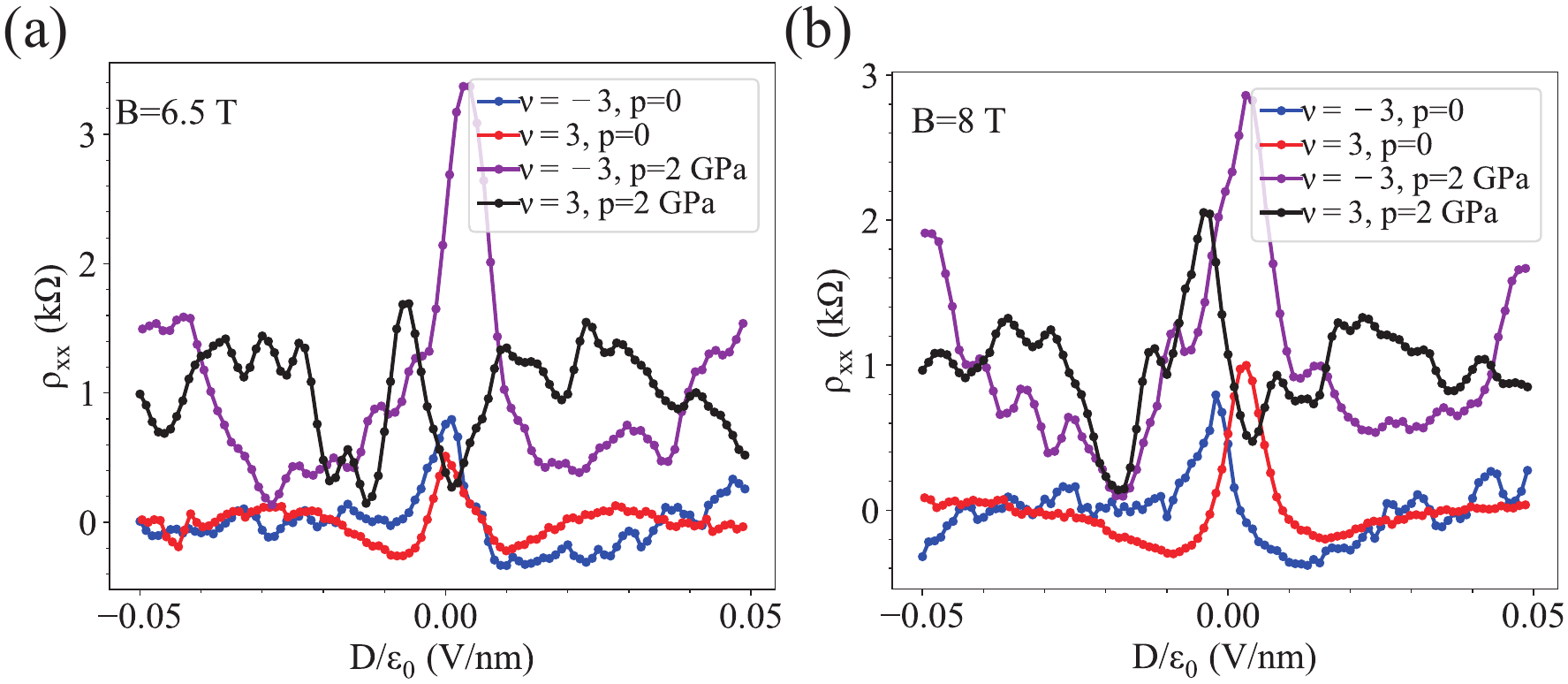}
\caption{Line-cuts of Fig.\,\ref{supfig_nupm3} along $D$. (a) $\rho_{xx}$ as a function of $D$ at $B=6.5$\,T along $\nu=\pm3$. (b) $\rho_{xx}$ as a function of $D$ at $B=8$\,T along $\nu=\pm3$}
\label{supfig_nu3_linecuts}
\end{center}
\end{figure}

We extract the position of the $\nu=\pm3$ crossing points in $D$ by two methods. On the one hand, we measure $\rho_{xx}$ maps as a function of $n$ and $D$ at a fixed magnetic field. Here, we take a line cut at $n=\nu e/h|B|$ and from the resistance peak we obtain the position of the crossing point. On the other hand, we perform measurements, where we measure maps as a function of $D$ and $B$ while keeping $\nu$ fixed and setting $n$ as $n=\nu e/h|B|$. These measurements are shown in Fig.~\ref{supfig_nupm3} for zero and 2 GPa pressures. Line cuts from  Fig.~\ref{supfig_nupm3} is shown in  Fig.~\ref{supfig_nu3_linecuts}. Here we obtain the crossing points from the peak in the resistance. For the error bars in the main text, the error from the uncertainty of the lever arms is less than $\Delta D/\epsilon_0=0.0015$\,V/nm the main contribution comes from the read-off error of the peaks. 

\section{Shubnikov\,--\,de Haas oscillations}
\begin{figure}[!htb]
\begin{center}
\includegraphics[width=.5\linewidth]{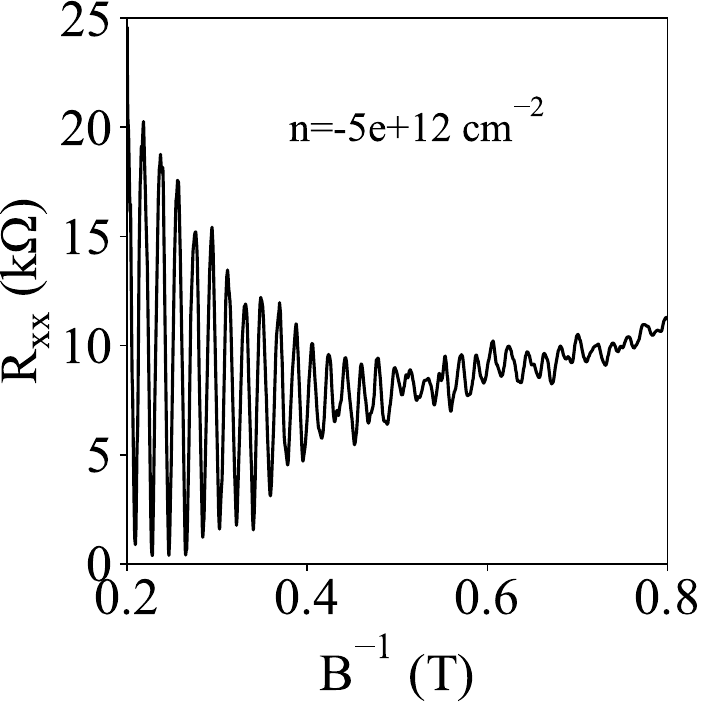}
\caption{Raw magnetoresistance data at $n=-5\cdot10^{12}$\,cm$^{-2}$, $D=0$, $T=1.4$\,K and $p=0$. }
\label{supfig_raw}
\end{center}
\end{figure}

\begin{figure}[!hbt]
\begin{center}
\includegraphics[width=.9\linewidth]{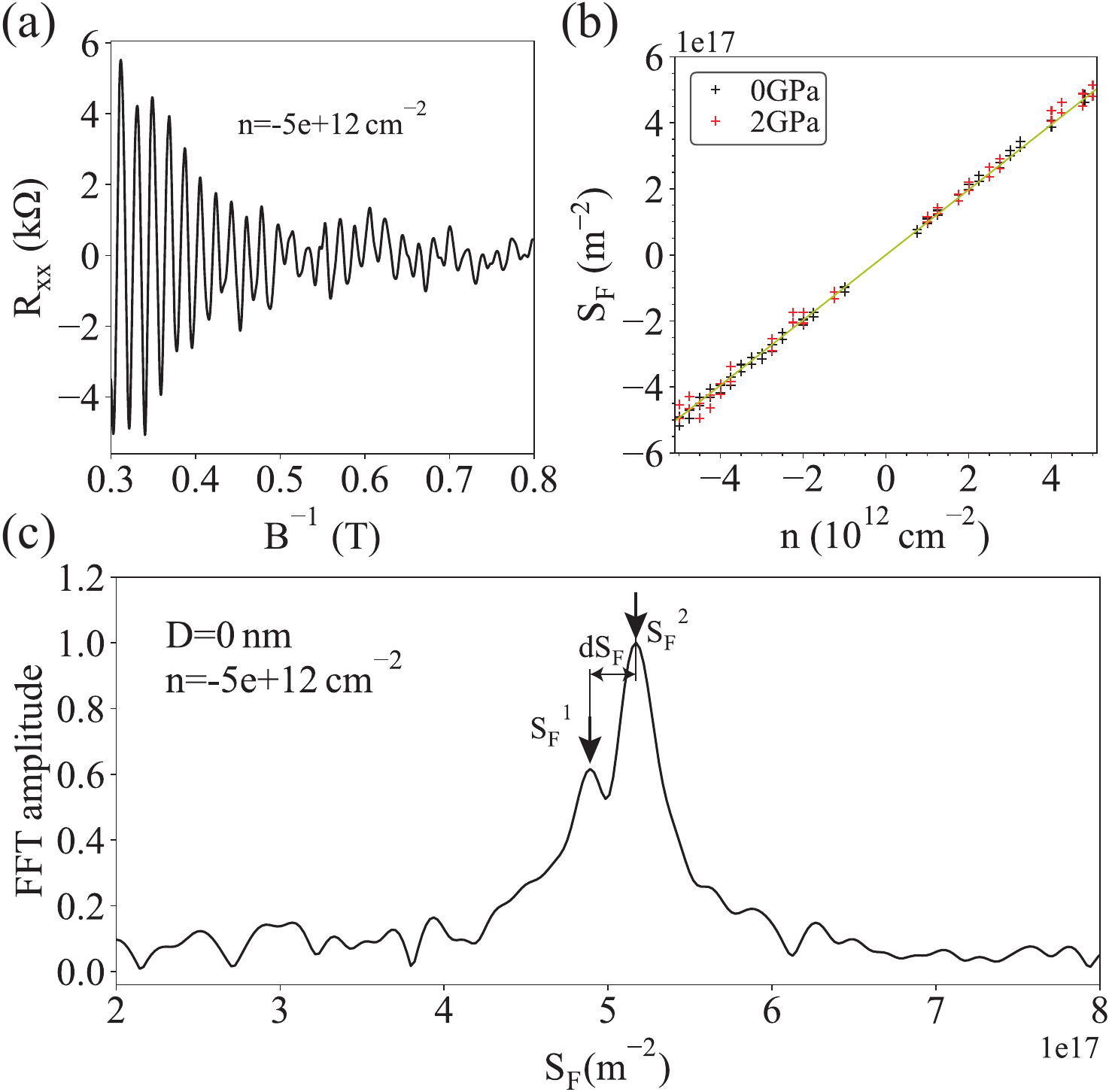}
\caption{Shubnikov\,--\,de Haas oscillations. (a) $\rho_{xx}$ after subtracting a quadratic background at $n=-5\cdot10^{12}$\,cm$^{-2}$ at $p=0$. (b) Fermi surface areas extracted from the Shubnikov\,--\,de Haas oscillations as a function of $n$. The black markers are measurements made at zero pressure and the red ones are made at $p=2$\,GPa. The yellow line would be the Fermi surface $S_F=\pi^2n$ if the spectrum would be spin degenerate. (c) Fourier-transform of (a) as a function of the Fermi-surface ($S_F$) converted from the frequency $f_B$ as $S_F=\frac{2\pi e}{\hbar}f_B$. The arrows shows the peaks ($S_F^{1,2}$) and their difference ($dS_F$) which is used to get $\lambda_R$. }
\label{supfig_fft}
\end{center}
\end{figure}
We measure Shubnikov\,--\,de Haas oscillations at 1.4\,K at different charge densities at $D=0$. We subtract a quadratic background and the result is depicted in Fig.\,\ref{supfig_fft}a and the raw data is plotted in Fig\,\ref{supfig_raw}. We take FFT on the signals and convert the obtained frequencies $f_B$ into Fermi-surface area as $S_F=\frac{2\pi e}{\hbar}f_B$. We extract the peaks shown in panel (c) with arrows. In Fig.\,\ref{supfig_fft}b we show the Fourier-transform of the SdH oscillations at different densities. At higher densities as the splitting is larger from the SOC, the peaks are more visible and separated. 
We fit the low-energy model with $\lambda_R$ as the only fitting parameter on the difference.

\clearpage
\section{Weak localization}
\begin{figure}[!ht]
\begin{center}
\includegraphics[width=.8\linewidth]{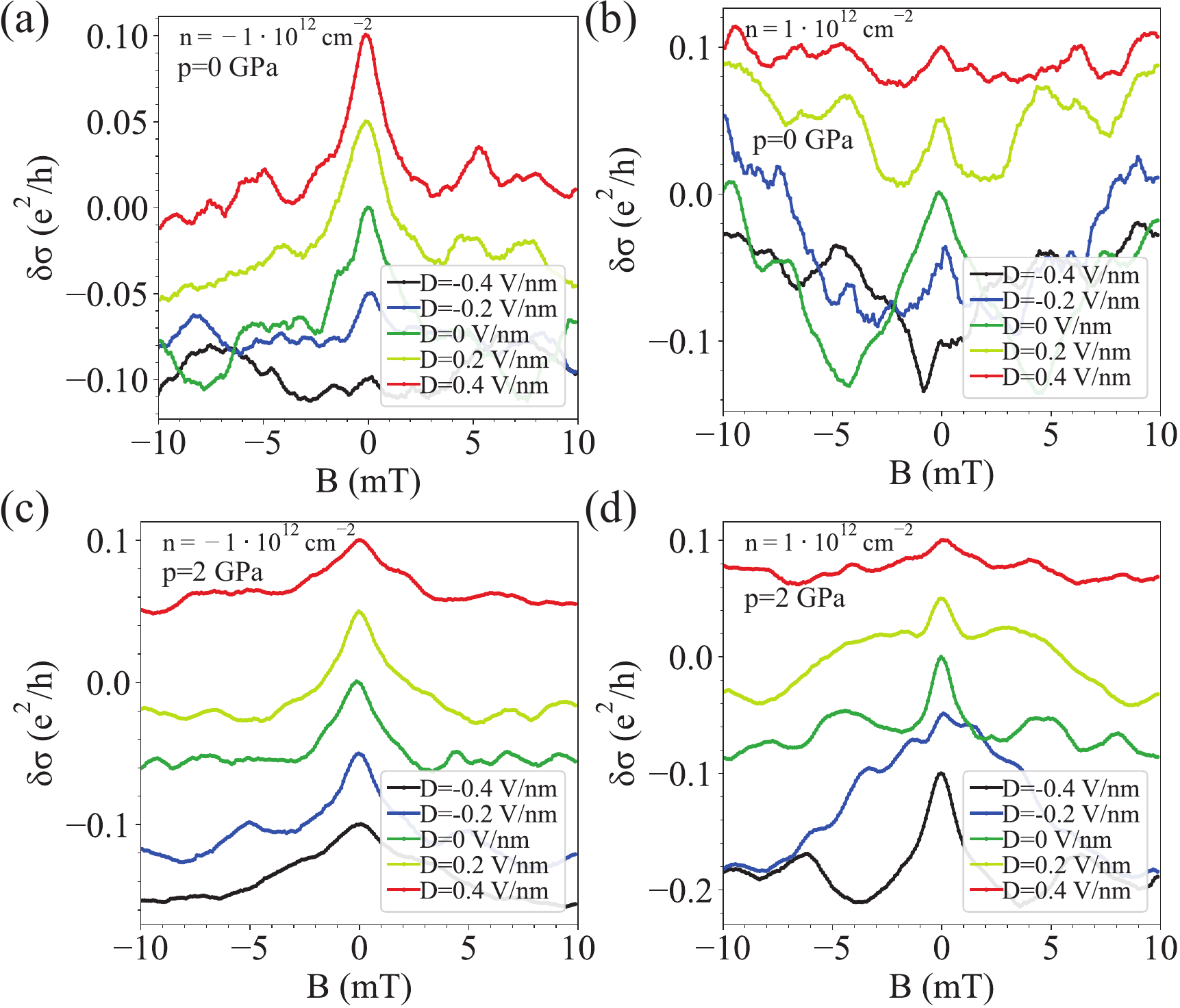}
\caption{Weak localization measurements at $n=\pm1\cdot10^{12}$\,cm$^{-2}$. (a-b) measurements at p=0\,GPa and (c-d) measurements at p=2\,GPa. Measurements made at different displacement fields at $T=1.5$\,K.}
\label{supfig_WAL}
\end{center}
\end{figure}

We performed weak localization measurements at $T=1.5$\,K at zero and at $p=2$\,GPa pressures. We average over 41 curves in a $2\cdot10^{11}$\,cm$^{-2}$ density range at fixed displacement fields. The magnetoconductivity $\delta\sigma=\sigma(B)-\sigma(B=0)$ at different $D$s and $p$s at $n=\pm1\cdot10^{12}$\,cm$^{-2}$ is shown in Fig.\,\ref{supfig_WAL}. Most of the curves show weak antilocalization (WAL) and the signal depends on $D$. The dependence is similar as in Ref.\cite{Amann2022} at ambient pressure. At $p=0$ at the hole side the WAL signal disappears by decreasing $D$, and at the electron side the signal decreases by increasing $|D|$. At $p=2$\,GPa at the hole side the signal decreases at large $|D|$. At the electron side the the WAL signal is decreasing by increasing $D$. Though this resembles the layer-tunable SOC effect, the tendency on a larger dataset is not that clear.

\clearpage
\section{Further devices}
\begin{figure}[!ht]
\begin{center}
\includegraphics[width=.8\linewidth]{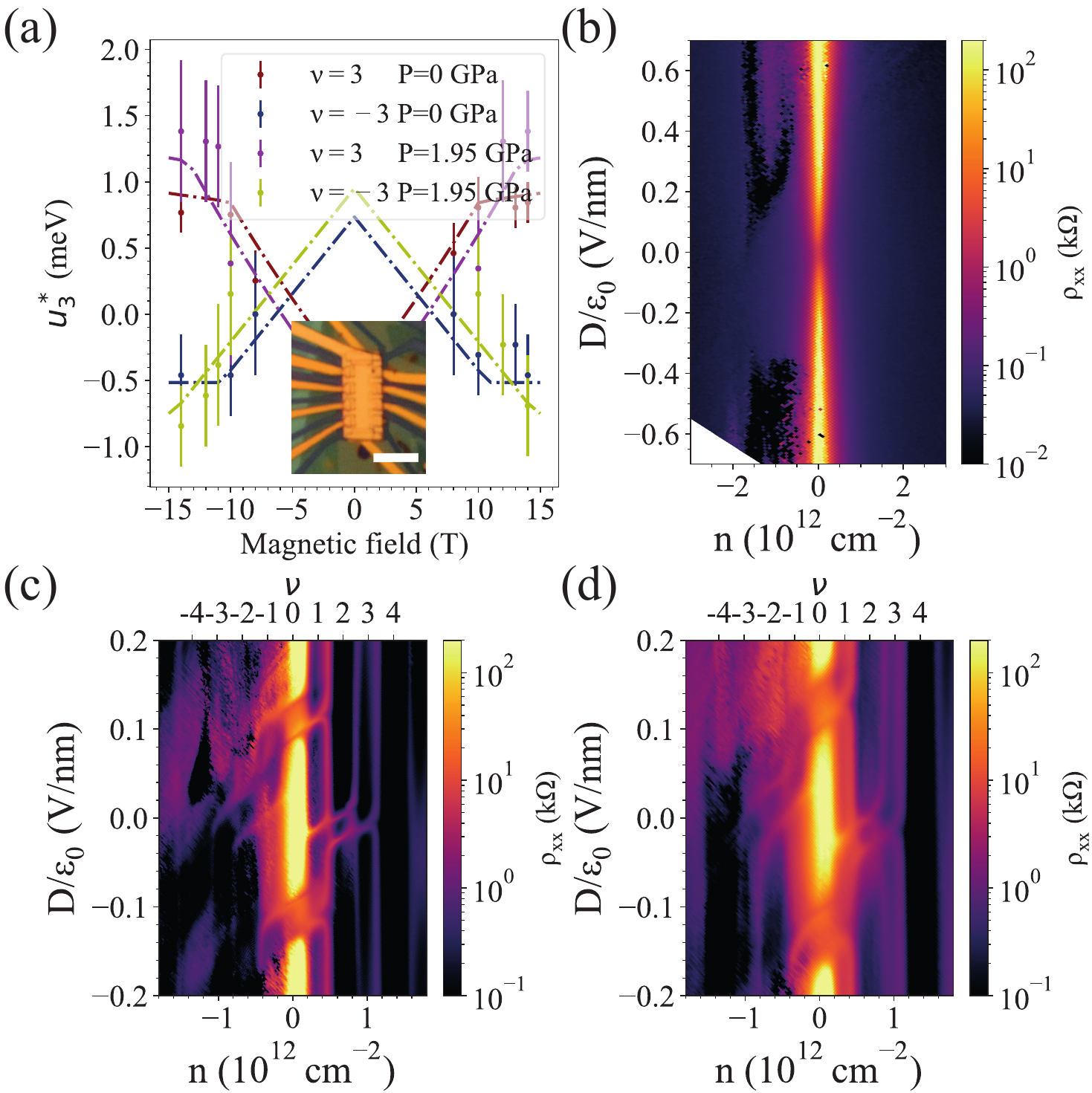}
\caption{Measurements on device B. (a) Extracted crossing points from the quantum Hall measurements at zero and at 1.95\,GPa. The dashed lines show the corresponding fits with $\lambda_I(p=0)=1.25$\,meV and $\lambda_I(p=1.95$\,GPa$)=1.7$\,meV respectively. Inset: optical microscope image of the sample. The scale bar is \SI{5}{\micro\metre}. (b) $\rho_{xx}$ as a function of $n$ and $D$ at $p=0$ at $T=4.2$\,K. (c) and (d) $\rho_{xx}$ as a function of $n$ and $D$ in $B=14\,T$ at $p=0$ and at $p=1.95$\,GPa respectively.  }
\label{supfig_szbst2}
\end{center}
\end{figure}

\begin{figure}[!ht]
\begin{center}
\includegraphics[width=.8\linewidth]{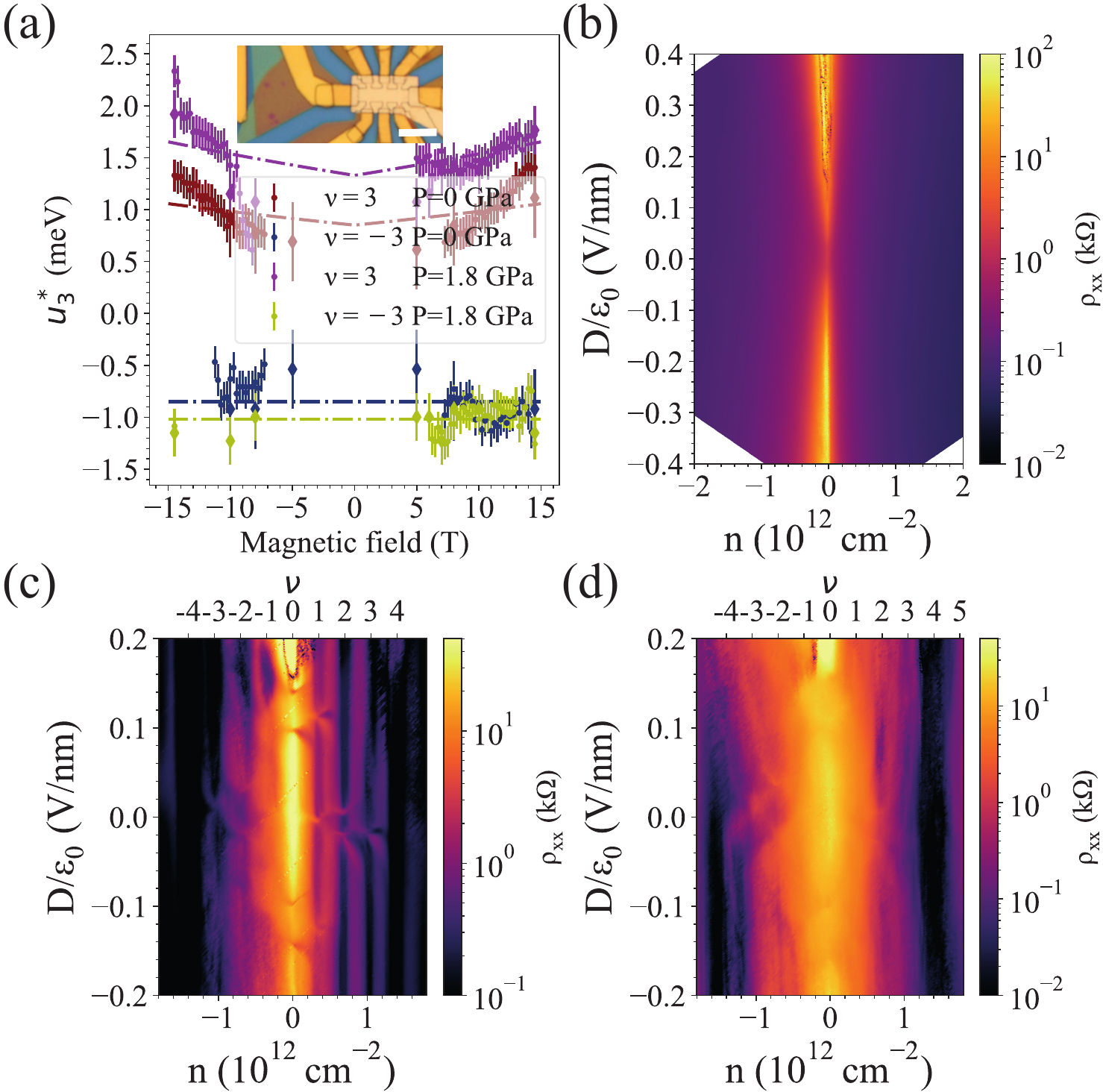}
\caption{Measurements on device C. (a) Extracted crossing points from the quantum Hall measurements at zero and at 1.8\,GPa. The dashed lines show the corresponding fits with $\lambda_I(p=0)=-1.7$\,meV and $\lambda_I(p=1.8$\,GPa$)=-2.35$\,meV respectively. Inset: optical microscope image of the sample. The scale bar is \SI{5}{\micro\metre}. (b) $\rho_{xx}$ as a function of $n$ and $D$ at $p=0$ at $T=4.2$\,K. (c) and (d) $\rho_{xx}$ as a function of $n$ and $D$ in $B=14.5\,T$ at $p=0$ and at $p=1.95$\,GPa, respectively.}
\label{supfig_szbst27}
\end{center}
\end{figure}

In Fig.\,\ref{supfig_szbst2}. and Fig.\,\ref{supfig_szbst27}. we show measurements on two more devices, labeled B and C. We performed similar measurements on these devices as on device A at $T=4.2$\,K. The $\rho_{xx}$ maps are similar and we use them to extract the crossing points to determine the Ising SOC strength. Here, during the fit we also include a sublattice asymmetry term $H_{asy}=u_{asy}\sigma_z$, which shifts all the  LLs with $u_{asy}$ without any mixing. This term take into account the on-site potential difference of the BLG. From the fitting on device B (see Fig.\,\ref{supfig_szbst2}) we obtain $\lambda_I(p=0)=1.25(10)$\,meV with $u_{asy}(p=0)=-0.22$\,meV and $\lambda_I(p=1.95$\,GPa$)=1.7(1)$\,meV with $u_{asy}(p=1.95$\,GPa$)=-0.42$\,meV. From the optical micrograph we estimate the twist angle as $\vartheta\sim0$° for device B.
For device C, we obtain $\lambda_I(p=0)=-1.7(1)$\,meV with $u_{asy}(p=0)=-0.03$\,meV and $\lambda_I(p=1.8$\,GPa$)=-2.35(10)$\,meV with $u_{asy}(p=1.8$\,GPa$)=-0.31$\,meV (see Fig.\,\ref{supfig_szbst27}). The increase of the SOC is similar and consistent with the obtained values of device A. The opposite sign is possible as it depends on the twist angle $\vartheta$ as $\lambda_I(\vartheta)=-\lambda_I(-\vartheta)$\cite{Naimer2021} and it was seen in previous measurements\cite{Wang2019}. From the optical micrograph we estimate the twist angle as $\vartheta\sim6$° for device C.

\begin{figure}[!ht]
\begin{center}
\includegraphics[width=.6\linewidth]{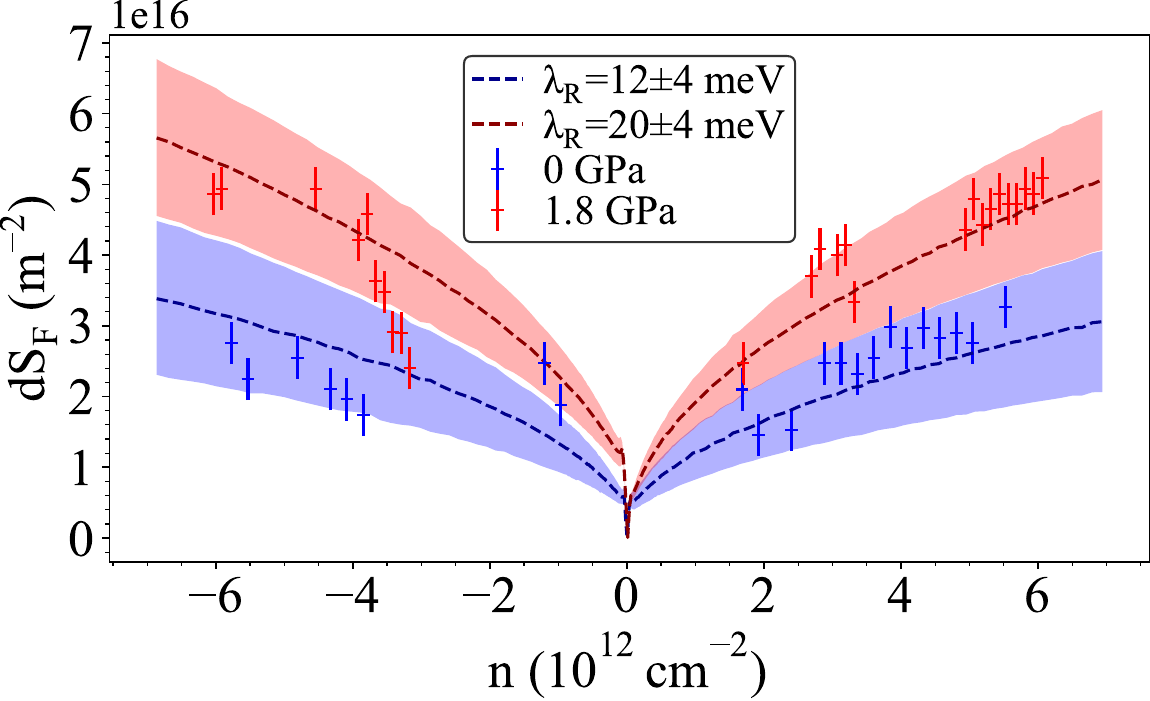}
\caption{The difference of the Fermi surfaces as a function of $n$ based on Shubnikov\,--\,de Haas oscillations on device C. The blue points are at $p=0$\,GPa and the red points are at $p=1.6$\,GPa. The dashed lines are the fitting of the model with the corresponding $\lambda_R$s. The colored ranges show the confidence interval of the fits.}
\label{supfig_szbst27_SdH}
\end{center}
\end{figure}

On device C we also performed Shubnikov\,--\,de Haas oscillation measurements and extracted $\lambda_R$ the same way as on device A: we subtracted a quadratic background from the magnetoresistance. We Fourier-transformed it as a function of $B^{-1}$. We converted the frequency peaks to Fermi-surface using $S_F=2\pi e f_B/\hbar$. We extracted the peaks which correspond to split Fermi-surfaces of the sample at a given $n$. We took their difference, which is plotted in Fig.\,\ref{supfig_szbst27_SdH}. We calculated the Fermi-surfaces from Eq. \ref{BLGHamiltonian}  with respect to $n$ and used $\lambda_R$ as the only fitting parameter to fit our data. We found that for this device $\lambda_R(p=0)=12\pm4$\,meV, which increased with pressure as $\lambda_R(p=1.8$\,GPa$)=20\pm4$\,meV, though the data quality is lower than for device A presented in the main text.

\newpage
\bibliography{Szentpeteri_et_al_2024_SupMat.bib}